\documentclass[12pt]{article}

\usepackage{graphicx}
\usepackage{amsmath}
\usepackage{tabularx}
\usepackage{lscape}

\begin{document}

\title{{\bf Leading vector coupling in baryon semileptonic decays in large-$N_c$ chiral perturbation theory revisited: Extension to decuplet baryons}}

\author{Rub\'en Flores-Mendieta \\
	{\it \normalsize Instituto de F{\'\i}sica, Universidad Aut\'onoma de San Luis Potos{\'\i}} \\
	{\it \normalsize \'Alvaro Obreg\'on 64, Zona Centro, San Luis Potos{\'\i}, 78000, San Luis Potos{\'\i}, M\'exico}
        \and
        Guillermo S\'anchez-Almanza \\
	{\it \normalsize Instituto de F{\'\i}sica, Universidad Aut\'onoma de San Luis Potos{\'\i}} \\
	{\it \normalsize \'Alvaro Obreg\'on 64, Zona Centro, San Luis Potos{\'\i}, 78000, San Luis Potos{\'\i}, M\'exico}
	}

\maketitle

\begin{abstract}
The leading vector coupling defined in the semileptonic decay of a baryon is revisited in the framework of large-$N_c$ baryon chiral perturbation. Aspects of $SU(3)$ flavor symmetry breaking are taken into account in two ways: Explicitly through perturbative symmetry breaking and implicitly through the integrals occurring in the one-loop corrections. The analysis of perturbative breaking is performed by using flavor projection operators which are exceptionally helpful in understanding previously unidentified issues. The approach is extended to theoretically evaluate the leading vector coupling in the transition $\Omega^-\to {\Xi^*}^0$. A fitting procedure to data reveals that second-order symmetry breaking corrections, while small, can have a significant impact on the evaluation of the baryon vector coupling, potentially increasing its overall strength due to the involved interplay of both explicit and implicit contributions.
\end{abstract}

\maketitle

\section{Introduction}

The {\it eightfold way} proposed independently by Gell-Mann and Ne'eman \cite{gell} in 1961 predicted the existence of the $\Omega^-$ baryon, which was a pivotal missing piece in a then-emerging classification scheme for elementary particles based on the $SU(3)$ flavor symmetry. It was proposed that this baryon would have a spin of $3/2$, a charge of $-1$, and a mass around $1.672\,\mathrm{GeV}$. The discovery of $\Omega^-$ in 1964 with the predicted properties represented a milestone in particle physics due to its confirmation of the quark model and the underlying $SU(3)$ flavor symmetry of particle interactions.

The $\Omega^-$ baryon decays via the weak interaction. The experimental information systematized in the {\it Review of Particle Physics} \cite{part} points that the three dominant decays are the non-leptonic ones $\Omega^-\to\Lambda K^-$ $(67.7\%)$, $\Omega^-\to\Xi^0 \pi^-$ $(24.3\%)$, and $\Omega^-\to\Xi^-\pi^0$ $(8.55\%)$. Rare decays comprise $\Omega^-\to \Xi^-\pi^+\pi^-$, $\Omega^-\to \Xi(1530)^0\pi^-$, $\Omega^-\to\Xi^0e^-\overline{\nu}_e$, and $\Omega^-\to \Xi^-\gamma$ with rather small branching ratios reported.

While important progress has been achieved in the analysis of non-leptonic $\Omega^-$ decays in both the phenomenological \cite{oh1,tan1} and experimental bent \cite{abli1,bou1}, rare three-body decays have been a subject of recent theoretical research to further refine our understanding of the weak interaction \cite{leu1,leu2}. Oppositely, experimental information for these rare decays is rather scarce.

The recent work by Bertilsson and Leupold \cite{leu2} draws attention to creating a research program to conduct Wu-type experiments, {\it i.e.}, $B_1\to B_2 \ell^-\overline{\nu}_\ell$, where $B_1$ and $B_2$ stand for spin-$3/2$ baryon states. Examples of energetically allowed processes are $\Delta^0 \to \Delta^+e^-\overline{\nu}_e$ and $\Omega^- \to {\Xi^*}^0\ell^-\overline{\nu}_\ell$. According to the authors, while the former process is practically inaccessible due to the fact that the $\Delta$ baryons are short-lived, the latter process, in contrast, is feasible to be within the reach of experiments like BESIII or LHCb in the near future. In order to reiterate the main message of their work, the authors conclude, through a simple leading-order calculation in baryon chiral perturbation theory, that if the branching ratio could be determined with a precision of around $10\%$, it would be possible to estimate the low-energy constant which provides the strength of the coupling of decuplet states to pseudo scalar mesons.

Motivated by the aforementioned facts, the leading axial vector couplings in the semileptonic decays $\Omega^- \to \Xi^0\ell^-\overline{\nu}_\ell$ and $\Omega^- \to {\Xi^*}^0\ell^-\overline{\nu}_\ell$ (among other axial couplings) were recently discussed in Ref.~\cite{rfm25} in the context of a computational scheme that combines $1/N_c$ and chiral corrections, hereafter loosely referred to as the combined formalism. Here, $N_c$ will denote the number of color charges. The combined formalism has been used to extract low-energy consequences of Quantum Chromodynamics (QCD) \cite{djm94,djm95} in a more effective way than either method separately. The pertinent $1/N_c$ chiral Lagrangian was introduced in Ref.~\cite{jen96} and a brief account of the relevant baryon static properties evaluated using this Lagrangian can be found in Ref.~\cite{rfm25}.

The leading vector coupling relevant in the semileptonic decays of octet baryons also contain information about hadron dynamics. Lots of efforts and a considerable number of methods have been devoted to study them. A selection of such methods is constituted by the $1/N_c$ expansion \cite{dai,rfm98,rfm04,rfm17}, chiral perturbation theory \cite{kra,and,villa,meiss,geng}, the combined formalism \cite{rfm14,fer}, the fast-growing lattice QCD \cite{gua,sas1,sas2,sas3}, to name but a few.

In the present analysis the leading vector coupling will be reexamined in the context of the combined formalism, with further applications to the semileptonic decay $\Omega^- \to {\Xi^*}^0\ell^-\overline{\nu}_\ell$. Including decuplet baryons into play offers an opportunity to scrutinize some aspects of the theory to advance into a new domain of understanding not discussed previously \cite{rfm98,rfm04,rfm17,rfm25}. The strategy adopted in order to reach the goal is based on the following two main steps: 1) The construction of the most general baryon operators containing first- and second-order perturbative $SU(3)$ flavor symmetry breaking (SB) corrections to the vector current operator \cite{rfm98,rfm04,rfm17}. The important properties exploited along this step are the non-renormalization of the baryon electric charge and the conserved vector current (CVC) hypothesis,\footnote{The weak vector current is assumed to be conserved with a universal coupling constant in an analogy to the electromagnetic vector current. This assumption is the CVC hypothesis.} which are the cornerstones of the Ademollo-Gatto theorem \cite{bs,ag}. The extensive use of flavor projection operators \cite{banda1,banda2} offers the uttermost assistance to classify $SU(3)$ flavor representations and identify those which yield nonvanishing corrections. 2) The evaluation of the vector current operator at one-loop order in large-$N_c$ baryon chiral perturbation theory. Loop graphs with octet and decuplet intermediate states are systematically incorporated along with the effects of the decuplet-octet mass difference, giving the full result to order $\mathcal{O}(p^2)$ in the chiral expansion \cite{rfm14}.

The plan of the paper is as follows. Section \ref{sec:sul} presents some elementary aspects of large-$N_c$ chiral perturbation theory in order to introduce notation and conventions. The most general $1/N_c$ expansion for an arbitrary operator sharing the same spin-flavor transformation properties with the baryon vector current in the $SU(3)$ symmetric case is first given. Next, the corresponding $1/N_c$ expansions for operators containing first- and second-order symmetry breaking effects are constructed with the help of flavor projection operators. Section \ref{sec:vc} focuses on applying the results of the preceding section specifically to the baryon vector operator. The non-renormalization of the baryon electric charge and the CVC hypothesis are employed to reach the conclusion determined by Ademollo and Gatto, which in turn allows one to identify non trivial contributions. Section \ref{sec:one} evaluates one-loop corrections to the leading vector coupling for the decay $\Omega^- \to {\Xi^*}^0\ell^-\overline{\nu}_\ell$. Section \ref{sec:num} provides a preliminary numerical analysis through a least-squares fit to the available data; salient factors are discussed. Section \ref{sec:cr} lists a few closing remarks. The article is complemented by three appendices. Appendix \ref{app:pab} shows explicit reductions of flavor projection operators acting on the operator basis defining first-order symmetry breaking. Appendix \ref{app:oabc} lists the operator basis defining second-order symmetry breaking whereas Appendix \ref{app:q1q2q3} lists the operator structures associated to strangeness-conserving and strangeness-changing processes. Supplementary material is also attached to this article where analytical expressions for baryon charges and vector couplings, along with matrix elements are given.

\section{\label{sec:sul}The $1/N_c$ expansion for a $(0,\mathbf{8})$ operator}

The baryon sector of QCD has a contracted $SU(2N_f)$ symmetry, where $N_f$ is the number of light quark flavors \cite{dm1,dm2,gs1,gs2}. Under the decomposition $SU(2N_f) \to SU(2) \otimes SU(N_f)$, the spin-flavor representation yields a tower of baryon flavor representations with spins $J=1/2,3/2,\ldots,N_c/2$ \cite{djm95,gs1}. The $SU(2N_f)$ spin-flavor generators, namely, the spin generators $J^k$, the flavor generators $T^c$, and the spin-flavor generator $G^{kc}$, can be written as 1-body quark operators acting on the $N_c$-quark baryon states,
\label{eq:su6gen}
\begin{eqnarray}
\begin{array}{llr}
\displaystyle J^k = \sum_\alpha^{N_c} q_\alpha^\dagger \left(\frac{\sigma^k}{2}\otimes {\mathcal I} \right) q_\alpha, & \qquad & (1,1) \\
\displaystyle T^c = \sum_\alpha^{N_c} q_\alpha^\dagger \left({\mathcal I} \otimes \frac{\lambda^c}{2} \right) q_\alpha, & & (0,\mathrm{adj}) \\
\displaystyle G^{kc} = \sum_\alpha^{N_c} q_\alpha^\dagger \left(\frac{\sigma^k}{2}\otimes \frac{\lambda^c}{2} \right) q_\alpha. & & (1,\mathrm{adj})
\end{array}
\end{eqnarray}
Here $q_\alpha^\dagger$ and $q_\alpha$ represent quark creation and annihilation operators, where $\alpha=1,\ldots,N_f$ denote the $N_f$ quark flavors with spin up and $\alpha=N_f+1,\ldots,2N_f$ the $N_f$ quark flavors with spin down, and $(j,\mathrm{dim})$ indicates the transformation properties under $SU(2) \otimes SU(N_f)$, where $\mathrm{dim}$ denotes the dimension of the $SU(N_f)$ representation. Hereafter, $N_f$ will be set to 3.

Baryon operators of interest here possess spin 0 and transform in the adjoint (octet) representation of $SU(3)$. In order to construct the $1/N_c$ expansion for an operator transforming as $(0,\mathbf{8})$ under $SU(2)\otimes SU(3)$ in a rigorous way, the operator basis $R^{(ij)(abc)}$ introduced in Ref.~\cite{banda2} becomes handy. $R^{(ij)(abc)}$ represents a complete basis of linearly independent $3$-body operators with spin 2 and three flavor indices. After contracting $R^{(ij)(abc)}$ with $\delta^{ij}$ and $\delta^{ab}$ to saturate appropriate spin and flavor indices and eliminating irrelevant constant factors, only three operators prevail, namely,
\begin{subequations}
\label{eq:oc}
\begin{eqnarray}
O_1^c & = & T^c, \\
O_2^c & = & \{J^r,G^{rc}\}, \\
O_3^c & = & \{J^2,T^c\}.
\end{eqnarray}
\end{subequations}
The operators $O_n^c$ listed in Eq.~(\ref{eq:oc}) are $T$-even. A naive large-$N_c$ power counting indicates that $O_n^c$ is of order $\mathcal{O}({N_c^n})$.

The $1/N_c$ expansion for a $(0,\mathbf{8})$ operator, retaining up to $3$-body operators, can therefore be written as
\begin{eqnarray}
O^c & = & \sum_{n=1}^3 \frac{1}{N_c^{n-1}} v_n O_n^c \nonumber \\
& = & v_1 T^c + \frac{1}{N_c} v_2 \{J^r,G^{rc}\} + \frac{1}{N_c^2} v_3 \{J^2,T^c\}, \label{eq:vc}
\end{eqnarray}
which is at most of order $\mathcal{O}(N_c)$. Although the operator coefficients $v_n$ are unknown, they are expected to be of order $\mathcal{O}(N_c^0)$.

\subsection{First-order SB to a $(0,\mathbf{8})$ operator}

An arbitrary baryon operator $O^c$ transforming as $(0,\mathbf{8})$ under $SU(2)\otimes SU(3)$ possesses a $1/N_c$ expansion of the form (\ref{eq:vc}). Flavor $SU(3)$ breaking in QCD is due to the light quark masses and transforms as a flavor octet \cite{djm95}. The $SU(3)$ SB correction to an operator $O^c$ is computed to linear order from the tensor product of $O^c$ and the perturbation. The tensor product $\mathbf{8}\otimes \mathbf{8}$ decomposes into a symmetric and an antisymmetric product, $(\mathbf{8}\otimes \mathbf{8})_S$ and $(\mathbf{8}\otimes \mathbf{8})_A$, respectively, which can be written as \cite{djm95}
\begin{subequations}
\label{eq:8x8}
\begin{eqnarray}
& & (\mathbf{8}\otimes \mathbf{8})_S = \mathbf{1} + \mathbf{8} + \mathbf{27}, \\
& & (\mathbf{8}\otimes \mathbf{8})_A = \mathbf{8} + \mathbf{10}+\overline{\mathbf{10}}.
\end{eqnarray}
\end{subequations}

The subsequent step is the construction of the $1/N_c$ expansions of the operators transforming in the $(0,\mathbf{1})$, $(0,\mathbf{8})$, $(0,\mathbf{10}+\overline{\mathbf{10}})$, and $(0,\mathbf{27})$ spin-flavor representations. This task facilitates considerably by employing the spin and flavor projection operator technique introduced in Ref.~\cite{banda1,banda2}, which builds on the decomposition of the tensor space formed by the product of the adjoint space with itself $n$ times, $\prod_{i=1}^n adj \otimes$, into subspaces labeled by a specific eigenvalue of the quadratic Casimir operator $C$ of $SU(3)$. The projection operators $\mathcal{P}^{(m)}$ constructed for each subspace read,
\begin{equation}
\label{eq:general_proj}
\mathcal{P}^{(m)} = \prod_{i=1}^k \left[ \frac{C - c_{n_i}{\mathcal I}}{c_{m} - c_{n_i}} \right], \qquad \qquad c_m \neq c_{n_i},
\end{equation}
where $k$ labels the number of different possible eigenvalues for $C$ and $c_{m}$ are its eigenvalues. Further details can be found in the original reference.

Thus, for the product of two $SU(3)$ adjoints, the flavor projectors $[\tilde{\mathcal{P}}^{(\mathrm{dim})}]^{a_1a_2a_3a_4}$ for the irreducible representation of dimension $\mathrm{dim}$ contained in (\ref{eq:8x8}) read \cite{banda1},
\begin{equation}
[\tilde{\mathcal{P}}^{(1)}]^{a_1a_2a_3a_4} = \frac{1}{N_f^2-1} \delta^{a_1a_2} \delta^{a_3a_4}, \label{eq:p1ab}
\end{equation}
\begin{equation}
[\tilde{\mathcal{P}}^{(8)}]^{a_1a_2a_3a_4} = \frac{N_f}{N_f^2-4} d^{a_1a_2e_1} d^{a_3a_4e_1}, \label{eq:p8ab}
\end{equation}
\begin{equation}
[\tilde{\mathcal{P}}^{(8_A)}]^{a_1a_2a_3a_4} = \frac{1}{N_f} f^{a_1a_2e_1} f^{a_3a_4e_1}, \label{eq:p8aab}
\end{equation}
\begin{equation}
[\tilde{\mathcal{P}}^{(10+\overline{10})}]^{a_1a_2a_3a_4} = \frac12 (\delta^{a_1a_3} \delta^{a_2a_4} - \delta^{a_2a_3} \delta^{a_1a_4}) - \frac{1}{N_f} f^{a_1a_2e_1} f^{a_3a_4e_1}, \label{eq:p10ab}
\end{equation}
and
\begin{equation}
[\tilde{\mathcal{P}}^{(27)}]^{a_1a_2a_3a_4} = \frac12 (\delta^{a_1a_3} \delta^{a_2a_4} + \delta^{a_2a_3} \delta^{a_1a_4}) - \frac{1}{N_f^2-1} \delta^{a_1a_2} \delta^{a_3a_4} - \frac{N_f}{N_f^2-4} d^{a_1a_2e_1} d^{a_3a_4e_1}, \label{eq:p27ab}
\end{equation}
where
\begin{equation}
[\tilde{\mathcal{P}}^{(1)} + \tilde{\mathcal{P}}^{(8)} + \tilde{\mathcal{P}}^{(8_A)} + \tilde{\mathcal{P}}^{(10+\overline{10})} + \tilde{\mathcal{P}}^{(27)}]^{a_1a_2a_3a_4} = \delta^{a_1a_3} \delta^{a_2a_4},
\end{equation}
which is a property satisfied by ordinary projection operators.

The next step is the construction of a complete operator basis of linearly independent $3$-body operators with spin 0 and two flavor indices, one of which will be set to 8 to account for first-order SB. Let $O^{ab}$ denote such basis. Starting from the $R^{(ij)(abc)}$ basis described in Sec.~\ref{sec:sul}, $O^{ab}$ can also be straightforwardly obtained with suitable contractions with $\delta^{ij}$ and either $d^{abe}$ or $if^{abe}$. In either case, ignoring irrelevant constant factors, the resultant basis is
\begin{equation}
\label{eq:onab}
O^{ab} = \{O_n^{ab}\},
\end{equation}
where
\begin{eqnarray}
\label{eq:oab}
\begin{array}{lcl}
O_1^{ab} = \delta^{ab}, & \qquad &
O_2^{ab} = i f^{abe} T^e, \\
O_3^{ab} = d^{abe} T^e, & \qquad &
O_4^{ab} = \delta^{ab} J^2, \\
O_5^{ab} = \{T^a,T^b\}, & \qquad &
O_6^{ab} = \{G^{ra},G^{rb}\}, \\
O_7^{ab} = i f^{abe} \{J^r,G^{re}\}, & \qquad &
O_8^{ab} = d^{abe} \{J^r,G^{re}\}, \\
O_9^{ab} = \{T^a,\{J^r,G^{rb}\}\}, & \qquad &
O_{10}^{ab} = \{T^b,\{J^r,G^{ra}\}\}, \\
O_{11}^{ab} = i f^{abe} \{J^2,T^e\}, & \qquad &
O_{12}^{ab} = d^{abe} \{J^2,T^e\}.
\end{array}
\end{eqnarray}
The operators $O_{n}^{ab}$ listed in Eq.~(\ref{eq:oab}) constitute a set of linearly independent operators. Except for $O_2^{ab}$, $O_7^{ab}$, and $O_{11}^{ab}$, which are $T$-odd operators, the remaining ones are $T$-even.

The flavor projection operator $[\tilde{\mathcal{P}}^{(\mathrm{dim})}]^{a_1a_2b_1b_2}$ applied to an operator $O_n^{b_1b_2}$, {\it i.e.} $[\tilde{\mathcal{P}}^{(\mathrm{dim})} O_n]^{a_1a_2}$, will effectively project out the component of $O_n^{a_1a_2}$ falling into representation $\mathrm{dim}$. The full analytic evaluations of $[\tilde{\mathcal{P}}^{(\mathrm{dim})} O_n]^{a_1a_2}$ are listed in Appendix \ref{app:pab}. Therefore, from the operator basis $O^{ab}$, the resultant operators falling into the flavor representations indicated in decomposition (\ref{eq:8x8}) read \\

\noindent
$\mathbf{1}$ flavor representation
\begin{equation}
S_1^{ab(1)} = \delta^{ab},
\end{equation}
\begin{equation}
S_2^{ab(1)} = \delta^{ab} J^2,
\end{equation}

\noindent
$\mathbf{8}$ flavor representation
\begin{equation}
S_1^{ab(8)} = d^{abe} T^e,
\end{equation}
\begin{equation}
S_2^{ab(8)} = d^{abe} \{J^r,G^{re}\},
\end{equation}
\begin{equation}
S_3^{ab(8)} = d^{abe} \{J^2,T^e\},
\end{equation}

\noindent
$\mathbf{27}$ flavor representation
\begin{eqnarray}
S_1^{ab(27)} & = & \{T^a,T^b\} - \frac{N_c(N_c+2N_f)(N_f-2)}{2N_f(N_f^2-1)} \delta^{ab} - \frac{2}{N_f^2-1} \delta^{ab} J^2 - \frac{(N_c+N_f)(N_f-4)}{N_f^2-4} d^{abe} T^e \nonumber \\
& & \mbox{} - \frac{2N_f}{N_f^2-4} d^{abe} \{J^r,G^{re}\},
\end{eqnarray}
\begin{eqnarray}
S_2^{ab(27)} & = & \{G^{ra},G^{rb}\} - \frac38 \frac{N_c(N_c+2N_f)}{N_f^2-1} \delta^{ab} + \frac{N_f+2}{2N_f(N_f^2-1)} \delta^{ab} J^2 - \frac34 \frac{(N_c+N_f)N_f}{N_f^2-4} d^{abe} T^e \nonumber \\
& & \mbox{} + \frac12 \frac{N_f+4}{N_f^2-4} d^{abe} \{J^r,G^{re}\},
\end{eqnarray}
\begin{eqnarray}
S_3^{ab(27)} & = & \frac12 \{T^a,\{J^r,G^{rb}\}\} + \frac12 \{T^b,\{J^r,G^{ra}\}\} - \frac{2(N_c+N_f)}{N_f(N_f+1)} \delta^{ab} J^2 - \frac{N_c+N_f}{N_f+2} d^{abe} \{J^r,G^{re}\} \nonumber \\
& & \mbox{} - \frac{1}{N_f+2} d^{abe} \{J^2,T^e\}.
\end{eqnarray}

\noindent
$\mathbf{8}_A$ flavor representation
\begin{equation}
S_1^{ab(8_A)} = i f^{abe} T^e,
\end{equation}
\begin{equation}
S_2^{ab(8_A)} = i f^{abe} \{J^r,G^{re}\},
\end{equation}
\begin{equation}
S_3^{ab(8_A)} = i f^{abe} \{J^2,T^e\},
\end{equation}

\noindent
$\mathbf{10}+\overline{\mathbf{10}}$ flavor representation
\begin{equation}
S_1^{ab(10+\overline{10})} = \frac12 \{T^a,\{J^r,G^{rb}\}\} - \frac12 \{T^b,\{J^r,G^{ra}\}\}.
\end{equation}

The corresponding $1/N_c$ expansion for each representation can be easily constructed as
\begin{equation}
N_c \tilde{c}_1^{(1)} S_1^{ab(1)} + \frac{1}{N_c} \tilde{c}_2^{(1)} S_2^{ab(1)}, \label{eq:sb1}
\end{equation}
\begin{equation}
\tilde{c}_1^{(8)} S_1^{ab(8)} + \frac{1}{N_c} \tilde{c}_2^{(8)} S_2^{ab(8)} + \frac{1}{N_c^2} \tilde{c}_3^{(8)} S_3^{ab(8)},
\end{equation}
\begin{equation}
\frac{1}{N_c} \tilde{c}_1^{(27)} S_1^{ab(27)} + \frac{1}{N_c} \tilde{c}_2^{(27)} S_2^{ab(27)} + \frac{1}{N_c^2} \tilde{c}_3^{(27)} S_3^{ab(27)},
\end{equation}
\begin{equation}
\tilde{c}_1^{(8_A)} S_1^{ab(8_A)} + \frac{1}{N_c} \tilde{c}_2^{(8_A)} S_2^{ab(8_A)} + \frac{1}{N_c^2} \tilde{c}_3^{(8_A)} S_3^{ab(8_A)},
\end{equation}
\begin{equation}
\frac{1}{N_c^2} \tilde{c}_1^{(10+\overline{10})} S_1^{ab(10+\overline{10})}. \label{eq:sb10}
\end{equation}
The 12 operator coefficients $\tilde{c}_{k}^{(\mathrm{dim})}$ are unknown.

\subsection{Second-order SB to a $(0,\mathbf{8})$ operator}

Second-order SB corrections to a $(0,\mathbf{8})$ operator can be evaluated with the tensor product of three adjoint representations. This tensor product decomposes as
\begin{equation}
\label{eq:8x8x8}
\mathbf{8}\otimes \mathbf{8} \otimes \mathbf{8} = 2 (\mathbf{1}) \oplus 8(\mathbf{8}) \oplus 4(\mathbf{10} \oplus \overline{\mathbf{10}}) \oplus 6 (\mathbf{27}) \oplus 2(\mathbf{35}\oplus \overline{\mathbf{35}}) \oplus \mathbf{64}.
\end{equation}

The analytical construction of flavor projection operators for the product of three adjoints (\ref{eq:8x8x8}), denoted by $[\mathcal{P}^{(\mathrm{dim})}]^{a_1a_2a_3b_1b_2b_3}$, is rather involved so a matrix method was proposed instead \cite{banda2}. For this purpose, the projection operator is replaced with a well defined $512\times 512$ matrix, ${\sf P}^{(m)}$, where
\begin{equation}
{\sf P}^{(m)} {\sf P}^{(m)} = {\sf P}^{(m)}, \qquad \qquad {\sf P}^{(m)} {\sf P}^{(n)} = 0, \qquad \qquad n\neq m,
\end{equation}
and
\begin{equation}
{\sf P}^{(1)} + {\sf P}^{(8)} + {\sf P}^{(10+\overline{10})} + {\sf P}^{(27)} + {\sf P}^{(35+\overline{35})} + {\sf P}^{(64)} = {\sf I}_{512},
\end{equation}
where ${\sf I}_{512}$ represents the identity matrix of order $512$. Further details can be found in Ref.~\cite{banda2}.

A complete operator basis of linearly independent $3$-body operators with spin 0 and three flavor indices is listed in Eq.~(10) of Ref.~\cite{rfm24}; this basis will be borrowed and adapted here. Let $O^{abc}$ denote this basis and provide its full structure in Appendix \ref{app:oabc} for completeness. Second-order SB corrections to a $(0,\mathbf{8})$ operator can thus be accounted for by using $[\mathcal{P}^{(\mathrm{dim})}O_n]^{c88}$. Thus, for each dimension $\mathrm{dim}$, the corresponding $1/N_c$ expansion can be cast into
\begin{eqnarray}
 \label{eq:m3abcI}
O^{c88(\mathrm{dim})} & = & N_c \sum_{k=1}^2 c_k^{(\mathrm{dim})} [\mathcal{P}^{(\mathrm{dim})} O_k]^{c88} + \sum_{k=3}^5 c_k^{(\mathrm{dim})} [\mathcal{P}^{(\mathrm{dim})} O_k]^{c88} + \frac{1}{N_c} \sum_{k=6}^{20} c_k^{(\mathrm{dim})} [\mathcal{P}^{(\mathrm{dim})} O_k]^{c88} \nonumber \\
& & \mbox{} + \frac{1}{N_c^2} \sum_{k=21}^{59} c_k^{(\mathrm{dim})} [\mathcal{P}^{(\mathrm{dim})} O_k]^{c88},
\end{eqnarray}
where $\mathrm{dim}=\mathbf{1}$, $\mathbf{8}$, $\mathbf{10}+\overline{\mathbf{10}}$, $\mathbf{27}$, $\mathbf{35}+\overline{\mathbf{35}}$ and $\mathbf{64}$. The explicit forms of the operator structures $[\mathcal{P}^{(\mathrm{dim})} Q_1Q_2Q_3]^{a_1a_2a_3}$, where $Q_j^{a_i}$ are flavor adjoints, are listed in Appendix \ref{app:q1q2q3} for the sake of completeness. From these structures, expressions like (\ref{eq:m3abcI}) can straightforwardly be obtained. Formally, there are 354 unknown operator coefficients $c_k^{(\mathrm{dim})}$ $(k=1,\ldots,59)$, so the $1/N_c$ expansion (\ref{eq:m3abcI}) seems to have minimal utility. Before drawing any conclusion, one can resort to the analysis of the baryon electric charge and the CVC hypothesis to gain some insight towards the determination of the baryon vector coupling.

\section{\label{sec:vc}The baryon vector current in the $1/N_c$ expansion}

The analysis presented in the preceding section is applied here to the baryon vector current operator, first by exploiting the limit of exact $SU(3)$ symmetry and then by introducing first- and second-order SB.

\subsection{The baryon vector current in the $SU(3)$ symmetric limit}

The $1/N_c$ expansion for the baryon vector current in the $SU(3)$ flavor symmetry limit, hereafter denoted by $V^c$, can be obtained straightforwardly from relation (\ref{eq:vc}). At $q^2=0$, the baryon matrix elements for the vector current are given by the matrix elements of the associated charge or $SU(3)$ generator. Thus, the peculiarity of $V^c$ as the generator of $SU(3)$ transformations imposes $v_1=1$ and $v_n=0$ for $n>1$, so the $1/N_c$ expansion for $V^c$ is given by
\begin{equation}
V^c = T^c,
\end{equation}
which is valid to all orders in the $1/N_c$ expansion \cite{rfm98}. The matrix elements of $V^c$ between $SU(6)$ baryon states $|B_1\rangle$ and $|B_2\rangle$ will be referred to as the leading vector coupling constant $g_V^{B_1B_2}$. For definiteness, a generic operator of the form $X^{1\pm i2}$ will stand for $X^1\pm iX^2$. Strangeness conserving and strangeness changing semileptonic baryon decays are characterized by $c=1\pm i2$ and $c=4\pm i5$, respectively.

The vector couplings in the limit of exact $SU(3)$ symmetry for some transitions read,
\begin{subequations}
\begin{equation}
g_V^{np[SU(3)]} = 1,
\end{equation}
\begin{equation}
g_V^{\Lambda p[SU(3)]} = - \sqrt{\frac32},
\end{equation}
\begin{equation}
g_V^{\Sigma^-n[SU(3)]} = - 1,
\end{equation}
\begin{equation}
g_V^{\Xi^-\Lambda[SU(3)]} = \sqrt{\frac32},
\end{equation}
\begin{equation}
g_V^{\Xi^-\Sigma^0[SU(3)]} = \frac{1}{\sqrt{2}},
\end{equation}
\begin{equation}
g_V^{\Xi^0\Sigma^+[SU(3)]} = 1,
\end{equation}
\begin{equation}
g_V^{\Omega^-{\Xi^*}^0[SU(3)]} = \sqrt{3}.
\end{equation}
\end{subequations}

\subsection{\label{sec:sb1}First-order SB to the leading vector coupling}

Some time ago, Behrends and Sirlin \cite{bs} and Ademollo and Gatto \cite{ag} independently pointed out that to first order in the SB interaction all the vector coupling constants are not renormalized. Formally, one thus expects that these corrections should be second order in SB. This important result, systematized by Ademollo and Gatto in a theorem named after them, built on the assumptions that (i) the vector current and the electromagnetic current belong to the same unitary octet,{\it i.e.}, it relied on the validity of the CVC hypothesis, and that (ii) the breaking of the unitary symmetry is due to a term behaving like the eighth component of an octet \cite{ag}.

The approach implemented by Ademollo and Gatto can be examined with clarity within the large-$N_c$ formalism. Much of this section is quite elementary, but the discussion may be enlightening. For this aim, the baryon electric charge operator, defined as
\begin{equation}
T^Q \equiv T^3 + \frac{1}{\sqrt{3}} T^8,
\end{equation}
leads to the electric charge of a baryon $B$, $Q_B$, given by
\begin{equation}
Q_B \equiv \langle B|T^Q|B \rangle.
\end{equation}

In light of the non-renormalization of the baryon electric charge, first-order SB corrections to the electric charge, denoted here by $\epsilon Q_B$ and obtained through the $1/N_c$ expansions (\ref{eq:sb1})-(\ref{eq:sb10}) for $N_c=3$, must vanish. From now on, $\epsilon$ should be regarded as a continuous real parameter introduced to keep track of the number of times the perturbation enters. The explicit expressions for $\epsilon Q_B$ read\footnote{A global factor of $1/\sqrt{3}$ was absorbed into $\tilde{c}_k^{(\mathrm{dim})}$ for ease of notation.}
\begin{equation}
\epsilon Q_{n} = 3 \tilde{c}_1^{(1)} + \frac14 \tilde{c}_2^{(1)} - \tilde{c}_1^{(8)} - \frac12 \tilde{c}_2^{(8)} - \frac16 \tilde{c}_3^{(8)} + \frac16 \tilde{c}_1^{(10+\overline{10})} - \frac{1}{20} \tilde{c}_1^{(27)} - \frac{1}{80} \tilde{c}_2^{(27)} - \frac{1}{60} \tilde{c}_3^{(27)}, \label{eq:eQn}
\end{equation}
\begin{equation}
\epsilon Q_{p} = 3 \tilde{c}_1^{(1)} + \frac14 \tilde{c}_2^{(1)} + \frac13 \tilde{c}_2^{(8)} - \frac16 \tilde{c}_1^{(10+\overline{10})} + \frac{7}{20} \tilde{c}_1^{(27)} + \frac{7}{80} \tilde{c}_2^{(27)} + \frac{7}{60} \tilde{c}_3^{(27)},
\end{equation}
\begin{equation}
\epsilon Q_{\Sigma^+} = 3 \tilde{c}_1^{(1)} + \frac14 \tilde{c}_2^{(1)} + \tilde{c}_1^{(8)} + \frac16 \tilde{c}_2^{(8)} + \frac16 \tilde{c}_3^{(8)} + \frac16 \tilde{c}_1^{(10+\overline{10})} - \frac{1}{20} \tilde{c}_1^{(27)} - \frac{1}{80} \tilde{c}_2^{(27)} - \frac{1}{60} \tilde{c}_3^{(27)},
\end{equation}
\begin{equation}
\epsilon Q_{\Sigma^0} = 3 \tilde{c}_1^{(1)} + \frac14 \tilde{c}_2^{(1)} - \frac16 \tilde{c}_2^{(8)} - \frac{1}{20} \tilde{c}_1^{(27)} - \frac{1}{80} \tilde{c}_2^{(27)} - \frac{1}{60} \tilde{c}_3^{(27)},
\end{equation}
\begin{equation}
\epsilon Q_{\Sigma^-} = 3 \tilde{c}_1^{(1)} + \frac14 \tilde{c}_2^{(1)} - \tilde{c}_1^{(8)} - \frac12 \tilde{c}_2^{(8)} - \frac16 \tilde{c}_3^{(8)} - \frac16 \tilde{c}_1^{(10+\overline{10})} - \frac{1}{20} \tilde{c}_1^{(27)} - \frac{1}{80} \tilde{c}_2^{(27)} - \frac{1}{60} \tilde{c}_3^{(27)},
\end{equation}
\begin{equation}
\epsilon Q_{\Xi^-} = 3 \tilde{c}_1^{(1)} + \frac14 \tilde{c}_2^{(1)} + \frac13 \tilde{c}_2^{(8)} + \frac16 \tilde{c}_1^{(10+\overline{10})} + \frac{7}{20} \tilde{c}_1^{(27)} + \frac{7}{80} \tilde{c}_2^{(27)} + \frac{7}{60} \tilde{c}_3^{(27)},
\end{equation}
\begin{equation}
\epsilon Q_{\Xi^0} = 3 \tilde{c}_1^{(1)} + \frac14 \tilde{c}_2^{(1)} + \tilde{c}_1^{(8)} + \frac16 \tilde{c}_2^{(8)} + \frac16 \tilde{c}_3^{(8)} - \frac16 \tilde{c}_1^{(10+\overline{10})} - \frac{1}{20} \tilde{c}_1^{(27)} - \frac{1}{80} \tilde{c}_2^{(27)} - \frac{1}{60} \tilde{c}_3^{(27)},
\end{equation}
\begin{equation}
\epsilon Q_{\Lambda} = 3 \tilde{c}_1^{(1)} + \frac14 \tilde{c}_2^{(1)} + \frac16 \tilde{c}_2^{(8)} - \frac{9}{20} \tilde{c}_1^{(27)} - \frac{9}{80} \tilde{c}_2^{(27)} - \frac{3}{20} \tilde{c}_3^{(27)},
\end{equation}
\begin{equation}
\epsilon Q_{\Delta^{++}} = 3 \tilde{c}_1^{(1)} + \frac54 \tilde{c}_2^{(1)} + \tilde{c}_1^{(8)} + \frac56 \tilde{c}_2^{(8)} + \frac56 \tilde{c}_3^{(8)} + \frac{9}{10} \tilde{c}_1^{(27)} + \frac{9}{40} \tilde{c}_2^{(27)} + \frac34 \tilde{c}_3^{(27)},
\end{equation}
\begin{equation}
\epsilon Q_{\Delta^+} = 3 \tilde{c}_1^{(1)} + \frac54 \tilde{c}_2^{(1)} + \frac12 \tilde{c}_1^{(27)} + \frac18 \tilde{c}_2^{(27)} + \frac{5}{12} \tilde{c}_3^{(27)},
\end{equation}
\begin{equation}
\epsilon Q_{\Delta^0} = 3 \tilde{c}_1^{(1)} + \frac54 \tilde{c}_2^{(1)} - \tilde{c}_1^{(8)} - \frac56 \tilde{c}_2^{(8)} - \frac56 \tilde{c}_3^{(8)} + \frac{1}{10} \tilde{c}_1^{(27)} + \frac{1}{40} \tilde{c}_2^{(27)} + \frac{1}{12} \tilde{c}_3^{(27)},
\end{equation}
\begin{equation}
\epsilon Q_{\Delta^-} = 3 \tilde{c}_1^{(1)} + \frac54 \tilde{c}_2^{(1)} - 2 \tilde{c}_1^{(8)} - \frac53 \tilde{c}_2^{(8)} - \frac53 \tilde{c}_3^{(8)} - \frac{3}{10} \tilde{c}_1^{(27)} - \frac{3}{40} \tilde{c}_2^{(27)} - \frac14 \tilde{c}_3^{(27)},
\end{equation}
\begin{equation}
\epsilon Q_{{\Sigma^*}^+} = 3 \tilde{c}_1^{(1)} + \frac54 \tilde{c}_2^{(1)} + \tilde{c}_1^{(8)} + \frac56 \tilde{c}_2^{(8)} + \frac56 \tilde{c}_3^{(8)} - \frac{11}{10} \tilde{c}_1^{(27)} - \frac{11}{40} \tilde{c}_2^{(27)} - \frac{11}{12} \tilde{c}_3^{(27)},
\end{equation}
\begin{equation}
\epsilon Q_{{\Sigma^*}^0} = 3 \tilde{c}_1^{(1)} + \frac54 \tilde{c}_2^{(1)} - \frac12 \tilde{c}_1^{(27)} - \frac18 \tilde{c}_2^{(27)} - \frac{5}{12} \tilde{c}_3^{(27)},
\end{equation}
\begin{equation}
\epsilon Q_{{\Sigma^*}^-} = 3 \tilde{c}_1^{(1)} + \frac54 \tilde{c}_2^{(1)} - \tilde{c}_1^{(8)} - \frac56 \tilde{c}_2^{(8)} - \frac56 \tilde{c}_3^{(8)} + \frac{1}{10} \tilde{c}_1^{(27)} + \frac{1}{40} \tilde{c}_2^{(27)} + \frac{1}{12} \tilde{c}_3^{(27)},
\end{equation}
\begin{equation}
\epsilon Q_{{\Xi^*}^-} = 3 \tilde{c}_1^{(1)} + \frac54 \tilde{c}_2^{(1)} + \frac12 \tilde{c}_1^{(27)} + \frac18 \tilde{c}_2^{(27)} + \frac{5}{12} \tilde{c}_3^{(27)},
\end{equation}
\begin{equation}
\epsilon Q_{{\Xi^*}^0} = 3 \tilde{c}_1^{(1)} + \frac54 \tilde{c}_2^{(1)} + \tilde{c}_1^{(8)} + \frac56 \tilde{c}_2^{(8)} + \frac56 \tilde{c}_3^{(8)} - \frac{11}{10} \tilde{c}_1^{(27)} - \frac{11}{40} \tilde{c}_2^{(27)} - \frac{11}{12} \tilde{c}_3^{(27)},
\end{equation}
\begin{equation}
\epsilon Q_{\Omega^-} = 3 \tilde{c}_1^{(1)} + \frac54 \tilde{c}_2^{(1)} + \tilde{c}_1^{(8)} + \frac56 \tilde{c}_2^{(8)} + \frac56 \tilde{c}_3^{(8)} + \frac{9}{10} \tilde{c}_1^{(27)} + \frac{9}{40} \tilde{c}_2^{(27)} + \frac34 \tilde{c}_3^{(27)}. \label{eq:eQo}
\end{equation}

Furthermore, the isospin relations
\begin{equation}
\epsilon Q_{\Sigma^+}^{(\mathrm{dim})} - 2 \epsilon Q_{\Sigma^0}^{(\mathrm{dim})} + \epsilon Q_{\Sigma^-}^{(\mathrm{dim})} = 0,
\end{equation}
\begin{equation}
\epsilon Q_{{\Sigma^*}^+}^{(\mathrm{dim})} - 2 \epsilon Q_{{\Sigma^*}^0}^{(\mathrm{dim})} + \epsilon Q_{{\Sigma^*}^-}^{(\mathrm{dim})} = 0,
\end{equation}
\begin{equation}
\epsilon Q_{\Delta^{++}}^{(\mathrm{dim})} - \epsilon Q_{\Delta^+}^{(\mathrm{dim})} - \epsilon Q_{\Delta^0}^{(\mathrm{dim})} + 
\epsilon Q_{\Delta^-}^{(\mathrm{dim})} = 0,
\end{equation}
\begin{equation}
\epsilon Q_{\Delta^{++}}^{(\mathrm{dim})} - 3\epsilon Q_{\Delta^+}^{(\mathrm{dim})} + 3\epsilon Q_{\Delta^0}^{(\mathrm{dim})} - 
\epsilon Q_{\Delta^-}^{(\mathrm{dim})} = 0,
\end{equation}
are fulfilled, which is a consistent with expectations.

The homogeneous system of linear equations $\{\epsilon Q_B^{(\mathrm{dim})}=0\}$ admits only the trivial solution
\begin{equation}
\tilde{c}_{k}^{(\mathrm{dim})} = 0, \label{eq:agt}
\end{equation}
except for the coefficients $\tilde{c}_{k}^{(8_A)}$ which come from $T$-odd operators and do not appear in expressions (\ref{eq:eQn})-(\ref{eq:eQo}).

First-order SB corrections to the vector couplings, on the other hand, read,
\begin{equation}
\sqrt{3} \epsilon g_V^{np} = \tilde{c}_1^{(8)} + \frac56 \tilde{c}_2^{(8)} + \frac16 \tilde{c}_3^{(8)} - \frac13 \tilde{c}_1^{(10+\overline{10})} + \frac25 \tilde{c}_1^{(27)} + \frac{1}{10} \tilde{c}_2^{(27)} + \frac{2}{15} \tilde{c}_3^{(27)}, \label{eq:gnp}
\end{equation}
\begin{equation}
\sqrt{2} \epsilon g_V^{\Sigma^\pm\Lambda} = \frac13 \tilde{c}_2^{(8)} - \frac25 \tilde{c}_1^{(27)} - \frac{1}{10} \tilde{c}_2^{(27)} - \frac{2}{15} \tilde{c}_3^{(27)},
\end{equation}
\begin{equation}
\sqrt{2} \epsilon g_V^{\Lambda p} = \frac12 \tilde{c}_1^{(8)} + \frac14 \tilde{c}_2^{(8)} + \frac{1}{12} \tilde{c}_3^{(8)} - \frac32 \tilde{c}_1^{(8_A)} - \frac34 \tilde{c}_2^{(8_A)} - \frac14 \tilde{c}_1^{(8_A)} + \frac16 \tilde{c}_1^{(10+\overline{10})} - \frac35 \tilde{c}_1^{(27)} - \frac{3}{20} \tilde{c}_2^{(27)} - \frac15 \tilde{c}_3^{(27)},
\end{equation}
\begin{equation}
\sqrt{3} \epsilon g_V^{\Sigma^-n} = \frac12 \tilde{c}_1^{(8)} - \frac{1}{12} \tilde{c}_2^{(8)} + \frac{1}{12} \tilde{c}_3^{(8)} - \frac32 \tilde{c}_1^{(8_A)} + \frac14 \tilde{c}_2^{(8_A)} - \frac14 \tilde{c}_1^{(8_A)} - \frac16 \tilde{c}_1^{(10+\overline{10})} - \frac15 \tilde{c}_1^{(27)} - \frac{1}{20} \tilde{c}_2^{(27)} - \frac{1}{15} \tilde{c}_3^{(27)},
\end{equation}
\begin{equation}
\sqrt{2} \epsilon g_V^{\Xi^-\Lambda} = - \frac12 \tilde{c}_1^{(8)} - \frac{1}{12} \tilde{c}_2^{(8)} - \frac{1}{12} \tilde{c}_3^{(8)} + \frac32 \tilde{c}_1^{(8_A)} + \frac14 \tilde{c}_2^{(8_A)} + \frac14 \tilde{c}_1^{(8_A)} - \frac16 \tilde{c}_1^{(10+\overline{10})} - \frac35 \tilde{c}_1^{(27)} - \frac{3}{20} \tilde{c}_2^{(27)} - \frac15 \tilde{c}_3^{(27)},
\end{equation}
\begin{eqnarray}
\sqrt{6} \epsilon g_V^{\Xi^-\Sigma^0} & = & - \frac12 \tilde{c}_1^{(8)} - \frac{5}{12} \tilde{c}_2^{(8)} - \frac{1}{12} \tilde{c}_3^{(8)} + \frac32 \tilde{c}_1^{(8_A)} + \frac54 \tilde{c}_2^{(8_A)} + \frac14 \tilde{c}_1^{(8_A)} + \frac16 \tilde{c}_1^{(10+\overline{10})} - \frac15 \tilde{c}_1^{(27)} \nonumber \\
& & \mbox{} - \frac{1}{20} \tilde{c}_2^{(27)} - \frac{1}{15} \tilde{c}_3^{(27)},
\end{eqnarray}
\begin{eqnarray}
\sqrt{3} \epsilon g_V^{\Xi^0\Sigma^+} & = & - \frac12 \tilde{c}_1^{(8)} - \frac{5}{12} \tilde{c}_2^{(8)} - \frac{1}{12} \tilde{c}_3^{(8)} + \frac32 \tilde{c}_1^{(8_A)} + \frac54 \tilde{c}_2^{(8_A)} + \frac14 \tilde{c}_1^{(8_A)} + \frac16 \tilde{c}_1^{(10+\overline{10})} - \frac15 \tilde{c}_1^{(27)} \nonumber \\
& & \mbox{} - \frac{1}{20} \tilde{c}_2^{(27)} - \frac{1}{15} \tilde{c}_3^{(27)},
\end{eqnarray}
\begin{equation}
 \epsilon g_V^{\Omega^-{\Xi^*}^0} = - \frac12 \tilde{c}_1^{(8)} - \frac{5}{12} \tilde{c}_2^{(8)} - \frac{5}{12} \tilde{c}_3^{(8)} + \frac32 \tilde{c}_1^{(8_A)} + \frac54 \tilde{c}_2^{(8_A)} + \frac54 \tilde{c}_3^{(8_A)} - \frac65 \tilde{c}_1^{(27)} - \frac{3}{10} \tilde{c}_2^{(27)} - \tilde{c}_3^{(27)}. \label{eq:gox}
\end{equation}

Once the constraints (\ref{eq:agt}) are imposed on expressions (\ref{eq:gnp})-(\ref{eq:gox}), $\epsilon g_V^{B_1B_2}$ in the $\Delta S=0$ sector vanish. In contrast, $\epsilon g_V^{B_1B_2}$ in the $|\Delta S|=1$ sector do not vanish unless $T$-odd operator contributions are left out, which is equivalent to demand that the vector couplings be real. Thus,
\begin{equation}
\epsilon g_V^{B_1B_2} = 0,
\end{equation}
which is the conclusion reached by Ademollo and Gatto in their seminal paper \cite{ag}.\footnote{Actually, the analysis by Ademollo and Gatto involved octet baryons only. Extending the analysis to include decuplet baryons is straightforward.}

\subsection{\label{eq:sb2}Second-order SB to the leading vector coupling}

In the spirit of the analysis of Sec.~\ref{sec:sb1}, second-order SB effects to the vector coupling will follow by exploiting the same working assumptions. Although second-order SB corrections to the baryon electric charge must be zero, their theoretical expressions provide useful information about some unknown operator coefficients introduced in Eq.~(\ref{eq:m3abcI}).

Let $\epsilon^2 Q_B^{(\mathrm{dim})}$ denote the second-order SB correction to the electric charge of a baryon $B$ coming from $SU(3)$ flavor representation of dimension $\mathrm{dim}$, given by
\begin{equation}
\epsilon^2 Q_B^{(\mathrm{dim})} = \left\langle B \left| O^{388(\mathrm{dim})} + \frac{1}{\sqrt{3}} O^{888(\mathrm{dim})} \right|B \right\rangle,
\end{equation}
where $O^{c88(\mathrm{dim})}$ is given by Eq.~(\ref{eq:m3abcI}). The matrix elements of the operators $[\mathcal{P}^{(\mathrm{dim})}O_k]^{388}$ and $[\mathcal{P}^{(\mathrm{dim})}O_k]^{888}$ can be found in Eqs.~(B1)-(B12) of Ref.~\cite{rfm24}. Evaluating $\epsilon^2 Q_B^{(\mathrm{dim})}$ is easily done by direct computation; however, due to the length and unilluminating nature of the resulting expressions, they have been relegated to the Supplementary Material associated to this work, listed from Eqs.~(1) to (108) of Sec.~I. A glance at those raw expressions indeed reveals some interesting findings worth mentioning. It can be noticed that (i) $T$-odd operators do not contribute to $\epsilon^2 Q_B^{(\mathrm{dim})}$ and (ii) the $\mathbf{10}+\overline{\mathbf{10}}$ representation does not contribute to decuplet baryons whereas the $\mathbf{64}$ representation does not contribute to octet baryons.

Furthermore, unlike first-order SB corrections for which the homogeneous system of linear equations $\{\epsilon Q_B=0\}$ yield only the trivial solution, second-order SB corrections for which the homogeneous system of linear equations $\{\epsilon^2 Q_B^{(\mathrm{dim})}=0\}$ yield a nontrivial solution and impose the constraints on some operator coefficients given by Eqs.~(114)-(127) of the Supplementary Material. By appealing again to the CVC hypothesis, second-order SB corrections to the vector couplings do not necessarily vanish.

On the other hand, the matrix elements of the operators $[\mathcal{P}^{(\mathrm{dim})}O_k]^{c88}$, with $c=1\pm i2$ and $c=4\pm i5$ for strangeness conserving and strangeness changing processes, respectively, are needed. The corresponding operator structures are listed in Appendix \ref{app:q1q2q3}. Thus, raw expressions for second-order corrections to the vector coupling are readily obtained and listed in Sec.~II of the Supplementary Material. Once the constraints on the operator coefficients due to the non-renormalization of the baryon electric charge are imposed on these raw expressions the coupling constants turn out to be the ones listed in Sec.~III of the Supplementary Material.

There are also some interesting findings that can be extracted from the latter expressions: For all processes, $\epsilon^2 g_V^{B_1B_2(\mathrm{1})} = \epsilon^2 g_V^{B_1B_2(35+\overline{35})}=0$. Additionally, for octet baryon-octet baryon transitions, $\epsilon^2 g_V^{B_1B_2(\mathrm{64})}=0$ whereas for decuplet baryon-decuplet baryon transitions, $\epsilon^2 g_V^{B_1B_2(\mathrm{10+\overline{10}})}=0$. Finally, for $\Delta S=0$ processes, $\epsilon^2 g_V^{np(\mathrm{dim})}=0$ and $\epsilon^2 g_V^{\Sigma^\pm\Lambda(\mathrm{10+\overline{10}})}\neq 0$.

There is another symmetry property to be imposed on the vector couplings. The operator basis $\{M_r^{a_1a_2a_3}\}$, Eq.~(\ref{eq:q1q2q3}), is constituted by both $T$-odd and $T$-even operators. The former are identified as $O_1^{abc}$, $O_6^{abc}$, $O_{14}^{abc}$, $O_{15}^{abc}$, $O_{16}^{abc}$, $O_{17}^{abc}$, $O_{21}^{abc}$, $O_{22}^{abc}$, $O_{32}^{abc}$, $O_{33}^{abc}$, $O_{34}^{abc}$, $O_{35}^{abc}$, $O_{36}^{abc}$, $O_{37}^{abc}$, $O_{51}^{abc}$, $O_{52}^{abc}$, $O_{53}^{abc}$, $O_{54}^{abc}$, $O_{55}^{abc}$, $O_{56}^{abc}$, $O_{57}^{abc}$, $O_{58}^{abc}$, and $O_{59}^{abc}$. $T$-odd operators cannot appear in the baryon vector current operator in the $SU(3)$ limit by time reversal invariance but are allowed in the case of broken symmetry. Furthermore, time reversal invariance demands that the vector couplings be real so $T$-odd operator contributions must be excluded. The reduced expressions $\epsilon^2 g_V^{B_1B_2(\mathrm{dim})}$ are thus listed in Eqs.~(224)-(234) of the Supplementary Material.

Two important results can be derived:
\begin{enumerate}
\item For $\Delta S=0$ processes, $\epsilon^2 g_V^{B_1B_2}=0$, which is consistent with the fact that isospin symmetry is not broken by the strange quark mass. Corrections should have an electromagnetic origin as it was highlighted by Fubini and Furlan \cite{fubini}.
\item For $|\Delta S|=1$ processes, $\epsilon^2 g_V^{B_1B_2(\mathrm{8})}$ and $\epsilon^2 g_V^{B_1B_2(\mathrm{10+\overline{10}})}$ are nonvanishing only for octet baryons and $\epsilon^2 g_V^{B_1B_2(\mathrm{8})}$ is nonvanishing only for decuplet baryons.
\end{enumerate}
Additional symmetry breaking induced by isospin and electromagnetic interactions should emerge from operator structures like $[\mathcal{P}^{(\mathrm{dim})}O_k]^{c33}$ and $[\mathcal{P}^{(\mathrm{dim})}O_k]^{c38}$ and naturally, from those operator structures emerging from the tensor product of four adjoints. The latter issue is beyond the scope of the present analysis.

As it was previously pointed out, the $1/N_c$ expansions (\ref{eq:m3abcI}) yield 354 unknown operator coefficients. After imposing all working assumptions, only 39 remain. This number of free parameters is still large in order for those expansions to have some predictive power. Fortunately, some of these operators coefficients can still be redefined to reduce them further. After some algebraic manipulation, the final expressions obtained read,
\begin{eqnarray}
\sqrt{\frac23} \epsilon^2 g_V^{\Lambda p(8)} & = & \frac{1}{N_c} v_1^{(8)} + \frac{1}{N_c} v_2^{(8)} + \frac{1}{N_c} v_3^{(8)} + \frac{1}{N_c^2} v_4^{(8)} + \frac{1}{N_c^2} v_5^{(8)} + \frac{1}{N_c^2} v_6^{(8)} + \frac{1}{N_c^2} v_7^{(8)} + \frac{1}{N_c^2} v_8^{(8)}  \nonumber \\
 & & \mbox{} + \frac{1}{N_c^2} v_9^{(8)} + \frac{1}{N_c^2} v_{10}^{(8)} + \frac{1}{N_c^2} v_{11}^{(8)}, \label{eq:lp}
\end{eqnarray}
\begin{equation}
\sqrt{\frac23} \epsilon^2 g_V^{\Lambda p(10+\overline{10})} = \frac{1}{N_c^2} v_1^{(10+\overline{10})},
\end{equation}
\begin{eqnarray}
\epsilon^2 g_V^{\Sigma^-n(8)} & = & - \frac{1}{7N_c} v_1^{(8)} - \frac{1}{3N_c} v_2^{(8)} + \frac{1}{N_c} v_3^{(8)} + \frac{1}{N_c^2} v_4^{(8)} + \frac{1}{N_c^2} v_5^{(8)} + \frac{1}{N_c^2} v_6^{(8)} + \frac{1}{N_c^2} v_7^{(8)} + \frac{1}{N_c^2} v_8^{(8)}  \nonumber \\
 & & \mbox{} + \frac{5}{53N_c^2} v_9^{(8)} + \frac{65}{161N_c^2} v_{10}^{(8)} - \frac{5}{91N_c^2} v_{11}^{(8)},
\end{eqnarray}
\begin{equation}
\epsilon^2 g_V^{\Sigma^-n (10+\overline{10})} = - \frac{1}{N_c^2} v_1^{(10+\overline{10})},
\end{equation}
\begin{eqnarray}
\sqrt{\frac23} \epsilon^2 g_V^{\Xi^-\Lambda(8)} & = & - \frac{3}{7N_c} v_1^{(8)} - \frac{1}{3N_c} v_2^{(8)} - \frac{1}{N_c} v_3^{(8)} - \frac{1}{N_c^2} v_4^{(8)} - \frac{1}{N_c^2} v_5^{(8)} - \frac{1}{N_c^2} v_6^{(8)} - \frac{1}{N_c^2} v_7^{(8)}  \nonumber \\
 & & \mbox{} - \frac{1}{N_c^2} v_8^{(8)} - \frac{29}{53N_c^2} v_9^{(8)} - \frac{113}{161N_c^2} v_{10}^{(8)} - \frac{43}{91N_c^2} v_{11}^{(8)},
\end{eqnarray}
\begin{equation}
\sqrt{\frac23} \epsilon^2 g_V^{\Xi^-\Lambda(10+\overline{10})} = - \frac{1}{N_c^2} v_1^{(10+\overline{10})},
\end{equation}
\begin{eqnarray}
\sqrt{2} \epsilon^2 g_V^{\Xi^-\Sigma^0(8)} & = & - \frac{11}{7N_c} v_1^{(8)} - \frac{5}{3N_c} v_2^{(8)} - \frac{1}{N_c} v_3^{(8)} - \frac{1}{N_c^2} v_4^{(8)} - \frac{1}{N_c^2} v_5^{(8)} - \frac{1}{N_c^2} v_6^{(8)} - \frac{1}{N_c^2} v_7^{(8)} \nonumber \\
 & & \mbox{} - \frac{1}{N_c^2} v_8^{(8)} - \frac{77}{53N_c^2} v_9^{(8)} - \frac{209}{161N_c^2} v_{10}^{(8)} - \frac{139}{91N_c^2} v_{11}^{(8)},
\end{eqnarray}
\begin{equation}
\sqrt{2} \epsilon^2 g_V^{\Xi^-\Sigma^0(10+\overline{10})} = \frac{1}{N_c^2} v_1^{(10+\overline{10})},
\end{equation}
\begin{eqnarray}
\epsilon^2 g_V^{\Xi^0\Sigma^+(8)} & = & - \frac{11}{7N_c} v_1^{(8)} - \frac{5}{3N_c} v_2^{(8)} - \frac{1}{N_c} v_3^{(8)} - \frac{1}{N_c^2} v_4^{(8)} - \frac{1}{N_c^2} v_5^{(8)} - \frac{1}{N_c^2} v_6^{(8)} - \frac{1}{N_c^2} v_7^{(8)} \nonumber \\
 & & \mbox{} - \frac{1}{N_c^2} v_8^{(8)} - \frac{77}{53N_c^2} v_9^{(8)} - \frac{209}{161N_c^2} v_{10}^{(8)} - \frac{139}{91N_c^2} v_{11}^{(8)}, \label{eq:xs}
\end{eqnarray}
\begin{equation}
\epsilon^2 g_V^{\Xi^0\Sigma^+(10+\overline{10})} = \frac{1}{N_c^2} v_1^{(10+\overline{10})},
\end{equation}
and
\begin{eqnarray}
\frac{1}{\sqrt{3}} \epsilon^2 g_V^{\Omega^-{\Xi^*}^0(8)} & = & - \frac{15}{7N_c} v_1^{(8)} - \frac{5}{3N_c} v_2^{(8)} - \frac{5}{N_c} v_3^{(8)}-\frac{9}{5N_c^2} v_4^{(8)} - \frac{1}{N_c^2} v_5^{(8)} - \frac{157}{85N_c^2} v_6^{(8)} \nonumber \\
 & & \mbox{} - \frac{83}{47N_c^2} v_7^{(8)} - \frac{527}{275N_c^2} v_8^{(8)} - \frac{391}{265N_c^2} v_9^{(8)} - \frac{187}{115N_c^2} v_{10}^{(8)} - \frac{157}{91N_c^2} v_{11}^{(8)}.
\end{eqnarray}
where
\begin{equation}
v_1^{(8)} = \frac{21}{32} c_{8}^{(8)} + \frac{21}{32} c_{9}^{(8)} - \frac{21}{16} c_{10}^{(8)},
\end{equation}
\begin{equation}
v_2^{(8)} = - \frac98 c_{11}^{(8)} - \frac98 c_{12}^{(8)},
\end{equation}
\begin{equation}
v_3^{(8)} = - \frac{1}{16} c_{13}^{(8)},
\end{equation}
\begin{equation}
v_4^{(8)} = - \frac38 c_{18}^{(8)} - \frac38 c_{19}^{(8)} + \frac34 c_{20}^{(8)},
\end{equation}
\begin{equation}
v_5^{(8)} = - \frac12 c_{26}^{(8)} + \frac14 c_{27}^{(8)} + \frac14 c_{28}^{(8)},
\end{equation}
\begin{equation}
v_6^{(8)} = \frac{17}{8} c_{29}^{(8)} - \frac{17}{16} c_{30}^{(8)} - \frac{17}{16} c_{31}^{(8)},
\end{equation}
\begin{equation}
v_7^{(8)} = - \frac{47}{32} c_{41}^{(8)} - \frac{47}{32} c_{44}^{(8)} + \frac{47}{16} c_{47}^{(8)},
\end{equation}
\begin{equation}
v_8^{(8)} = \frac{55}{32} c_{42}^{(8)} + \frac{55}{32} c_{46}^{(8)},
\end{equation}
\begin{equation}
v_9^{(8)} = \frac{53}{16} c_{43}^{(8)} + \frac{53}{16} c_{45}^{(8)},
\end{equation}
\begin{equation}
v_{10}^{(8)} = - \frac{161}{32} c_{48}^{(8)} - \frac{161}{32} c_{49}^{(8)},
\end{equation}
\begin{equation}
v_{11}^{(8)} = \frac{91}{48} c_{50}^{(8)},
\end{equation}
\begin{eqnarray}
v_1^{(10+\overline{10})} & = & - \frac32 c_{26}^{(10+\overline{10})} - \frac32 c_{27}^{(10+\overline{10})} - \frac32 c_{28}^{(10+\overline{10})} + \frac{7}{8} c_{29}^{(10+\overline{10})} + \frac18 c_{30}^{(10+\overline{10})} + \frac18 c_{31}^{(10+\overline{10})}  \nonumber \\
& & \mbox{} + \frac12 c_{38}^{(10+\overline{10})} + \frac12 c_{39}^{(10+\overline{10})} + \frac54 c_{40}^{(10+\overline{10})} + \frac34 c_{41}^{(10+\overline{10})} - \frac34 c_{43}^{(10+\overline{10})} + \frac34 c_{44}^{(10+\overline{10})}  \nonumber \\
& & \mbox{} - \frac34 c_{45}^{(10+\overline{10})} + \frac{21}{16} c_{47}^{(10+\overline{10})} + \frac{3}{16} c_{48}^{(10+\overline{10})} + \frac{3}{16} c_{49}^{(10+\overline{10})}.
\end{eqnarray}

At the end, one is left with 11 free parameters $v_r^{(8)}$ and only one $v_1^{(10+\overline{10})}$. This apparent large number of free parameters may suggest a degree of limited applicability of the approach and thus one should exercise some caution when performing an actual comparison to data. In a conservative scenario, one might keep the first three terms in the octet piece $v_1^{(8)}$, $v_2^{(8)}$, and $v_3^{(8)}$ and the only term $v_1^{(10+\overline{10})}$.

To close this section, relations (\ref{eq:lp})-(\ref{eq:xs}) can be tested against the set of benchmark sum rules obtained in Ref.~\cite{rfm17} to assess their performance. These sum rules yield
\begin{equation}
\frac14 \frac{\epsilon^2 g_V^{\Xi^-\Sigma^0}}{g_V^{\Xi^-\Sigma^0[SU(3)]}} + \frac34 \frac{\epsilon^2 g_V^{\Xi^-\Lambda}}{g_V^{\Xi^-\Lambda[SU(3)]}} - \frac14 \frac{\epsilon^2 g_V^{\Sigma^-n}}{g_V^{\Sigma^-n[SU(3)]}} - \frac34 \frac{\epsilon^2 g_V^{\Lambda p}}{g_V^{\Lambda p[SU(3)]}} = 0,
\end{equation}
and
\begin{equation}
\frac34 \frac{\epsilon^2 g_V^{\Xi^-\Sigma^0}}{g_V^{\Xi^-\Sigma^0[SU(3)]}} - \frac34 \frac{\epsilon^2 g_V^{\Xi^-\Lambda}}{g_V^{\Xi^-\Lambda[SU(3)]}} + \frac34 \frac{\epsilon^2 g_V^{\Sigma^-n}}{g_V^{\Sigma^-n[SU(3)]}} - \frac34 \frac{\epsilon^2 g_V^{\Lambda p}}{g_V^{\Lambda p[SU(3)]}} = \frac{3}{N_c^2} v_1^{(10+\overline{10})}. \label{eq:sm2}
\end{equation}
Violations to relation (\ref{eq:sm2}) come from $\mathbf{10}+\overline{\mathbf{10}}$ representation.

\section{\label{sec:one}One-loop corrections to the baryon vector operator}

\begin{figure}[ht]
\scalebox{0.95}{\includegraphics{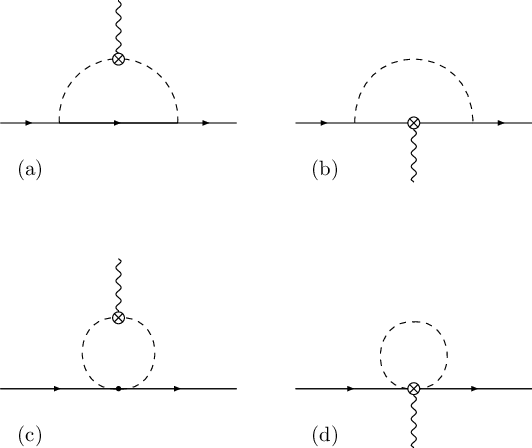}}
\caption{\label{fig:vcloop}Feynman diagrams that yield one-loop corrections to the baryon vector current. Dashed lines and solid lines denote mesons and baryons, respectively.}
\end{figure}

$SU(3)$ flavor symmetry is also explicitly broken by the meson mass in loops. The Feynman diagrams that yield one-loop corrections to the baryon vector current are displayed in Fig.~\ref{fig:vcloop}. The analytical structure of these corrections have been discussed in detail in Ref.~\cite{rfm14} so only a few salient facts will be discussed here.

Specifically, the one-loop contribution to the baryon vector current from the Feynman diagram of Fig.~\ref{fig:vcloop}(a) can be written as
\begin{equation}
\delta V_{\textrm{(a)}}^c = \sum_{\textsf{j}} A^{ia} \mathcal{P}_{\textsf{j}} A^{ib} P^{abc}(\Delta_{\textsf{j}}), \label{eq:loop1a}
\end{equation}
where the meson-baryon vertex is specified by the baryon axial vector operator $A^{kc}$, which is written as \cite{djm95}
\begin{equation}
A^{kc} = a_1 G^{kc} + \frac{1}{N_c} b_2 \mathcal{D}_2^{kc} + \frac{1}{N_c^2} b_3 \mathcal{D}_3^{kc} + \frac{1}{N_c^2} c_3 \mathcal{D}_2^{kc}, \label{eq:akc}
\end{equation}
where the series has been truncated at $N_c=3$, and
\begin{subequations}
\begin{eqnarray}
\mathcal{D}_2^{kc} & = & J^kT^c, \\
\mathcal{D}_3^{kc} & = & \{J^k,\{J^r,G^{rc}\}\}, \\
\mathcal{O}_3^{kc} & = & \{J^2,G^{kc}\} - \frac12 \{J^k,\{J^r,G^{rc}\}\}.
\end{eqnarray}
\end{subequations}
Unlike operators $\mathcal{D}_2^{kc}$ and $\mathcal{D}_3^{kc}$, operator $\mathcal{O}_3^{kc}$ connects baryon states of different spins.

In the chiral limit, the baryon propagator is diagonal in spin and is written as \cite{jen96}
\begin{equation}
\frac{i\mathcal{P}_{\mathsf{j}}}{k^0-\Delta_{\mathsf{j}}}, \label{eq:barprop}
\end{equation}
where $\mathcal{P}_{\mathsf{j}}$ is a spin projection operator for spin $J=\mathsf{j}$ and $\Delta_{\mathsf{j}}$ stands for the difference of the hyperfine mass splitting between the intermediate baryon with spin $J=\mathsf{j}$ and the external baryon, namely,
\begin{equation}
\Delta_{\mathsf{j}} = \mathcal{M}_{\textrm{hyperfine}}|_{J^2=\mathsf{j}(\mathsf{j}+1)}-\mathcal{M}_{\textrm{hyperfine}}|_{J^2=\mathsf{j}_{\textrm{ext}}(\mathsf{j}_{\textrm{ext}}+1)}.
\end{equation}

The explicit expressions for $\mathcal{P}_{\mathsf{j}}$ and $\Delta_{\mathsf{j}}$ required here are
\begin{subequations}
\label{eq:projnc3}
\begin{eqnarray}
\mathcal{P}_\frac12 & = & -\frac13 \left(J^2-\frac{15}{4}\right), \\
\mathcal{P}_\frac32 & = & \frac13 \left(J^2-\frac{3}{4}\right),
\end{eqnarray}
\end{subequations}
and
\begin{subequations}
\begin{equation}
\Delta_\frac12 = \left\{
\begin{array}{ll}
\displaystyle 0, & \mathsf{j}_{\textrm{ext}}=\frac12, \\[2mm]
\displaystyle -\Delta, & \mathsf{j}_{\textrm{ext}}=\frac32,
\end{array}
\right.
\end{equation}
\begin{equation}
\Delta_\frac32 = \left\{
\begin{array}{ll}
\displaystyle \Delta, & \mathsf{j}_{\textrm{ext}}=\frac12, \\[2mm]
\displaystyle 0, & \mathsf{j}_{\textrm{ext}}=\frac32. \\[2mm]
\end{array}
\right.
\end{equation}
\end{subequations}

Additionally, $P^{abc}(\Delta_{\textsf{j}})$ is an antisymmetric tensor expressed as \cite{rfm14}
\begin{equation}
P^{abc}(\Delta_{\textsf{j}}) = A_{\mathbf{8}}(\Delta_{\textsf{j}}) if^{acb} + A_{\mathbf{10}+\overline{\mathbf{10}}}(\Delta_{\textsf{j}}) i(f^{aec}d^{be8} - f^{bec}d^{ae8} - f^{abe}d^{ec8}), \label{eq:pabc}
\end{equation}
where $if^{acb}$ and $i(f^{aec}d^{be8}-f^{bec}d^{ae8}-f^{abe}d^{ec8})$ break $SU(3)$ as $\mathbf{8}$ and $\mathbf{10}+\overline{\mathbf{10}}$, respectively. The tensor contains a dependence on the Feynman loop integral, denoted as $I_a(m_1,m_2,\Delta_{\textsf{j}},\mu;q^2)$, through
\begin{subequations}
\label{eq:ais}
\begin{eqnarray}
A_{\mathbf{8}}(\Delta_{\textsf{j}}) & = & \frac12 [ I_a(m_\pi,m_K,\Delta_{\textsf{j}},\mu;0) + I_a(m_K,m_\eta,\Delta_{\textsf{j}},\mu;0) ], \\
A_{\mathbf{10}+\overline{\mathbf{10}}}(\Delta_{\textsf{j}}) & = & -\frac{\sqrt{3}}{2}[ I_a(m_\pi,m_K,\Delta_{\textsf{j}},\mu;0) - I_a(m_K,m_\eta,\Delta_{\textsf{j}},\mu;0) ].
\end{eqnarray}
\end{subequations}
All the necessary expressions to evaluate the contribution ({\ref{eq:loop1a}) can be found in Ref.~\cite{rfm14}, specialized to transitions between octet baryons.

As for transitions involving decuplet baryons, the analysis can be performed in a similar fashion. Accordingly, the contribution to the baryon vector current operator from Fig.~\ref{fig:vcloop}(a) can be written as
\begin{equation}
\delta V_{\textrm(a)}^c = \mathcal{P}_\frac32 A^{ia} \mathcal{P}_\frac12 A^{ib} \mathcal{P}_\frac32 P^{abc}(-\Delta) + \mathcal{P}_\frac32 A^{ia} \mathcal{P}_\frac32 A^{ib} \mathcal{P}_\frac32 P^{abc}(0). \label{eq:vc1a}
\end{equation}

For concreteness, for the $\Omega^-\to {\Xi^*}^0$ transition one finds,
\begin{eqnarray}
\frac{1}{\sqrt{3}} \delta g_{V(a)}^{\Omega^-{\Xi^*}^0} & = & \left[ \frac{5}{16} a_1^2 + \frac58 a_1b_2 + \frac{25}{24} a_1b_3 + \frac{5}{16}
b_2^2 + \frac{25}{24} b_2b_3 + \frac{125}{144} b_3^2 \right] I_a(m_\pi,m_K,0,\mu;0) \nonumber \\
& & \mbox{} + \left[ \frac{5}{16} a_1^2 + \frac58 a_1b_2 + \frac{25}{24} a_1b_3 + \frac{5}{16} b_2^2 + \frac{25}{24} b_2b_3 + \frac{125}{144} b_3^2 \right] I_a(m_K,m_\eta,0,\mu;0) \nonumber \\
& & \mbox{} + \left[ \frac14 a_1^2 + \frac14 a_1c_3 + \frac{1}{16}c_3^2 \right] I_a(m_\pi,m_K,-\Delta,\mu;0) \nonumber \\
& & \mbox{} + \left[ \frac14 a_1^2 + \frac14 a_1c_3 + \frac{1}{16}c_3^2 \right] I_a(m_K,m_\eta,-\Delta,\mu;0).
\end{eqnarray}

The contributions from diagrams coming from Fig.~\ref{fig:vcloop}({b,c,d}) can also obtained from the expressions presented in Ref.~\cite{rfm14}. Without further ado, the resultant expressions read,
\begin{eqnarray}
& & \frac{1}{\sqrt{3}} \delta g_{V(b)}^{\Omega^-{\Xi^*}^0} \nonumber \\
& & \mbox{\hglue0.2truecm} = \left[ \frac{9}{32} a_1^2 + \frac{5}{16} a_1b_2 + \frac{25}{48} a_1b_3 + \frac18 a_1c_3 + \frac{5}{32} b_2^2 + \frac{25}{48} b_2b_3 + \frac{125}{288} b_3^2 + \frac{1}{32} c_3^2 \right] I_b^{(1)}(m_\pi,\mu) \nonumber \\
& & \mbox{\hglue0.6truecm} - \left[ \frac18 a_1^2 + \frac18 a_1c_3 + \frac{1}{32}c_3^2 \right] \Delta I_b^{(2)}(m_\pi,\mu) + \left[ \frac{1}{16} a_1^2 + \frac{1}{16} a_1c_3 + \frac{1}{64}c_3^2 \right] \Delta ^2 I_b^{(3)}(m_\pi,\mu) \nonumber \\
& & \mbox{\hglue0.6truecm} + \left[ \frac{9}{16} a_1^2 + \frac58 a_1b_2 + \frac{25}{24} a_1b_3 + \frac14 a_1c_3 + \frac{5}{16} b_2^2 + \frac{25}{24} b_2b_3 + \frac{125}{144} b_3^2 + \frac{1}{16}c_3^2 \right] I_b^{(1)}(m_K,\mu) \nonumber \\
& & \mbox{\hglue0.6truecm} - \left[ \frac14 a_1^2 + \frac14 a_1c_3 + \frac{1}{16}c_3^2 \right] \Delta I_b^{(2)}(m_K,\mu) + 
 \left[ \frac18 a_1^2 + \frac18 a_1c_3 + \frac{1}{32}c_3^2 \right] \Delta ^2 I_b^{(3)}(m_K,\mu) \nonumber \\
& & \mbox{\hglue0.6truecm} + \left[ \frac{9}{32} a_1^2 + \frac{5}{16} a_1b_2 + \frac{25}{48} a_1b_3 + \frac18 a_1c_3 + \frac{5}{32} b_2^2 + \frac{25}{48} b_2b_3 + \frac{125}{288} b_3^2 + \frac{1}{32}c_3^2 \right] I_b^{(1)}(m_\eta,\mu) \nonumber \\
& & \mbox{\hglue0.6truecm} - \left[ \frac18 a_1^2 + \frac18 a_1c_3 + \frac{1}{32}c_3^2 \right]\Delta I_b^{(2)}(m_\eta,\mu) + 
 \left[ \frac{1}{16} a_1^2 + \frac{1}{16} a_1c_3 + \frac{1}{64}c_3^2 \right] \Delta ^2 I_b^{(3)}(m_\eta,\mu),
\end{eqnarray}
\begin{equation}
\frac{1}{\sqrt{3}} \delta g_{V(c)}^{\Omega^-{\Xi^*}^0} = \frac34 I_c(m_\pi,m_K,\mu;0) + \frac34 I_c(m_K,m_\eta,\mu;0),
\end{equation}
and
\begin{equation}
\frac{1}{\sqrt{3}} \delta g_{V(d)}^{\Omega^-{\Xi^*}^0} = - \frac38 I_d(m_\pi,\mu) - \frac34 I_d(m_K,\mu) - \frac38 I_d(m_\eta,\mu).
\end{equation}

Suitable combinations of diagrams yield,
\begin{eqnarray}
& & \frac{1}{\sqrt{3}} \left[ \delta g_{V(a)}^{\Omega^-{\Xi^*}^0} + \delta g_{V(b)}^{\Omega^-{\Xi^*}^0} \right] \nonumber \\
& & \mbox{\hglue0.2truecm} = \left[ \frac{5}{32} a_1^2 + \frac{5}{16} a_1b_2 + \frac{25}{48} a_1b_3 + \frac{5}{32} b_2^2 + \frac{25}{48} b_2b_3 + \frac{125}{288} b_3^2 \right] \left[ H(m_\pi,m_K) + H(m_K,m_\eta) \right] \nonumber \\
& & \mbox{\hglue0.6truecm} + \left[ \frac18 a_1^2 + \frac18 a_1c_3 + \frac{1}{32} c_3^2 \right] \left[ K(m_\pi,m_K,-\Delta) + K(m_K,m_\eta,-\Delta) \right]. \label{eq:apb}
\end{eqnarray}
and
\begin{equation}
\frac{1}{\sqrt{3}} \left[ \delta g_{V(c)}^{\Omega^-{\Xi^*}^0} + \delta g_{V(d)}^{\Omega^-{\Xi^*}^0} \right] = \frac38 H(m_\pi,m_K) + \frac38 H(m_K,m_\eta),
\end{equation}

The explicit form of the function $K(m_1,m_2,\Delta)$ is
\begin{eqnarray}
& & 16 f^2\pi^2 K(m_1,m_2,\Delta) = \nonumber \\
& & \mbox{\hglue0.6truecm} - \frac12 m_1^2 - \frac12 m_2^2 + \frac43 \Delta^2 + \frac13 \frac{3m_1^2m_2^2-6(m_1^2+m_2^2) \Delta^2 + 8\Delta^4}{m_1^2-m_2^2} \ln \frac{m_1^2}{m_2^2} \nonumber \\
& & \mbox{\hglue0.6truecm} + \left\{ \begin{array}{ll}
\displaystyle \frac43 \Delta \frac{m_1^2+3 m_2^2-4 \Delta^2}{m_1^2-m_2^2} \sqrt{m_1^2-\Delta^2} \left[ \frac{\pi}{2} - \tan^{-1} \left[\frac{\Delta}{\sqrt{m_1^2-\Delta^2}} \right] \right] & \nonumber \\
\displaystyle - \frac43 \Delta \frac{3m_1^2+m_2^2-4\Delta^2}{m_1^2-m_2^2} \sqrt{m_2^2-\Delta^2} \left[ \frac{\pi}{2} - \tan^{-1} \left[ \frac{\Delta}{\sqrt{m_2^2-\Delta^2}} \right] \right], & \Delta < m_1,m_2 \\
\displaystyle + \frac23 \Delta \frac{m_1^2+3 m_2^2-4 \Delta^2}{m_1^2-m_2^2} \sqrt{\Delta^2-m_1^2} \ln \left[ \frac{\Delta-\sqrt{\Delta^2-m_1^2}}{\Delta+\sqrt{\Delta^2-m_1^2}}\right] & \nonumber \\
\displaystyle - \frac43 \Delta \frac32 \frac{m_1^2+m_2^2-4\Delta^2}{m_1^2-m_2^2} \sqrt{m_2^2-\Delta^2} \left[ \frac{\pi}{2} - \tan^{-1} \left[ \frac{\Delta}{\sqrt{m_2^2-\Delta^2}} \right] \right], & m_1 < \Delta < m_2
\end{array}
\right.
\end{eqnarray}
from which
\begin{eqnarray}
H(m_1,m_2) & = & \lim_{\Delta\to 0} K(m_1,m_2,\Delta) \nonumber \\
& = & \frac{1}{16f^2\pi^2} \left[ - \frac12 m_1^2 - \frac12 m_2^2 + \frac{m_1^2m_2^2}{m_1^2-m_2^2} \ln \frac{m_1^2}{m_2^2} \right].
\end{eqnarray}

Finally, one can proceed further and rewrite expression (\ref{eq:apb}) in terms of the $SU(3)$ invariant couplings $D$, $F$, $\mathcal{C}$ and $\mathcal{H}$ introduced in heavy baryon chiral perturbation theory \cite{jm255,jm259}. These couplings are related to the operator coefficients $a_1$, $b_2$, $b_3$ and $c_3$ defined in Eq.~(\ref{eq:akc}) as \cite{jen96}
\begin{subequations}
\label{eq:rel1}
\begin{eqnarray}
& & D = \frac12 a_1 + \frac16 b_3, \\
& & F = \frac13 a_1 + \frac16 b_2 + \frac19 b_3, \\
& & \mathcal{C} = - a_1 - \frac12 c_3, \\
& & \mathcal{H} = -\frac32 a_1 - \frac32 b_2 - \frac52 b_3.
\end{eqnarray}
\end{subequations}

The final expression for the one-loop correction to the vector coupling of the $\Omega^-\to {\Xi^*}^0$ transition
\begin{equation}
\delta g_V^{\Omega^-{\Xi^*}^0} = \delta g_{V(a)}^{\Omega^-{\Xi^*}^0} + \delta g_{V(b)}^{\Omega^-{\Xi^*}^0} + \delta g_{V(c)}^{\Omega^-{\Xi^*}^0} + \delta g_{V(d)}^{\Omega^-{\Xi^*}^0},
\end{equation}
becomes
\begin{equation}
\frac{\delta g_V^{\Omega^-{\Xi^*}^0}}{g_V^{\Omega^-{\Xi^*}^0[SU(3)]}} = \left[ \frac38 + \frac{5}{72} \mathcal{H}^2 \right] \left[ H(m_\pi,m_K) + H(m_K,m_\eta) \right] + \frac18\mathcal{C}^2 \left[ K(m_\pi,m_K,-\Delta) + K(m_K,m_\eta,-\Delta) \right].
\end{equation}
Numerically,
\begin{subequations}
\begin{eqnarray}
K(m_\pi,m_K,-\Delta) & = & -0.1320, \\
K(m_K,m_\eta,-\Delta) & = & -0.0037, \\
H(m_\pi,m_K) & = & -0.0572, \\
H(m_K,m_\eta) & = & -0.0014,
\end{eqnarray}
\end{subequations}
for $f=0.093\,\mathrm{GeV}$ and $\Delta=0.237\,\mathrm{GeV}$ \cite{rfm24}.

Thus, one-loop corrections yield $\delta g_V^{\Omega^-{\Xi^*}^0}/g_V^{\Omega^-{\Xi^*}^0[SU(3)]} < 0$. An accurate evaluation of the reduction requires a precise determination of $\mathcal{C}$ and $\mathcal{H}$. At this point, the Wu-type experiments suggested by Bertilsson and Leupold can play an important role to determine these elusive coupling constants.

\section{\label{sec:num}A preliminary numerical analysis}

A numerical analysis through a least-squares fit to the available experimental data to determine the unknown parameters in the approach can be performed following the lines of Ref.~\cite{rfm25}. The analysis of that reference, for brevity, consisted of determining the operator coefficients $a_1$, $b_2$, $b_3$, and $c_3$ (or equivalently $D$, $F$, $\mathcal{C}$, and $\mathcal{H}$) from the axial vector current operator at tree-level Eq.~(\ref{eq:akc}), along with 10 relevant operator coefficients from first-order perturbative SB.

The set of experimental information used, for octet baryons, was constituted by the decay rates $R$, the angular correlation coefficients $\alpha_{e\nu}$, and the spin-asymmetry coefficients $\alpha_e$, $\alpha_\nu$, $\alpha_B$, $A$, and $B$ \cite{part}.\footnote{The ratios $g_A/g_V$ were not used in the fit. In essence, while spin-asymmetry coefficients and $g_A/g_V$ are distinct, they are not independent quantities.} Radiative corrections in $R$ and the $q^2$-dependence of the form factors were considered. For decuplet baryons, the axial couplings $g$ for the processes $\Delta \to N\pi$, $\Sigma^*\to\Lambda\pi$, $\Sigma^*\to\Sigma\pi$, and $\Xi^*\to\Xi\pi$ were used. Additional inputs were $f_\pi = 93\,\mathrm{MeV}$, $\mu = 1\, \mathrm{GeV}$, $\Delta = 0.237\, \mathrm{GeV}$ \cite{rfm24} and the pseudoscalar meson masses \cite{part}.

In the present analysis, second-order perturbative SB contributions to the leading vector couplings are incorporated as well. However, due to the numerous independent variables being considered, a pragmatic approach would be to ignore terms of relative order $1/N_c^2$ in Eqs.~(\ref{eq:lp})-(\ref{eq:xs}). One is thus left with three additional parameters $v_1^{(8)}$, $v_2^{(8)}$, and $v_3^{(8)}$. A 17-parameter fit for a set of 29 measured quantities may still provide an optimal solution. For this objective, the criteria imposed on the fits of Ref.~\cite{rfm25} are replicated here, namely, {\it i)} the value of $\chi^2/\mathrm{dof}$ should be around 1, which can be achieved with the inclusion of theoretical uncertainties added in quadrature to the experimental errors; {\it ii)} the best-fit parameters should yield consistent values of the $SU(3)$ invariants $D$, $F$, $\mathcal{C}$, and $\mathcal{H}$; and {\it iii)} a positive definite error matrix must be obtained. 

The fitting procedure lead to the best-fit parameters listed in Table \ref{t:bestf}, alongside the $SU(3)$ invariants $D$, $F$, $\mathcal{C}$, and $\mathcal{H}$ that follow them. Two scenarios were under consideration: one where $\Delta$ is equal to 0, and another where $\Delta$ is equal to $0.237 \, \mathrm{GeV}$. They are designated as fit A and fit B, respectively.

The best-fit parameters obtained in both scenarios are, on general grounds, in a good agreement with large-$N_c$ expectations, i.e., the leading order ones are roughly $\mathcal{O}(N_c^0)$ whereas the ones related to SB are properly suppressed by factors of $1/N_c$. The overall outputs are comparable to their counterpart from fit 3 of Ref.~\cite{rfm25}, listed in the fourth column in Table \ref{t:bestf}. There are naturally some perceptible shifts in the best-fit values due to the inclusion of second-order SB in $g_V$. The most notorious change is found in the prediction of $F/D$ in fit B, which moves away below its naive quark model prediction $F/D = 2/3$, which causes some concern. The probable sources of these noticeably changes may arise from the poor determinations of some parameters, mainly $v_3^{(8)}$ (which is higher than expected) and $b_2$ and $z_{2,8}$, which get large errors compared to their central values. Notice that effects of the inclusion of $\Delta\neq 0$ slightly improve the overall value of $\chi^2/\mathrm{dof}$ of fitB compared to fit A.

\begin{table}
\caption{\label{t:bestf}Best-fit parameters obtained in this work and a comparison with Fit 3 of Ref.~\cite{rfm25}.}
\begin{center}
\begin{tabular}{lrrr}
\hline
                  	  &            Fit A &            Fit B & Fit 3 of Ref.~\cite{rfm25} \\
\hline
$v_1^{(8)}$              & $-0.02 \pm 0.10$ & $-0.02 \pm 0.10$ &                  \\
$v_2^{(8)}$              & $ 0.01 \pm 0.10$ & $ 0.01 \pm 0.10$ &                  \\
$v_3^{(8)}$              & $-0.14 \pm 0.07$ & $-0.17 \pm 0.07$ &                  \\
$a_1$                    & $ 0.87 \pm 0.10$ & $ 1.01 \pm 0.16$ & $ 0.92 \pm 0.19$ \\
$b_2$                    & $ 0.08 \pm 0.24$ & $-1.33 \pm 1.16$ & $-0.81 \pm 0.95$ \\
$b_3$                    & $ 0.41 \pm 0.30$ & $ 0.87 \pm 0.56$ & $ 0.73 \pm 0.49$ \\
$c_3$                    & $-0.03 \pm 0.10$ & $-0.04 \pm 0.10$ & $-0.04 \pm 0.10$ \\
$z_{1,8}$                & $ 0.02 \pm 0.35$ & $ 0.06 \pm 0.45$ & $-0.24 \pm 0.45$ \\
$z_{2,8}$                & $ 1.32 \pm 1.63$ & $ 1.46 \pm 2.47$ & $ 2.58 \pm 2.56$ \\
$z_{3,8}$                & $-1.24 \pm 0.46$ & $-1.53 \pm 0.50$ & $-1.74 \pm 0.52$ \\
$z_{1,10+\overline{10}}$ & $ 0.14 \pm 0.10$ & $ 0.11 \pm 0.10$ & $ 0.13 \pm 0.10$ \\
$z_{2,10+\overline{10}}$ & $ 0.14 \pm 0.10$ & $ 0.12 \pm 0.10$ & $ 0.16 \pm 0.10$ \\
$z_{3,10+\overline{10}}$ & $ 0.06 \pm 0.10$ & $ 0.04 \pm 0.10$ & $ 0.04 \pm 0.10$ \\
$z_{1,27}$               & $ 0.07 \pm 0.10$ & $ 0.03 \pm 0.10$ & $ 0.06 \pm 0.20$ \\
$z_{2,27}$               & $-0.06 \pm 0.10$ & $-0.07 \pm 0.10$ & $-0.09 \pm 0.10$ \\
$z_{4,27}$               & $-0.13 \pm 0.10$ & $-0.09 \pm 0.10$ & $-0.09 \pm 0.10$ \\
$z_{5,27}$               & $ 0.11 \pm 0.10$ & $ 0.07 \pm 0.10$ & $ 0.06 \pm 0.10$ \\
\hline
$D$                      & $ 0.50 \pm 0.03$ & $ 0.65 \pm 0.14$ & $ 0.58 \pm 0.11$ \\
$F$                      & $ 0.35 \pm 0.03$ & $ 0.21 \pm 0.13$ & $ 0.25 \pm 0.13$ \\
$\mathcal{C}$            & $-0.85 \pm 0.11$ & $-0.99 \pm 0.14$ & $-0.90 \pm 0.19$ \\
$\mathcal{H}$            & $-2.45 \pm 0.54$ & $-1.71 \pm 0.28$ & $-1.98 \pm 0.39$ \\
$F/D$                    & $ 0.69 \pm 0.08$ & $ 0.33 \pm 0.26$ & $ 0.43 \pm 0.28$ \\
$3F-D$                   & $ 0.54 \pm 0.10$ & $-0.01 \pm 0.51$ & $ 0.18 \pm 0.46$ \\
\hline
$\chi^2/\mathrm{dof}$    &	       $1.8$ &            $1.5$ &            $1.6$ \\
\hline
\end{tabular}
\end{center}
\end{table}

For completeness, the pattern of symmetry breaking $\delta g_V/g_V^{SU(3)}-1$ obtained from the best-fit parameters for some observed processes is displayed in Table \ref{t:patt}. In that table, the total correction is obtained by adding up the SB and one-loop corrections. Separately, these contributions are small and consistent with the Ademollo-Gatto theorem. As it has been observed in several works (see Ref.~\cite{rfm14} and references therein) one-loop corrections systematically leads to a reduction in the $SU(3)$ symmetric value by few percent. On the contrary, SB corrections, as it was observed in Refs.~\cite{rfm98,rfm04}, leads to an increase, also by a few percent. In the present analysis, the combined effect produced a partial cancellation between these two contributions, so the net effect is even lower than each quantity alone. This observation is certainty unexpected because tree-level SB effects are estimated to be order $\mathcal{O}(p^4)$ while chiral one-loop effects are $\mathcal{O}(p^2)$ and thus dominant.

To close this section, a word of caution is in order. SB corrections to $g_V^{\Omega^-{\Xi^*}^0}$ are larger than expectations. This originates from the $v_3^{(8)}$ term, which produces a sizable outcome so the predicted value should be interpreted with care.

\begin{table}
\caption{\label{t:patt}Pattern of symmetry breaking $\delta g_V/g_V^{SU(3)}-1$.}
\tiny
\begin{center}
\begin{tabular}{lrrrrrrr}
\hline
                    & \multicolumn{3}{c}{Fit A} & \multicolumn{3}{c}{Fit B} & Fit 3, Ref.~\cite{rfm25} \\
$B_1B_2$            & Total   &     SB  & One-loop  & Total   &     SB  & One-loop  & \\
\hline
$\Lambda p$         & $0.001(0.048)$ & $0.050(0.047)$ & $-0.050(0.002)$ & $0.011(0.047)$ & $0.063(0.048)$ & $-0.052(0.006)$ & $-0.050(0.005$ \\
$\Sigma^- n$        & $0.030(0.027)$ & $0.047(0.027)$ & $-0.018(0.002)$ & $0.036(0.028)$ & $0.058(0.028)$ & $-0.022(0.006)$ & $-0.019(0.003$ \\
$\Xi^-\Lambda$      & $0.010(0.028)$ & $0.049(0.028)$ & $-0.039(0.001)$ & $0.015(0.028)$ & $0.061(0.029)$ & $-0.046(0.010)$ & $-0.042(0.006$ \\
$\Xi^-\Sigma^0$     & $0.001(0.072)$ & $0.052(0.071)$ & $-0.051(0.003)$ & $0.026(0.074)$ & $0.066(0.072)$ & $-0.040(0.009)$ & $-0.043(0.010$ \\
$\Xi^0\Sigma^+$     & $0.001(0.072)$ & $0.052(0.071)$ & $-0.051(0.003)$ & $0.026(0.074)$ & $0.066(0.072)$ & $-0.040(0.009)$ & $-0.043(0.010$ \\
$\Omega^-{\Xi^*}^0$ & $0.185(0.139)$ & $0.244(0.138)$ & $-0.059(0.009)$ & $0.253(0.145)$ & $0.303(0.144)$ & $-0.050(0.003)$ & \\
\hline
\end{tabular}
\end{center}
\end{table}

\section{\label{sec:cr}Concluding remarks}

The present study addressed to revisit the issue of symmetry breaking in the leading vector coupling introduced in the semileptonic decay of a baryon in the context of a combined expansion in $1/N_c$ and chiral corrections. The formalism is general enough that it can be applied to any light-flavor baryon (no $c$ or $b$ quark). The lowest-lying baryon states fall into a representation of the spin-flavor group $SU(6)$. At the physical value $N_c=3$, this is the well-known $\mathbf{56}$ dimensional representation of $SU(6)$, which contains the spin-1/2 octet- and spin-3/2 decuplet-baryon states.

The analysis involved two aspects of symmetry breaking: Explicit (perturbative) breaking and implicit breaking which emerges from the integrals occurring in the chiral one-loop corrections.

Features of perturbative SB were reexamined by using flavor projection operators, which were particularly useful to classify those $SU(3)$ representations participating in the breaking mechanism. First- and second-order SB effects were implemented to the baryon electric charge. Owing to the non renormalization of the electric charge {\it and} the validity of the CVC hypothesis, which are the building blocks of the Ademollo-Gatto theorem, critical constraints could be imposed on the operator coefficients introduced in the $1/N_c$ expansions of the baryon vector current operator. The matrix elements of this operator between $SU(6)$ states yield the baryon vector couplings. First-order SB effects vanish in both $\Delta S=0$ and $|\Delta S|=1$ sectors provided that $T$-odd operators are eliminated from the corresponding $1/N_c$ expansions. $T$-odd operators are ruled out from the vector current operator in the limit of exact $SU(3)$ symmetry but may be allowed for a broken symmetry. Rephrasing the previous arguments, vector couplings are required to be real.

The technical background gained from first-order SB allowed one to evaluate second-order SB in a close analogy. Thus, appropriate projection operators were used to first construct the corresponding $1/N_c$ expansions to the baryon electric charge, one for each flavor representation participating in the breaking; the use of working assumptions helped to find out restrictions on the operator coefficients, which, when implemented to the expansions of the vector current operator, provided a better understanding of the breaking mechanism. Thus, in the $\Delta S=0$ sector, the vector coupling in the $n\to p$ transition vanished identically whereas in the $\Sigma^\pm\to\Lambda$ transition, contributions from $T$-odd operators must be ruled out to achieve a vanishing vector coupling. Once more, the vector coupling is required to be real. In the $|\Delta S|=1$ sector, once all the working assumptions are implemented, two $SU(3)$ representation are responsible for the breaking: $\mathbf{8}$ and $\mathbf{10}+\overline{\mathbf{10}}$. This conclusion could have not been reached without the help of projection operators \cite{rfm98,rfm04}.

As for chiral one-loop effects, the approach was used to specifically compute corrections to the vector coupling in the $\Omega^-\to{\Xi^*}^0$ transition. The final expression depends on $\mathcal{H}^2$ and $\mathcal{C}^2$, with no interference terms. The functions obtained by combining the different loop integrals are negative for the physical inputs (meson mass and $\Delta$ values), so one-loop corrections the vector coupling for this transition are negative.

A least-squares fit to data allowed one to extract information on the free parameters in the theory. The outcome revealed that the best-fit parameters attained values in agreement with expectations, except for a few values off. The pattern of symmetry breaking in the vector coupling also revealed some other interesting facts. Both explicit and implicit breaking corrections are consistent with the Ademollo-Gatto theorem. The first kind results in positive corrections (values greater than zero), while the second kind results in negative corrections (values less than zero). The combined effect produces (small) values greater than zero (around 4\% for $\Sigma^-\to n$ transition and even smaller for the other ones). Unfortunately, for the $\Omega^-\to{\Xi^*}^0$, perturbative SB effects are rather larger than expected so the pattern of symmetry breaking is substantial. The poor determination of $\mathcal{H}$, which only appears in chiral loops, might be mostly responsible for this. Hence, the importance of performing Wu-type experiments, as it was pointed out in Ref.~\cite{leu2}, becomes more evident for this purpose.

In conclusion, second-order symmetry breaking corrections, while small, can have a significant impact on the evaluation of the baryon vector coupling, potentially increasing its overall strength due to the involved interplay of both explicit and implicit contributions.

\section*{Acknowledgment}
The authors are grateful to Consejo Nacional de Ciencia y Tecnolog{\'\i}a (Mexico) for support through the {\it Ciencia de Frontera} project CF-2023-I-162.

\appendix

\section{\label{app:pab}Explicit expressions for $[\mathcal{P}^{(\mathrm{dim})}O_n]^{ab}$}

The nonvanishing actions of projection operators $[\mathcal{P}^{(\mathrm{dim})}]^{abcd}$ acting on the basis $\{O_n^{cd}\}$, Eq.~(\ref{eq:onab}), for each dimension $\mathrm{dim}$ read, \\

\noindent
$\mathbf{1}$ flavor representation
\begin{equation}
[\mathcal{P}^{(1)} O_1]^{ab} = \delta^{ab},
\end{equation}
\begin{equation}
[\mathcal{P}^{(1)} O_4]^{ab} = \delta^{ab} J^2,
\end{equation}
\begin{equation}
[\mathcal{P}^{(1)} O_5]^{ab} = \frac12 \frac{N_c(N_c+2N_f)(N_f-2)}{N_f(N_f^2-1)} \delta^{ab} + \frac{2}{N_f^2-1} \delta^{ab} J^2,
\end{equation}
\begin{equation}
[\mathcal{P}^{(1)} O_6]^{ab} = \frac38 \frac{N_c(N_c+2N_f)}{N_f^2-1} \delta^{ab} - \frac12 \frac{N_f+2}{N_f(N_f-1)} \delta^{ab} J^2,
\end{equation}
\begin{equation}
[\mathcal{P}^{(1)} O_9]^{ab} = \frac{2(N_c+N_f)}{N_f(N_f+1)} \delta^{ab} J^2,
\end{equation}
\begin{equation}
[\mathcal{P}^{(1)} O_{10}]^{ab} = \frac{2(N_c+N_f)}{N_f(N_f+1)} \delta^{ab} J^2,
\end{equation}

\noindent
$\mathbf{8}$ flavor representation
\begin{equation}
[\mathcal{P}^{(8)} O_3]^{ab} = d^{abe} T^e,
\end{equation}
\begin{equation}
[\mathcal{P}^{(8)} O_5]^{ab} = \frac{(N_c+N_f)(N_f-4)}{N_f^2-4} d^{abe} T^e + \frac{2N_f}{N_f^2-4} d^{abe} \{J^r,G^{re}\},
\end{equation}
\begin{equation}
[\mathcal{P}^{(8)} O_6]^{ab} = \frac34 \frac{(N_c+N_f)N_f}{N_f^2-4} d^{abe} T^e - \frac12 \frac{N_f+4}{N_f^2-4} d^{abe} \{J^r,G^{re}\},
\end{equation}
\begin{equation}
[\mathcal{P}^{(8)} O_8]^{ab} = d^{abe} \{J^r,G^{re}\},
\end{equation}
\begin{equation}
[\mathcal{P}^{(8)} O_9]^{ab} = \frac{N_c+N_f}{N_f+2} d^{abe} \{J^r,G^{re}\} + \frac{1}{N_f+2} d^{abe} \{J^2,T^e\},
\end{equation}
\begin{equation}
[\mathcal{P}^{(8)} O_{10}]^{ab} = \frac{N_c+N_f}{N_f+2} d^{abe} \{J^r,G^{re}\} + \frac{1}{N_f+2} d^{abe} \{J^2,T^e\},
\end{equation}
\begin{equation}
[\mathcal{P}^{(8)} O_{12}]^{ab} = d^{abe} \{J^2,T^e\},
\end{equation}

\noindent
$\mathbf{27}$ flavor representation
\begin{eqnarray}
[\mathcal{P}^{(27)} O_5]^{ab} & = & \{T^a,T^b\} - \frac{2N_f}{N_f^2-4} d^{abe} \{J^r,G^{re}\} - \frac12 \frac{N_c(N_c+2N_f)(N_f-2)}{N_f(N_f^2-1)} \delta^{ab} \nonumber \\
& & \mbox{} - \frac{(N_c+N_f)(N_f-4)}{N_f^2-4} d^{abe} T^e - \frac{2}{N_f^2-1} \delta^{ab} J^2,
\end{eqnarray}
\begin{eqnarray}
[\mathcal{P}^{(27)} O_6]^{ab} & = & \{G^{ra},G^{rb}\} - \frac38 \frac{N_c(N_c+2N_f)}{N_f^2-1} \delta^{ab} - \frac34 \frac{(N_c+N_f)N_f}{N_f^2-4} d^{abe} T^e \nonumber \\
& & \mbox{} + \frac12 \frac{N_f+4}{N_f^2-4} d^{abe} \{J^r,G^{re}\} + \frac12 \frac{N_f+2}{N_f(N_f-1)} \delta^{ab} J^2,
\end{eqnarray}
\begin{eqnarray}
[\mathcal{P}^{(27)} O_9]^{ab} & = & \frac12 \{T^a,\{J^r,G^{rb}\}\} + \frac12 \{T^b,\{J^r,G^{ra}\}\} - \frac{N_c+N_f}{N_f+2} d^{abe} \{J^r,G^{re}\} \nonumber \\
& & \mbox{} - \frac{2(N_c+N_f)}{N_f (N_f+1)} \delta^{ab} J^2 - \frac{1}{N_f+2} d^{abe} \{J^2,T^e\},
\end{eqnarray}
\begin{eqnarray}
[\mathcal{P}^{(27)} O_{10}]^{ab} & = & \frac12 \{T^a,\{J^r,G^{rb}\}\} + \frac12 \{T^b,\{J^r,G^{ra}\}\} - \frac{N_c+N_f}{N_f+2} d^{abe} \{J^r,G^{re}\} \nonumber \\
& & \mbox{} - \frac{2(N_c+N_f)}{N_f(N_f+1)} \delta^{ab} J^2 - \frac{1}{N_f+2} d^{abe} \{J^2,T^e\}, 
\end{eqnarray}

\noindent
$\mathbf{8}_A$ flavor representation
\begin{equation}
[\mathcal{P}^{(8_A)} O_2]^{ab} = i f^{abe} T^e,
\end{equation}
\begin{equation}
[\mathcal{P}^{(8_A)} O_7]^{ab} = i f^{abe} \{J^r,G^{re}\},
\end{equation}
\begin{equation}
[\mathcal{P}^{(8_A)} O_{11}]^{ab} = i f^{abe} \{J^2,T^e\},
\end{equation}

\noindent
$\mathbf{10}+\overline{\mathbf{10}}$ flavor representation
\begin{equation}
[\mathcal{P}^{(10+\overline{10})} O_9]^{ab} = \frac12 \{T^a,\{J^r,G^{rb}\}\}-\frac12 \{T^b,\{J^r,G^{ra}\}\},
\end{equation}
\begin{equation}
[\mathcal{P}^{(10+\overline{10})} O_{10}]^{ab} = -\frac12 \{T^a,\{J^r,G^{rb}\}\}+\frac12 \{T^b,\{J^r,G^{ra}\}\}.
\end{equation}

\section{\label{app:oabc}Operator basis participating in second-order SB}

The $O^{a_1a_2a_3}$ basis is obtained from $R^{(ij)(a_1a_2a_3)}$ provided in Ref.~\cite{banda2} by simply contracting the spin indices with $\delta^{ij}$. After removing redundant operators, the basis becomes
\begin{equation}
O^{a_1a_2a_3} = \{O_i^{a_1a_2a_3}\}, \label{eq:mabc}
\end{equation}
where
\begin{eqnarray}
\label{eq:q1q2q3}
\begin{array}{lll}
O_1^{a_1a_2a_3} = i f^{a_1a_2a_3}, & \qquad &
O_2^{a_1a_2a_3} = d^{a_1a_2a_3}, \\
O_3^{a_1a_2a_3} = \delta^{a_1a_2} T^{a_3}, & \qquad &
O_4^{a_1a_2a_3} = \delta^{a_1a_3} T^{a_2}, \\
O_5^{a_1a_2a_3} = \delta^{a_2a_3} T^{a_1}, & \qquad &
O_6^{a_1a_2a_3} = i f^{a_1a_2a_3} J^2, \\
O_7^{a_1a_2a_3} = d^{a_1a_2a_3} J^2, & \qquad &
O_8^{a_1a_2a_3} = \delta^{a_1a_2} \{J^r,G^{ra_3}\}, \\
O_9^{a_1a_2a_3} = \delta^{a_1a_3} \{J^r,G^{ra_2}\}, & \qquad &
O_{10}^{a_1a_2a_3} = \delta^{a_2a_3} \{J^r,G^{ra_1}\}, \\
O_{11}^{a_1a_2a_3} = f^{a_1a_2e_1} f^{a_3e_1g_1} \{J^r,G^{rg_1}\}, & \qquad &
O_{12}^{a_1a_2a_3} = f^{a_1a_3e_1} f^{a_2e_1g_1} \{J^r,G^{rg_1}\}, \\
O_{13}^{a_1a_2a_3} = d^{a_1a_2e_1} d^{a_3e_1g_1} \{J^r,G^{rg_1}\}, & \qquad &
O_{14}^{a_1a_2a_3} = i f^{a_1a_2e_1} d^{a_3e_1g_1} \{J^r,G^{rg_1}\}, \\
O_{15}^{a_1a_2a_3} = i f^{a_1a_3e_1} d^{a_2e_1g_1} \{J^r,G^{rg_1}\}, & \qquad &
O_{16}^{a_1a_2a_3} = i d^{a_1e_1g_1} f^{a_2a_3e_1} \{J^r,G^{rg_1}\}, \\
O_{17}^{a_1a_2a_3} = i f^{a_1a_2e_1} \{T^{a_3},T^{e_1}\}, & \qquad &
O_{18}^{a_1a_2a_3} = d^{a_1a_2e_1} \{T^{a_3},T^{e_1}\}, \\
O_{19}^{a_1a_2a_3} = d^{a_1a_3e_1} \{T^{a_2},T^{e_1}\}, & \qquad &
O_{20}^{a_1a_2a_3} = d^{a_2a_3e_1} \{T^{a_1},T^{e_1}\}, \\
O_{21}^{a_1a_2a_3} = [T^{a_1},\{T^{a_2},T^{a_3}\}], & \qquad &
O_{22}^{a_1a_2a_3} = [T^{a_3},\{T^{a_1},T^{a_2}\}], \\
O_{23}^{a_1a_2a_3} = \delta^{a_1a_2} \{J^2,T^{a_3}\}, & \qquad &
O_{24}^{a_1a_2a_3} = \delta^{a_1a_3} \{J^2,T^{a_2}\}, \\
O_{25}^{a_1a_2a_3} = \delta^{a_2a_3} \{J^2,T^{a_1}\}, & \qquad &
O_{26}^{a_1a_2a_3} = \{T^{a_1},\{T^{a_2},T^{a_3}\}\}, \\
O_{27}^{a_1a_2a_3} = \{T^{a_2},\{T^{a_1},T^{a_3}\}\}, & \qquad &
O_{28}^{a_1a_2a_3} = \{T^{a_3},\{T^{a_1},T^{a_2}\}\}, \\
O_{29}^{a_1a_2a_3} = \{T^{a_1},\{G^{ra_2},G^{ra_3}\}\}, & \qquad &
O_{30}^{a_1a_2a_3} = \{T^{a_2},\{G^{ra_1},G^{ra_3}\}\}, \\
O_{31}^{a_1a_2a_3} = \{T^{a_3},\{G^{ra_1},G^{ra_2}\}\}, & \qquad &
O_{32}^{a_1a_2a_3} = i f^{a_1a_2e_1} \{T^{a_3},\{J^r,G^{re_1}\}\}, \\
O_{33}^{a_1a_2a_3} = i f^{a_1a_3e_1} \{T^{a_2},\{J^r,G^{re_1}\}\}, & \qquad &
O_{34}^{a_1a_2a_3} = i f^{a_2a_3e_1} \{T^{a_1},\{J^r,G^{re_1}\}\}, \\
O_{35}^{a_1a_2a_3} = i f^{a_1a_2e_1} \{T^{e_1},\{J^r,G^{ra_3}\}\}, & \qquad &
O_{36}^{a_1a_2a_3} = i f^{a_1a_3e_1} \{T^{e_1},\{J^r,G^{ra_2}\}\}, \\
O_{37}^{a_1a_2a_3} = i f^{a_2a_3e_1} \{T^{e_1},\{J^r,G^{ra_1}\}\}, & \qquad &
O_{38}^{a_1a_2a_3} = d^{a_1a_2e_1} \{T^{e_1},\{J^r,G^{ra_3}\}\}, \\
O_{39}^{a_1a_2a_3} = d^{a_1a_3e_1} \{T^{e_1},\{J^r,G^{ra_2}\}\}, & \qquad &
O_{40}^{a_1a_2a_3} = d^{a_2a_3e_1} \{T^{e_1},\{J^r,G^{ra_1}\}\}, \\
O_{41}^{a_1a_2a_3} = f^{a_1e_1g_1} f^{a_2e_1h_1} \{T^{a_3},\{G^{rg_1},G^{rh_1}\}\}, & \qquad &
O_{42}^{a_1a_2a_3} = f^{a_1e_1g_1} f^{a_2e_1h_1} \{T^{g_1},\{G^{ra_3},G^{rh_1}\}\}, \\
O_{43}^{a_1a_2a_3} = f^{a_1e_1g_1} f^{a_2e_1h_1} \{T^{h_1},\{G^{ra_3},G^{rg_1}\}\}, & \qquad &
O_{44}^{a_1a_2a_3} = f^{a_1e_1h_1} f^{a_3e_1g_1} \{T^{a_2},\{G^{rg_1},G^{rh_1}\}\}, \\
O_{45}^{a_1a_2a_3} = f^{a_1e_1h_1} f^{a_3e_1g_1} \{T^{g_1},\{G^{ra_2},G^{rh_1}\}\}, & \qquad &
O_{46}^{a_1a_2a_3} = f^{a_1e_1h_1} f^{a_3e_1g_1} \{T^{h_1},\{G^{ra_2},G^{rg_1}\}\}, \\
O_{47}^{a_1a_2a_3} = f^{a_2e_1g_1} f^{a_3e_1h_1} \{T^{a_1},\{G^{rg_1},G^{rh_1}\}\}, & \qquad &
O_{48}^{a_1a_2a_3} = f^{a_2e_1g_1} f^{a_3e_1h_1} \{T^{g_1},\{G^{ra_1},G^{rh_1}\}\}, \\
O_{49}^{a_1a_2a_3} = f^{a_2e_1g_1} f^{a_3e_1h_1} \{T^{h_1},\{G^{ra_1},G^{rg_1}\}\}, & \qquad &
O_{50}^{a_1a_2a_3} = d^{a_1e_1g_1} d^{a_2e_1h_1} \{T^{h_1},\{G^{ra_3},G^{rg_1}\}\}, \\
O_{51}^{a_1a_2a_3} = i d^{a_1e_1g_1} f^{a_2e_1h_1} \{T^{g_1},\{G^{ra_3},G^{rh_1}\}\}, & \qquad &
O_{52}^{a_1a_2a_3} = i d^{a_1e_1g_1} f^{a_2e_1h_1} \{T^{h_1},\{G^{ra_3},G^{rg_1}\}\}, \\
O_{53}^{a_1a_2a_3} = i f^{a_1e_1h_1} d^{a_3e_1g_1} \{T^{g_1},\{G^{ra_2},G^{rh_1}\}\}, & \qquad &
O_{54}^{a_1a_2a_3} = i f^{a_1e_1h_1} d^{a_3e_1g_1} \{T^{h_1},\{G^{ra_2},G^{rg_1}\}\}, \\
O_{55}^{a_1a_2a_3} = i d^{a_2e_1g_1} f^{a_3e_1h_1} \{T^{g_1},\{G^{ra_1},G^{rh_1}\}\}, & \qquad &
O_{56}^{a_1a_2a_3} = i d^{a_2e_1g_1} f^{a_3e_1h_1} \{T^{h_1},\{G^{ra_1},G^{rg_1}\}\}, \\
O_{57}^{a_1a_2a_3} = i f^{a_1e_1h_1} d^{a_2e_1g_1} \{T^{h_1},\{G^{ra_3},G^{rg_1}\}\}, & \qquad &
O_{58}^{a_1a_2a_3} = i f^{a_2e_1h_1} d^{a_3e_1g_1} \{T^{h_1},\{G^{ra_1},G^{rg_1}\}\}, \\
O_{59}^{a_1a_2a_3} = i d^{a_1e_1g_1} f^{a_3e_1h_1} \{T^{h_1},\{G^{ra_2},G^{rg_1}\}\}. & \qquad &
\end{array}
\end{eqnarray}

\section{\label{app:q1q2q3} Explicit expressions of projection operators acting on the product of three adjoints}

\begin{equation}
[\mathcal{P}^{(1)} Q_1Q_2Q_3]^{188} = 0,
\end{equation}
\begin{equation}
[\mathcal{P}^{(1)} Q_1Q_2Q_3]^{288} = 0,
\end{equation}
\begin{equation}
[\mathcal{P}^{(1)} Q_1Q_2Q_3]^{488} = 0,
\end{equation}
\begin{equation}
[\mathcal{P}^{(1)} Q_1Q_2Q_3]^{588} = 0,
\end{equation}
\begin{eqnarray}
& & [\mathcal{P}^{(8)} Q_1Q_2Q_3]^{188} \nonumber \\
& & \mbox{\hglue0.2truecm} = \frac{1}{10} Q_1^1Q_2^1Q_3^1 + \frac{7}{90} Q_1^1Q_2^2Q_3^2 + \frac{7}{90} Q_1^1Q_2^3Q_3^3 + \frac{13}{90} Q_1^1Q_2^4Q_3^4 + \frac{13}{90} Q_1^1Q_2^5Q_3^5 + \frac{13}{90} Q_1^1Q_2^6Q_3^6 \nonumber \\
& & \mbox{\hglue0.6truecm}  + \frac{13}{90} Q_1^1Q_2^7Q_3^7 + \frac16 Q_1^1Q_2^8Q_3^8 + \frac{1}{90} Q_1^2Q_2^1Q_3^2 + \frac{1}{90} Q_1^2Q_2^2Q_3^1 + \frac{1}{90} Q_1^3Q_2^1Q_3^3 + \frac{1}{90} Q_1^3Q_2^3Q_3^1 \nonumber \\
& & \mbox{\hglue0.6truecm}  - \frac{1}{45} Q_1^4Q_2^1Q_3^4 - \frac{1}{30} Q_1^4Q_2^2Q_3^5 + \frac{1}{30} Q_1^4Q_2^3Q_3^6 - \frac{1}{45} Q_1^4Q_2^4Q_3^1 - \frac{1}{30} Q_1^4Q_2^5Q_3^2 + \frac{1}{30} Q_1^4Q_2^6Q_3^3 \nonumber \\
& & \mbox{\hglue0.6truecm} + \frac{1}{30\sqrt{3}} Q_1^4Q_2^6Q_3^8 + \frac{1}{30\sqrt{3}} Q_1^4Q_2^8Q_3^6 - \frac{1}{45} Q_1^5Q_2^1Q_3^5 + \frac{1}{30} Q_1^5Q_2^2Q_3^4 + \frac{1}{30} Q_1^5Q_2^3Q_3^7 \nonumber \\
& & \mbox{\hglue0.6truecm}  + \frac{1}{30} Q_1^5Q_2^4Q_3^2 - \frac{1}{45} Q_1^5Q_2^5Q_3^1 + \frac{1}{30} Q_1^5Q_2^7Q_3^3 + \frac{1}{30\sqrt{3}} Q_1^5Q_2^7Q_3^8 + \frac{1}{30\sqrt{3}} Q_1^5Q_2^8Q_3^7 \nonumber \\
& & \mbox{\hglue0.6truecm}  - \frac{1}{45} Q_1^6Q_2^1Q_3^6 + \frac{1}{30} Q_1^6Q_2^2Q_3^7 - \frac{1}{30} Q_1^6Q_2^3Q_3^4 - \frac{1}{30} Q_1^6Q_2^4Q_3^3 + \frac{1}{30\sqrt{3}} Q_1^6Q_2^4Q_3^8 \nonumber \\
& & \mbox{\hglue0.6truecm} - \frac{1}{45} Q_1^6Q_2^6Q_3^1 + \frac{1}{30} Q_1^6Q_2^7Q_3^2 + \frac{1}{30\sqrt{3}} Q_1^6Q_2^8Q_3^4 - \frac{1}{45} Q_1^7Q_2^1Q_3^7 - \frac{1}{30} Q_1^7Q_2^2Q_3^6 \nonumber \\
& & \mbox{\hglue0.6truecm}  - \frac{1}{30} Q_1^7Q_2^3Q_3^5 - \frac{1}{30} Q_1^7Q_2^5Q_3^3 + \frac{1}{30\sqrt{3}} Q_1^7Q_2^5Q_3^8 - \frac{1}{30} Q_1^7Q_2^6Q_3^2 - \frac{1}{45} Q_1^7Q_2^7Q_3^1 \nonumber \\
& & \mbox{\hglue0.6truecm} + \frac{1}{30\sqrt{3}} Q_1^7Q_2^8Q_3^5 - \frac{1}{30} Q_1^8Q_2^1Q_3^8 - \frac{1}{15\sqrt{3}} Q_1^8Q_2^4Q_3^6 - \frac{1}{15\sqrt{3}} Q_1^8Q_2^5Q_3^7 - \frac{1}{15\sqrt{3}} Q_1^8Q_2^6Q_3^4 \nonumber \\
& & \mbox{\hglue0.6truecm} - \frac{1}{15\sqrt{3}} Q_1^8Q_2^7Q_3^5 - \frac{1}{30} Q_1^8Q_2^8Q_3^1,
\end{eqnarray}
\begin{eqnarray}
& & [\mathcal{P}^{(8)} Q_1Q_2Q_3]^{288} \nonumber \\
& & \mbox{\hglue0.2truecm} = \frac{1}{90} Q_1^1Q_2^1Q_3^2 + \frac{1}{90} Q_1^1Q_2^2Q_3^1 + \frac{7}{90} Q_1^2Q_2^1Q_3^1 + \frac{1}{10} Q_1^2Q_2^2Q_3^2 + \frac{7}{90} Q_1^2Q_2^3Q_3^3 + \frac{13}{90} Q_1^2Q_2^4Q_3^4 \nonumber \\
& & \mbox{\hglue0.6truecm} + \frac{13}{90} Q_1^2Q_2^5Q_3^5 + \frac{13}{90} Q_1^2Q_2^6Q_3^6 + \frac{13}{90} Q_1^2Q_2^7Q_3^7 + \frac16 Q_1^2Q_2^8Q_3^8 + \frac{1}{90} Q_1^3Q_2^2Q_3^3 + \frac{1}{90} Q_1^3Q_2^3Q_3^2 \nonumber \\
& & \mbox{\hglue0.6truecm} + \frac{1}{30} Q_1^4Q_2^1Q_3^5 - \frac{1}{45} Q_1^4Q_2^2Q_3^4 - \frac{1}{30} Q_1^4Q_2^3Q_3^7 - \frac{1}{45} Q_1^4Q_2^4Q_3^2 + \frac{1}{30} Q_1^4Q_2^5Q_3^1 - \frac{1}{30} Q_1^4Q_2^7Q_3^3 \nonumber \\
& & \mbox{\hglue0.6truecm} - \frac{1}{30\sqrt{3}} Q_1^4Q_2^7Q_3^8 - \frac{1}{30\sqrt{3}} Q_1^4Q_2^8Q_3^7 - \frac{1}{30} Q_1^5Q_2^1Q_3^4 - \frac{1}{45} Q_1^5Q_2^2Q_3^5 + \frac{1}{30} Q_1^5Q_2^3Q_3^6 \nonumber \\
& & \mbox{\hglue0.6truecm} - \frac{1}{30} Q_1^5Q_2^4Q_3^1 - \frac{1}{45} Q_1^5Q_2^5Q_3^2 + \frac{1}{30} Q_1^5Q_2^6Q_3^3 + \frac{1}{30\sqrt{3}} Q_1^5Q_2^6Q_3^8 + \frac{1}{30\sqrt{3}} Q_1^5Q_2^8Q_3^6 \nonumber \\
& & \mbox{\hglue0.6truecm}  - \frac{1}{30} Q_1^6Q_2^1Q_3^7 - \frac{1}{45} Q_1^6Q_2^2Q_3^6 - \frac{1}{30} Q_1^6Q_2^3Q_3^5 - \frac{1}{30} Q_1^6Q_2^5Q_3^3 + \frac{1}{30\sqrt{3}} Q_1^6Q_2^5Q_3^8 \nonumber \\
& & \mbox{\hglue0.6truecm} - \frac{1}{45} Q_1^6Q_2^6Q_3^2 - \frac{1}{30} Q_1^6Q_2^7Q_3^1 + \frac{1}{30\sqrt{3}} Q_1^6Q_2^8Q_3^5 + \frac{1}{30} Q_1^7Q_2^1Q_3^6 - \frac{1}{45} Q_1^7Q_2^2Q_3^7 \nonumber \\
& & \mbox{\hglue0.6truecm} + \frac{1}{30} Q_1^7Q_2^3Q_3^4 + \frac{1}{30} Q_1^7Q_2^4Q_3^3 - \frac{1}{30\sqrt{3}} Q_1^7Q_2^4Q_3^8 + \frac{1}{30} Q_1^7Q_2^6Q_3^1 - \frac{1}{45} Q_1^7Q_2^7Q_3^2 \nonumber \\
& & \mbox{\hglue0.6truecm} - \frac{1}{30\sqrt{3}} Q_1^7Q_2^8Q_3^4 - \frac{1}{30} Q_1^8Q_2^2Q_3^8 + \frac{1}{15\sqrt{3}} Q_1^8Q_2^4Q_3^7 - \frac{1}{15\sqrt{3}} Q_1^8Q_2^5Q_3^6 - \frac{1}{15\sqrt{3}} Q_1^8Q_2^6Q_3^5 \nonumber \\
& & \mbox{\hglue0.6truecm} + \frac{1}{15\sqrt{3}} Q_1^8Q_2^7Q_3^4 - \frac{1}{30} Q_1^8Q_2^8Q_3^2,
\end{eqnarray}
\begin{eqnarray}
& & [\mathcal{P}^{(8)} Q_1Q_2Q_3]^{488} \nonumber \\
& & \mbox{\hglue0.2truecm} = \frac{1}{60} Q_1^1Q_2^1Q_3^4 + \frac{1}{20} Q_1^1Q_2^2Q_3^5 - \frac{1}{20} Q_1^1Q_2^3Q_3^6 + \frac{1}{60} Q_1^1Q_2^4Q_3^1 + \frac{1}{20} Q_1^1Q_2^5Q_3^2 - \frac{1}{20} Q_1^1Q_2^6Q_3^3 \nonumber \\
& & \mbox{\hglue0.6truecm} - \frac{1}{20\sqrt{3}} Q_1^1Q_2^6Q_3^8 - \frac{1}{20\sqrt{3}} Q_1^1Q_2^8Q_3^6 - \frac{1}{20} Q_1^2Q_2^1Q_3^5 + \frac{1}{60} Q_1^2Q_2^2Q_3^4 + \frac{1}{20} Q_1^2Q_2^3Q_3^7 \nonumber \\
& & \mbox{\hglue0.6truecm} + \frac{1}{60} Q_1^2Q_2^4Q_3^2 - \frac{1}{20} Q_1^2Q_2^5Q_3^1 + \frac{1}{20} Q_1^2Q_2^7Q_3^3 + \frac{1}{20\sqrt{3}} Q_1^2Q_2^7Q_3^8 + \frac{1}{20\sqrt{3}} Q_1^2Q_2^8Q_3^7 \nonumber \\
& & \mbox{\hglue0.6truecm} + \frac{1}{20} Q_1^3Q_2^1Q_3^6 - \frac{1}{20} Q_1^3Q_2^2Q_3^7 + \frac{1}{60} Q_1^3Q_2^3Q_3^4 + \frac{1}{60} Q_1^3Q_2^4Q_3^3 - \frac{1}{20\sqrt{3}} Q_1^3Q_2^4Q_3^8 \nonumber \\
& & \mbox{\hglue0.6truecm} + \frac{1}{20} Q_1^3Q_2^6Q_3^1 - \frac{1}{20} Q_1^3Q_2^7Q_3^2 - \frac{1}{20\sqrt{3}} Q_1^3Q_2^8Q_3^4 + \frac{1}{15} Q_1^4Q_2^1Q_3^1 + \frac{1}{15} Q_1^4Q_2^2Q_3^2 \nonumber \\
& & \mbox{\hglue0.6truecm} + \frac{1}{15} Q_1^4Q_2^3Q_3^3 + \frac{1}{10} Q_1^4Q_2^4Q_3^4 + \frac16 Q_1^4Q_2^5Q_3^5 + \frac16 Q_1^4Q_2^6Q_3^6 + \frac16 Q_1^4Q_2^7Q_3^7 + \frac15 Q_1^4Q_2^8Q_3^8 \nonumber \\
& & \mbox{\hglue0.6truecm} - \frac{1}{30} Q_1^5Q_2^4Q_3^5 - \frac{1}{30} Q_1^5Q_2^5Q_3^4 - \frac{1}{30} Q_1^6Q_2^4Q_3^6 - \frac{1}{30} Q_1^6Q_2^6Q_3^4 - \frac{1}{30} Q_1^7Q_2^4Q_3^7 - \frac{1}{30} Q_1^7Q_2^7Q_3^4 \nonumber \\
& & \mbox{\hglue0.6truecm} + \frac{1}{20\sqrt{3}} Q_1^8Q_2^1Q_3^6 - \frac{1}{20\sqrt{3}} Q_1^8Q_2^2Q_3^7 + \frac{1}{20\sqrt{3}} Q_1^8Q_2^3Q_3^4 + \frac{1}{20\sqrt{3}} Q_1^8Q_2^4Q_3^3 - \frac{1}{20} Q_1^8Q_2^4Q_3^8 \nonumber \\
& & \mbox{\hglue0.6truecm} + \frac{1}{20\sqrt{3}} Q_1^8Q_2^6Q_3^1 - \frac{1}{20\sqrt{3}} Q_1^8Q_2^7Q_3^2 - \frac{1}{20} Q_1^8Q_2^8Q_3^4,
\end{eqnarray}
\begin{eqnarray}
& & [\mathcal{P}^{(8)} Q_1Q_2Q_3]^{588} \nonumber \\
& & \mbox{\hglue0.2truecm} = \frac{1}{60} Q_1^1Q_2^1Q_3^5 - \frac{1}{20} Q_1^1Q_2^2Q_3^4 - \frac{1}{20} Q_1^1Q_2^3Q_3^7 - \frac{1}{20} Q_1^1Q_2^4Q_3^2 + \frac{1}{60} Q_1^1Q_2^5Q_3^1 - \frac{1}{20} Q_1^1Q_2^7Q_3^3 \nonumber \\
& & \mbox{\hglue0.6truecm} - \frac{1}{20\sqrt{3}} Q_1^1Q_2^7Q_3^8 - \frac{1}{20\sqrt{3}} Q_1^1Q_2^8Q_3^7 + \frac{1}{20} Q_1^2Q_2^1Q_3^4 + \frac{1}{60} Q_1^2Q_2^2Q_3^5 - \frac{1}{20} Q_1^2Q_2^3Q_3^6 \nonumber \\
& & \mbox{\hglue0.6truecm} + \frac{1}{20} Q_1^2Q_2^4Q_3^1 + \frac{1}{60} Q_1^2Q_2^5Q_3^2 - \frac{1}{20} Q_1^2Q_2^6Q_3^3 - \frac{1}{20\sqrt{3}} Q_1^2Q_2^6Q_3^8 - \frac{1}{20\sqrt{3}} Q_1^2Q_2^8Q_3^6 \nonumber \\
& & \mbox{\hglue0.6truecm} + \frac{1}{20} Q_1^3Q_2^1Q_3^7 + \frac{1}{20} Q_1^3Q_2^2Q_3^6 + \frac{1}{60} Q_1^3Q_2^3Q_3^5 + \frac{1}{60} Q_1^3Q_2^5Q_3^3 - \frac{1}{20\sqrt{3}} Q_1^3Q_2^5Q_3^8 \nonumber \\
& & \mbox{\hglue0.6truecm} + \frac{1}{20} Q_1^3Q_2^6Q_3^2 + \frac{1}{20} Q_1^3Q_2^7Q_3^1 - \frac{1}{20\sqrt{3}} Q_1^3Q_2^8Q_3^5 - \frac{1}{30} Q_1^4Q_2^4Q_3^5 - \frac{1}{30} Q_1^4Q_2^5Q_3^4 \nonumber \\
& & \mbox{\hglue0.6truecm} + \frac{1}{15} Q_1^5Q_2^1Q_3^1 + \frac{1}{15} Q_1^5Q_2^2Q_3^2 + \frac{1}{15} Q_1^5Q_2^3Q_3^3 + \frac16 Q_1^5Q_2^4Q_3^4 + \frac{1}{10} Q_1^5Q_2^5Q_3^5 + \frac16 Q_1^5Q_2^6Q_3^6 \nonumber \\
& & \mbox{\hglue0.6truecm} + \frac16 Q_1^5Q_2^7Q_3^7 + \frac15 Q_1^5Q_2^8Q_3^8 - \frac{1}{30} Q_1^6Q_2^5Q_3^6 - \frac{1}{30} Q_1^6Q_2^6Q_3^5 - \frac{1}{30} Q_1^7Q_2^5Q_3^7 - \frac{1}{30} Q_1^7Q_2^7Q_3^5 \nonumber \\
& & \mbox{\hglue0.6truecm} + \frac{1}{20\sqrt{3}} Q_1^8Q_2^1Q_3^7 + \frac{1}{20\sqrt{3}} Q_1^8Q_2^2Q_3^6 + \frac{1}{20\sqrt{3}} Q_1^8Q_2^3Q_3^5 + \frac{1}{20\sqrt{3}} Q_1^8Q_2^5Q_3^3 - \frac{1}{20} Q_1^8Q_2^5Q_3^8 \nonumber \\
& & \mbox{\hglue0.6truecm} + \frac{1}{20\sqrt{3}} Q_1^8Q_2^6Q_3^2 + \frac{1}{20\sqrt{3}} Q_1^8Q_2^7Q_3^1 - \frac{1}{20} Q_1^8Q_2^8Q_3^5,
\end{eqnarray}
\begin{eqnarray}
& & [\mathcal{P}^{(10+\overline{10})} Q_1Q_2Q_3]^{188} \nonumber \\
& & \mbox{\hglue0.2truecm} = - \frac16 Q_1^1Q_2^1Q_3^1 - \frac{1}{18} Q_1^1Q_2^2Q_3^2 - \frac{1}{18} Q_1^1Q_2^3Q_3^3 + \frac{1}{36} Q_1^1Q_2^4Q_3^4 + \frac{1}{36} Q_1^1Q_2^5Q_3^5 + \frac{1}{36} Q_1^1Q_2^6Q_3^6 \nonumber \\
& & \mbox{\hglue0.6truecm} + \frac{1}{36} Q_1^1Q_2^7Q_3^7 + \frac16 Q_1^1Q_2^8Q_3^8 - \frac{1}{18} Q_1^2Q_2^1Q_3^2 - \frac{1}{18} Q_1^2Q_2^2Q_3^1 - \frac{1}{18} Q_1^3Q_2^1Q_3^3 - \frac{1}{18} Q_1^3Q_2^3Q_3^1 \nonumber \\
& & \mbox{\hglue0.6truecm} + \frac{1}{36} Q_1^4Q_2^1Q_3^4 + \frac{1}{36} Q_1^4Q_2^4Q_3^1 + \frac{1}{12\sqrt{3}} Q_1^4Q_2^6Q_3^8 + \frac{1}{12\sqrt{3}} Q_1^4Q_2^8Q_3^6 + \frac{1}{36} Q_1^5Q_2^1Q_3^5 \nonumber \\
& & \mbox{\hglue0.6truecm} + \frac{1}{36} Q_1^5Q_2^5Q_3^1 + \frac{1}{12\sqrt{3}} Q_1^5Q_2^7Q_3^8 + \frac{1}{12\sqrt{3}} Q_1^5Q_2^8Q_3^7  + \frac{1}{36} Q_1^6Q_2^1Q_3^6 + \frac{1}{12\sqrt{3}} Q_1^6Q_2^4Q_3^8 \nonumber \\
& & \mbox{\hglue0.6truecm} + \frac{1}{36} Q_1^6Q_2^6Q_3^1 + \frac{1}{12\sqrt{3}} Q_1^6Q_2^8Q_3^4 + \frac{1}{36} Q_1^7Q_2^1Q_3^7 + \frac{1}{12\sqrt{3}} Q_1^7Q_2^5Q_3^8 + \frac{1}{36} Q_1^7Q_2^7Q_3^1 \nonumber \\
& & \mbox{\hglue0.6truecm} + \frac{1}{12\sqrt{3}} Q_1^7Q_2^8Q_3^5 + \frac16 Q_1^8Q_2^1Q_3^8 + \frac{1}{12\sqrt{3}} Q_1^8Q_2^4Q_3^6 + \frac{1}{12\sqrt{3}} Q_1^8Q_2^5Q_3^7 + \frac{1}{12\sqrt{3}} Q_1^8Q_2^6Q_3^4 \nonumber \\
& & \mbox{\hglue0.6truecm} + \frac{1}{12\sqrt{3}} Q_1^8Q_2^7Q_3^5 + \frac16 Q_1^8Q_2^8Q_3^1,
\end{eqnarray}
\begin{eqnarray}
& & [\mathcal{P}^{(10+\overline{10})} Q_1Q_2Q_3]^{288} \nonumber \\ 
& & \mbox{\hglue0.2truecm} = - \frac{1}{18} Q_1^1Q_2^1Q_3^2 - \frac{1}{18} Q_1^1Q_2^2Q_3^1 - \frac{1}{18} Q_1^2Q_2^1Q_3^1 - \frac16 Q_1^2Q_2^2Q_3^2 - \frac{1}{18} Q_1^2Q_2^3Q_3^3 + \frac{1}{36} Q_1^2Q_2^4Q_3^4 \nonumber \\
& & \mbox{\hglue0.6truecm} + \frac{1}{36} Q_1^2Q_2^5Q_3^5 + \frac{1}{36} Q_1^2Q_2^6Q_3^6 + \frac{1}{36} Q_1^2Q_2^7Q_3^7 + \frac16 Q_1^2Q_2^8Q_3^8 - \frac{1}{18} Q_1^3Q_2^2Q_3^3 - \frac{1}{18} Q_1^3Q_2^3Q_3^2 \nonumber \\
& & \mbox{\hglue0.6truecm} + \frac{1}{36} Q_1^4Q_2^2Q_3^4 + \frac{1}{36} Q_1^4Q_2^4Q_3^2 - \frac{1}{12\sqrt{3}} Q_1^4Q_2^7Q_3^8 - \frac{1}{12\sqrt{3}} Q_1^4Q_2^8Q_3^7 + \frac{1}{36} Q_1^5Q_2^2Q_3^5 \nonumber \\
& & \mbox{\hglue0.6truecm} + \frac{1}{36} Q_1^5Q_2^5Q_3^2 + \frac{1}{12\sqrt{3}} Q_1^5Q_2^6Q_3^8 + \frac{1}{12 \sqrt{3}} Q_1^5Q_2^8Q_3^6  + \frac{1}{36} Q_1^6Q_2^2Q_3^6 + \frac{1}{12\sqrt{3}} Q_1^6Q_2^5Q_3^8 \nonumber \\
& & \mbox{\hglue0.6truecm} + \frac{1}{36} Q_1^6Q_2^6Q_3^2 + \frac{1}{12\sqrt{3}} Q_1^6Q_2^8Q_3^5 + \frac{1}{36} Q_1^7Q_2^2Q_3^7 - \frac{1}{12\sqrt{3}} Q_1^7Q_2^4Q_3^8 + \frac{1}{36} Q_1^7Q_2^7Q_3^2 \nonumber \\
& & \mbox{\hglue0.6truecm} - \frac{1}{12 \sqrt{3}} Q_1^7Q_2^8Q_3^4 + \frac16 Q_1^8Q_2^2Q_3^8 - \frac{1}{12\sqrt{3}} Q_1^8Q_2^4Q_3^7 + \frac{1}{12\sqrt{3}} Q_1^8Q_2^5Q_3^6 + \frac{1}{12\sqrt{3}} Q_1^8Q_2^6Q_3^5 \nonumber \\
& & \mbox{\hglue0.6truecm} - \frac{1}{12 \sqrt{3}} Q_1^8Q_2^7Q_3^4 + \frac16 Q_1^8Q_2^8Q_3^2,
\end{eqnarray}
\begin{eqnarray}
& & [\mathcal{P}^{(10+\overline{10})} Q_1Q_2Q_3]^{488} \nonumber \\
& & \mbox{\hglue0.2truecm} = \frac{1}{48} Q_1^1Q_2^1Q_3^4 - \frac{1}{16} Q_1^1Q_2^2Q_3^5 + \frac{1}{16} Q_1^1Q_2^3Q_3^6 + \frac{1}{48} Q_1^1Q_2^4Q_3^1 - \frac{1}{16} Q_1^1Q_2^5Q_3^2 + \frac{1}{16} Q_1^1Q_2^6Q_3^3 \nonumber \\
& & \mbox{\hglue0.6truecm} - \frac{1}{16\sqrt{3}} Q_1^1Q_2^6Q_3^8 - \frac{1}{16\sqrt{3}} Q_1^1Q_2^8Q_3^6 + \frac{1}{16} Q_1^2Q_2^1Q_3^5 + \frac{1}{48} Q_1^2Q_2^2Q_3^4 - \frac{1}{16} Q_1^2Q_2^3Q_3^7 \nonumber \\
& & \mbox{\hglue0.6truecm} + \frac{1}{48} Q_1^2Q_2^4Q_3^2 + \frac{1}{16} Q_1^2Q_2^5Q_3^1 - \frac{1}{16} Q_1^2Q_2^7Q_3^3 + \frac{1}{16\sqrt{3}} Q_1^2Q_2^7Q_3^8 + \frac{1}{16\sqrt{3}} Q_1^2Q_2^8Q_3^7 \nonumber \\
& & \mbox{\hglue0.6truecm} - \frac{1}{16} Q_1^3Q_2^1Q_3^6 + \frac{1}{16} Q_1^3Q_2^2Q_3^7 + \frac{1}{48} Q_1^3Q_2^3Q_3^4 + \frac{1}{48} Q_1^3Q_2^4Q_3^3 - \frac{1}{16\sqrt{3}} Q_1^3Q_2^4Q_3^8 \nonumber \\
& & \mbox{\hglue0.6truecm} - \frac{1}{16} Q_1^3Q_2^6Q_3^1 + \frac{1}{16} Q_1^3Q_2^7Q_3^2 - \frac{1}{16\sqrt{3}} Q_1^3Q_2^8Q_3^4 - \frac{1}{24} Q_1^4Q_2^1Q_3^1 - \frac{1}{24} Q_1^4Q_2^2Q_3^2 \nonumber \\
& & \mbox{\hglue0.6truecm} - \frac{1}{24} Q_1^4Q_2^3Q_3^3 + \frac{1}{12} Q_1^4Q_2^5Q_3^5 - \frac{1}{24}
Q_1^4Q_2^6Q_3^6 - \frac{1}{24} Q_1^4Q_2^7Q_3^7 + \frac18 Q_1^4Q_2^8Q_3^8 - \frac{1}{24} Q_1^5Q_2^4Q_3^5 \nonumber \\
& & \mbox{\hglue0.6truecm} - \frac{1}{24} Q_1^5Q_2^5Q_3^4 + \frac{1}{48} Q_1^6Q_2^4Q_3^6 + \frac{1}{16} Q_1^6Q_2^5Q_3^7 + \frac{1}{48} Q_1^6Q_2^6Q_3^4 + \frac{1}{16} Q_1^6Q_2^7Q_3^5 + \frac{1}{48} Q_1^7Q_2^4Q_3^7 \nonumber \\
& & \mbox{\hglue0.6truecm} - \frac{1}{16} Q_1^7Q_2^5Q_3^6 - \frac{1}{16} Q_1^7Q_2^6Q_3^5 + \frac{1}{48} Q_1^7Q_2^7Q_3^4 + \frac{1}{16\sqrt{3}} Q_1^8Q_2^1Q_3^6 - \frac{1}{16\sqrt{3}} Q_1^8Q_2^2Q_3^7 \nonumber \\
& & \mbox{\hglue0.6truecm} + \frac{1}{16\sqrt{3}} Q_1^8Q_2^3Q_3^4 + \frac{1}{16\sqrt{3}} Q_1^8Q_2^4Q_3^3 - \frac{1}{16} Q_1^8Q_2^4Q_3^8 + \frac{1}{16\sqrt{3}} Q_1^8Q_2^6Q_3^1 - \frac{1}{16\sqrt{3}} Q_1^8Q_2^7Q_3^2 \nonumber \\
& & \mbox{\hglue0.6truecm} - \frac{1}{16} Q_1^8Q_2^8Q_3^4,
\end{eqnarray}
\begin{eqnarray}
& & [\mathcal{P}^{(10+\overline{10})} Q_1Q_2Q_3]^{588} \nonumber \\
& & \mbox{\hglue0.2truecm} = \frac{1}{48} Q_1^1Q_2^1Q_3^5 + \frac{1}{16} Q_1^1Q_2^2Q_3^4 + \frac{1}{16} Q_1^1Q_2^3Q_3^7 + \frac{1}{16} Q_1^1Q_2^4Q_3^2 + \frac{1}{48} Q_1^1Q_2^5Q_3^1 + \frac{1}{16} Q_1^1Q_2^7Q_3^3 \nonumber \\
& & \mbox{\hglue0.6truecm} - \frac{1}{16\sqrt{3}} Q_1^1Q_2^7Q_3^8 - \frac{1}{16\sqrt{3}} Q_1^1Q_2^8Q_3^7 - \frac{1}{16} Q_1^2Q_2^1Q_3^4 + \frac{1}{48} Q_1^2Q_2^2Q_3^5 + \frac{1}{16} Q_1^2Q_2^3Q_3^6 \nonumber \\
& & \mbox{\hglue0.6truecm} - \frac{1}{16} Q_1^2Q_2^4Q_3^1 + \frac{1}{48} Q_1^2Q_2^5Q_3^2 + \frac{1}{16} Q_1^2Q_2^6Q_3^3 - \frac{1}{16\sqrt{3}} Q_1^2Q_2^6Q_3^8 - \frac{1}{16\sqrt{3}} Q_1^2Q_2^8Q_3^6 \nonumber \\
& & \mbox{\hglue0.6truecm} - \frac{1}{16} Q_1^3Q_2^1Q_3^7 - \frac{1}{16} Q_1^3Q_2^2Q_3^6 + \frac{1}{48} Q_1^3Q_2^3Q_3^5 + \frac{1}{48} Q_1^3Q_2^5Q_3^3 - \frac{1}{16\sqrt{3}} Q_1^3Q_2^5Q_3^8 \nonumber \\
& & \mbox{\hglue0.6truecm} - \frac{1}{16} Q_1^3Q_2^6Q_3^2 - \frac{1}{16} Q_1^3Q_2^7Q_3^1 - \frac{1}{16\sqrt{3}} Q_1^3Q_2^8Q_3^5 - \frac{1}{24} Q_1^4Q_2^4Q_3^5 - \frac{1}{24} Q_1^4Q_2^5Q_3^4 \nonumber \\
& & \mbox{\hglue0.6truecm} - \frac{1}{24} Q_1^5Q_2^1Q_3^1 - \frac{1}{24} Q_1^5Q_2^2Q_3^2 - \frac{1}{24}
Q_1^5Q_2^3Q_3^3 + \frac{1}{12} Q_1^5Q_2^4Q_3^4 - \frac{1}{24} Q_1^5Q_2^6Q_3^6 - \frac{1}{24} Q_1^5Q_2^7Q_3^7 \nonumber \\
& & \mbox{\hglue0.6truecm} + \frac18 Q_1^5Q_2^8Q_3^8 - \frac{1}{16} Q_1^6Q_2^4Q_3^7 + \frac{1}{48} Q_1^6Q_2^5Q_3^6 + \frac{1}{48} Q_1^6Q_2^6Q_3^5 - \frac{1}{16} Q_1^6Q_2^7Q_3^4 + \frac{1}{16} Q_1^7Q_2^4Q_3^6 \nonumber \\
& & \mbox{\hglue0.6truecm} + \frac{1}{48} Q_1^7Q_2^5Q_3^7 + \frac{1}{16} Q_1^7Q_2^6Q_3^4 + \frac{1}{48} Q_1^7Q_2^7Q_3^5 + \frac{1}{16\sqrt{3}} Q_1^8Q_2^1Q_3^7 + \frac{1}{16\sqrt{3}} Q_1^8Q_2^2Q_3^6 \nonumber \\
& & \mbox{\hglue0.6truecm} + \frac{1}{16\sqrt{3}} Q_1^8Q_2^3Q_3^5 + \frac{1}{16\sqrt{3}} Q_1^8Q_2^5Q_3^3 - \frac{1}{16} Q_1^8Q_2^5Q_3^8 + \frac{1}{16\sqrt{3}} Q_1^8Q_2^6Q_3^2 + \frac{1}{16\sqrt{3}} Q_1^8Q_2^7Q_3^1 \nonumber \\
& & \mbox{\hglue0.6truecm} - \frac{1}{16} Q_1^8Q_2^8Q_3^5,
\end{eqnarray}
\begin{eqnarray}
& & [\mathcal{P}^{(27)} Q_1Q_2Q_3]^{188} \nonumber \\
& & \mbox{\hglue0.2truecm} = \frac{3}{70} Q_1^1Q_2^1Q_3^1 - \frac{3}{35} Q_1^1Q_2^2Q_3^2 - \frac{3}{35} Q_1^1Q_2^3Q_3^3 - \frac{3}{140} Q_1^1Q_2^4Q_3^4 - \frac{3}{140} Q_1^1Q_2^5Q_3^5 - \frac{3}{140} Q_1^1Q_2^6Q_3^6 \nonumber \\
& & \mbox{\hglue0.6truecm} - \frac{3}{140} Q_1^1Q_2^7Q_3^7 + \frac{3}{14} Q_1^1Q_2^8Q_3^8 + \frac{9}{140} Q_1^2Q_2^1Q_3^2 + \frac{9}{140} Q_1^2Q_2^2Q_3^1 + \frac{9}{140} Q_1^3Q_2^1Q_3^3 \nonumber \\
& & \mbox{\hglue0.6truecm} + \frac{9}{140} Q_1^3Q_2^3Q_3^1 - \frac{3}{140} Q_1^4Q_2^1Q_3^4 + \frac{3}{40} Q_1^4Q_2^2Q_3^5 - \frac{3}{40} Q_1^4Q_2^3Q_3^6 - \frac{3}{140} Q_1^4Q_2^4Q_3^1 \nonumber \\
& & \mbox{\hglue0.6truecm} + \frac{3}{40} Q_1^4Q_2^5Q_3^2 - \frac{3}{40} Q_1^4Q_2^6Q_3^3 + \frac{1}{35} \sqrt{3} Q_1^4Q_2^6Q_3^8 + \frac{1}{35} \sqrt{3} Q_1^4Q_2^8Q_3^6 - \frac{3}{140} Q_1^5Q_2^1Q_3^5 \nonumber \\
& & \mbox{\hglue0.6truecm} - \frac{3}{40} Q_1^5Q_2^2Q_3^4 - \frac{3}{40} Q_1^5Q_2^3Q_3^7 - \frac{3}{40} Q_1^5Q_2^4Q_3^2 - \frac{3}{140} Q_1^5Q_2^5Q_3^1 - \frac{3}{40} Q_1^5Q_2^7Q_3^3 \nonumber \\
& & \mbox{\hglue0.6truecm} + \frac{1}{35} \sqrt{3} Q_1^5Q_2^7Q_3^8 + \frac{1}{35} \sqrt{3} Q_1^5Q_2^8Q_3^7 - \frac{3}{140} Q_1^6Q_2^1Q_3^6 - \frac{3}{40} Q_1^6Q_2^2Q_3^7 + \frac{3}{40} Q_1^6Q_2^3Q_3^4 \nonumber \\
& & \mbox{\hglue0.6truecm} + \frac{3}{40} Q_1^6Q_2^4Q_3^3 + \frac{1}{35} \sqrt{3} Q_1^6Q_2^4Q_3^8 - \frac{3}{140} Q_1^6Q_2^6Q_3^1 - \frac{3}{40} Q_1^6Q_2^7Q_3^2 + \frac{1}{35} \sqrt{3} Q_1^6Q_2^8Q_3^4 \nonumber \\
& & \mbox{\hglue0.6truecm} - \frac{3}{140} Q_1^7Q_2^1Q_3^7 + \frac{3}{40} Q_1^7Q_2^2Q_3^6 + \frac{3}{40} Q_1^7Q_2^3Q_3^5 + \frac{3}{40} Q_1^7Q_2^5Q_3^3 + \frac{1}{35} \sqrt{3} Q_1^7Q_2^5Q_3^8 \nonumber \\
& & \mbox{\hglue0.6truecm} + \frac{3}{40} Q_1^7Q_2^6Q_3^2 - \frac{3}{140} Q_1^7Q_2^7Q_3^1 + \frac{1}{35} \sqrt{3} Q_1^7Q_2^8Q_3^5 - \frac{3}{35} Q_1^8Q_2^1Q_3^8 - \frac{3}{140} \sqrt{3} Q_1^8Q_2^4Q_3^6 \nonumber \\
& & \mbox{\hglue0.6truecm} - \frac{3}{140} \sqrt{3} Q_1^8Q_2^5Q_3^7 - \frac{3}{140} \sqrt{3} Q_1^8Q_2^6Q_3^4 - \frac{3}{140} \sqrt{3} Q_1^8Q_2^7Q_3^5 - \frac{3}{35} Q_1^8Q_2^8Q_3^1,
\end{eqnarray}
\begin{eqnarray}
& & [\mathcal{P}^{(27)} Q_1Q_2Q_3]^{288} \nonumber \\
& & \mbox{\hglue0.2truecm} = \frac{9}{140} Q_1^1Q_2^1Q_3^2 + \frac{9}{140} Q_1^1Q_2^2Q_3^1 - \frac{3}{35} Q_1^2Q_2^1Q_3^1 + \frac{3}{70} Q_1^2Q_2^2Q_3^2 - \frac{3}{35} Q_1^2Q_2^3Q_3^3 - \frac{3}{140} Q_1^2Q_2^4Q_3^4 \nonumber \\
& & \mbox{\hglue0.6truecm} - \frac{3}{140} Q_1^2Q_2^5Q_3^5 - \frac{3}{140} Q_1^2Q_2^6Q_3^6 - \frac{3}{140} Q_1^2Q_2^7Q_3^7 + \frac{3}{14} Q_1^2Q_2^8Q_3^8 + \frac{9}{140} Q_1^3Q_2^2Q_3^3 \nonumber \\
& & \mbox{\hglue0.6truecm} + \frac{9}{140} Q_1^3Q_2^3Q_3^2 - \frac{3}{40} Q_1^4Q_2^1Q_3^5 - \frac{3}{140} Q_1^4Q_2^2Q_3^4 + \frac{3}{40} Q_1^4Q_2^3Q_3^7 - \frac{3}{140} Q_1^4Q_2^4Q_3^2 \nonumber \\
& & \mbox{\hglue0.6truecm} - \frac{3}{40} Q_1^4Q_2^5Q_3^1 + \frac{3}{40} Q_1^4Q_2^7Q_3^3 - \frac{1}{35} \sqrt{3} Q_1^4Q_2^7Q_3^8 - \frac{1}{35} \sqrt{3} Q_1^4Q_2^8Q_3^7 + \frac{3}{40} Q_1^5Q_2^1Q_3^4 \nonumber \\
& & \mbox{\hglue0.6truecm} - \frac{3}{140} Q_1^5Q_2^2Q_3^5 - \frac{3}{40} Q_1^5Q_2^3Q_3^6 + \frac{3}{40} Q_1^5Q_2^4Q_3^1 - \frac{3}{140} Q_1^5Q_2^5Q_3^2 - \frac{3}{40} Q_1^5Q_2^6Q_3^3 \nonumber \\
& & \mbox{\hglue0.6truecm} + \frac{1}{35} \sqrt{3} Q_1^5Q_2^6Q_3^8 + \frac{1}{35} \sqrt{3} Q_1^5Q_2^8Q_3^6 + \frac{3}{40} Q_1^6Q_2^1Q_3^7 - \frac{3}{140} Q_1^6Q_2^2Q_3^6 + \frac{3}{40} Q_1^6Q_2^3Q_3^5 \nonumber \\
& & \mbox{\hglue0.6truecm} + \frac{3}{40} Q_1^6Q_2^5Q_3^3 + \frac{1}{35} \sqrt{3} Q_1^6Q_2^5Q_3^8 - \frac{3}{140} Q_1^6Q_2^6Q_3^2 + \frac{3}{40} Q_1^6Q_2^7Q_3^1 + \frac{1}{35} \sqrt{3} Q_1^6Q_2^8Q_3^5 \nonumber \\
& & \mbox{\hglue0.6truecm} - \frac{3}{40} Q_1^7Q_2^1Q_3^6 - \frac{3}{140} Q_1^7Q_2^2Q_3^7 - \frac{3}{40} Q_1^7Q_2^3Q_3^4 - \frac{3}{40} Q_1^7Q_2^4Q_3^3 - \frac{1}{35} \sqrt{3} Q_1^7Q_2^4Q_3^8 \nonumber \\
& & \mbox{\hglue0.6truecm} - \frac{3}{40} Q_1^7Q_2^6Q_3^1 - \frac{3}{140} Q_1^7Q_2^7Q_3^2 - \frac{1}{35} \sqrt{3} Q_1^7Q_2^8Q_3^4 - \frac{3}{35} Q_1^8Q_2^2Q_3^8 + \frac{3}{140} \sqrt{3} Q_1^8Q_2^4Q_3^7 \nonumber \\
& & \mbox{\hglue0.6truecm} - \frac{3}{140} \sqrt{3} Q_1^8Q_2^5Q_3^6 - \frac{3}{140} \sqrt{3} Q_1^8Q_2^6Q_3^5 + \frac{3}{140} \sqrt{3} Q_1^8Q_2^7Q_3^4 - \frac{3}{35} Q_1^8Q_2^8Q_3^2,
\end{eqnarray}
\begin{eqnarray}
& & [\mathcal{P}^{(27)} Q_1Q_2Q_3]^{488} \nonumber \\
& & \mbox{\hglue0.2truecm} = - \frac{57}{1120} Q_1^1Q_2^1Q_3^4 - \frac{3}{160} Q_1^1Q_2^2Q_3^5 + \frac{3}{160} Q_1^1Q_2^3Q_3^6 - \frac{57}{1120} Q_1^1Q_2^4Q_3^1 - \frac{3}{160} Q_1^1Q_2^5Q_3^2 \nonumber \\
& & \mbox{\hglue0.6truecm} + \frac{3}{160} Q_1^1Q_2^6Q_3^3 - \frac{43}{1120} \sqrt{3} Q_1^1Q_2^6Q_3^8 - \frac{43}{1120} \sqrt{3} Q_1^1Q_2^8Q_3^6 + \frac{3}{160} Q_1^2Q_2^1Q_3^5 - \frac{57}{1120} Q_1^2Q_2^2Q_3^4 \nonumber \\
& & \mbox{\hglue0.6truecm} - \frac{3}{160} Q_1^2Q_2^3Q_3^7 - \frac{57}{1120} Q_1^2Q_2^4Q_3^2 + \frac{3}{160} Q_1^2Q_2^5Q_3^1 - \frac{3}{160} Q_1^2Q_2^7Q_3^3 + \frac{43}{1120} \sqrt{3} Q_1^2Q_2^7Q_3^8 \nonumber \\
& & \mbox{\hglue0.6truecm} + \frac{43}{1120} \sqrt{3} Q_1^2Q_2^8Q_3^7 - \frac{3}{160} Q_1^3Q_2^1Q_3^6 + \frac{3}{160} Q_1^3Q_2^2Q_3^7 - \frac{57}{1120} Q_1^3Q_2^3Q_3^4 - \frac{57}{1120} Q_1^3Q_2^4Q_3^3 \nonumber \\
& & \mbox{\hglue0.6truecm} - \frac{43}{1120} \sqrt{3} Q_1^3Q_2^4Q_3^8 - \frac{3}{160} Q_1^3Q_2^6Q_3^1 + \frac{3}{160} Q_1^3Q_2^7Q_3^2 - \frac{43}{1120} \sqrt{3} Q_1^3Q_2^8Q_3^4 - \frac{39}{560} Q_1^4Q_2^1Q_3^1 \nonumber \\
& & \mbox{\hglue0.6truecm} - \frac{39}{560} Q_1^4Q_2^2Q_3^2 - \frac{39}{560} Q_1^4Q_2^3Q_3^3 + \frac{1}{56} \sqrt{3} Q_1^4Q_2^3Q_3^8 + \frac{3}{70} Q_1^4Q_2^4Q_3^4 - \frac{9}{56} Q_1^4Q_2^5Q_3^5 \nonumber \\
& & \mbox{\hglue0.6truecm} + \frac{3}{112} Q_1^4Q_2^6Q_3^6 + \frac{3}{112} Q_1^4Q_2^7Q_3^7 + \frac{1}{56} \sqrt{3} Q_1^4Q_2^8Q_3^3 + \frac{153}{560} Q_1^4Q_2^8Q_3^8 + \frac{57}{560} Q_1^5Q_2^4Q_3^5 \nonumber \\
& & \mbox{\hglue0.6truecm} + \frac{57}{560} Q_1^5Q_2^5Q_3^4 + \frac{1}{56} \sqrt{3} Q_1^6Q_2^1Q_3^8 + \frac{9}{1120} Q_1^6Q_2^4Q_3^6 - \frac{3}{32} Q_1^6Q_2^5Q_3^7 + \frac{9}{1120} Q_1^6Q_2^6Q_3^4 \nonumber \\
& & \mbox{\hglue0.6truecm} - \frac{3}{32} Q_1^6Q_2^7Q_3^5 + \frac{1}{56} \sqrt{3} Q_1^6Q_2^8Q_3^1 - \frac{1}{56} \sqrt{3} Q_1^7Q_2^2Q_3^8 + \frac{9}{1120} Q_1^7Q_2^4Q_3^7 + \frac{3}{32} Q_1^7Q_2^5Q_3^6 \nonumber \\
& & \mbox{\hglue0.6truecm} + \frac{3}{32} Q_1^7Q_2^6Q_3^5 + \frac{9}{1120} Q_1^7Q_2^7Q_3^4 - \frac{1}{56} \sqrt{3} Q_1^7Q_2^8Q_3^2 - \frac{57}{1120} \sqrt{3} Q_1^8Q_2^1Q_3^6 + \frac{57}{1120} \sqrt{3} Q_1^8Q_2^2Q_3^7 \nonumber \\
& & \mbox{\hglue0.6truecm} - \frac{57}{1120} \sqrt{3} Q_1^8Q_2^3Q_3^4 - \frac{57}{1120} \sqrt{3} Q_1^8Q_2^4Q_3^3 - \frac{9}{1120} Q_1^8Q_2^4Q_3^8 - \frac{57}{1120} \sqrt{3} Q_1^8Q_2^6Q_3^1 \nonumber \\
& & \mbox{\hglue0.6truecm} + \frac{57}{1120} \sqrt{3} Q_1^8Q_2^7Q_3^2 - \frac{9}{1120} Q_1^8Q_2^8Q_3^4,
\end{eqnarray}
\begin{eqnarray}
& & [\mathcal{P}^{(27)} Q_1Q_2Q_3]^{588} \nonumber \\
& & \mbox{\hglue0.2truecm} = - \frac{57}{1120} Q_1^1Q_2^1Q_3^5 + \frac{3}{160} Q_1^1Q_2^2Q_3^4 + \frac{3}{160}
Q_1^1Q_2^3Q_3^7 + \frac{3}{160} Q_1^1Q_2^4Q_3^2 - \frac{57}{1120} Q_1^1Q_2^5Q_3^1 \nonumber \\
& & \mbox{\hglue0.6truecm} + \frac{3}{160} Q_1^1Q_2^7Q_3^3 - \frac{43}{1120} \sqrt{3} Q_1^1Q_2^7Q_3^8 - \frac{43}{1120} \sqrt{3} Q_1^1Q_2^8Q_3^7 - \frac{3}{160} Q_1^2Q_2^1Q_3^4 - \frac{57}{1120} Q_1^2Q_2^2Q_3^5 \nonumber \\
& & \mbox{\hglue0.6truecm} + \frac{3}{160} Q_1^2Q_2^3Q_3^6 - \frac{3}{160} Q_1^2Q_2^4Q_3^1 - \frac{57}{1120} Q_1^2Q_2^5Q_3^2 + \frac{3}{160} Q_1^2Q_2^6Q_3^3 - \frac{43}{1120} \sqrt{3} Q_1^2Q_2^6Q_3^8 \nonumber \\
& & \mbox{\hglue0.6truecm} - \frac{43}{1120} \sqrt{3} Q_1^2Q_2^8Q_3^6 - \frac{3}{160} Q_1^3Q_2^1Q_3^7 - \frac{3}{160} Q_1^3Q_2^2Q_3^6 - \frac{57}{1120} Q_1^3Q_2^3Q_3^5 - \frac{57}{1120} Q_1^3Q_2^5Q_3^3 \nonumber \\
& & \mbox{\hglue0.6truecm} - \frac{43}{1120} \sqrt{3} Q_1^3Q_2^5Q_3^8 - \frac{3}{160} Q_1^3Q_2^6Q_3^2 - \frac{3}{160} Q_1^3Q_2^7Q_3^1 - \frac{43}{1120} \sqrt{3} Q_1^3Q_2^8Q_3^5 + \frac{57}{560} Q_1^4Q_2^4Q_3^5 \nonumber \\
& & \mbox{\hglue0.6truecm} + \frac{57}{560} Q_1^4Q_2^5Q_3^4 - \frac{39}{560} Q_1^5Q_2^1Q_3^1 - \frac{39}{560} Q_1^5Q_2^2Q_3^2 - \frac{39}{560} Q_1^5Q_2^3Q_3^3 + \frac{1}{56} \sqrt{3} Q_1^5Q_2^3Q_3^8 \nonumber \\
& & \mbox{\hglue0.6truecm} - \frac{9}{56} Q_1^5Q_2^4Q_3^4 + \frac{3}{70} Q_1^5Q_2^5Q_3^5 + \frac{3}{112} Q_1^5Q_2^6Q_3^6 + \frac{3}{112} Q_1^5Q_2^7Q_3^7 + \frac{1}{56} \sqrt{3} Q_1^5Q_2^8Q_3^3 \nonumber \\
& & \mbox{\hglue0.6truecm} + \frac{153}{560} Q_1^5Q_2^8Q_3^8 + \frac{1}{56} \sqrt{3} Q_1^6Q_2^2Q_3^8 + \frac{3}{32} Q_1^6Q_2^4Q_3^7 + \frac{9}{1120} Q_1^6Q_2^5Q_3^6 + \frac{9}{1120} Q_1^6Q_2^6Q_3^5 \nonumber \\
& & \mbox{\hglue0.6truecm} + \frac{3}{32} Q_1^6Q_2^7Q_3^4 + \frac{1}{56} \sqrt{3} Q_1^6Q_2^8Q_3^2 + \frac{1}{56} \sqrt{3} Q_1^7Q_2^1Q_3^8 - \frac{3}{32} Q_1^7Q_2^4Q_3^6 + \frac{9}{1120} Q_1^7Q_2^5Q_3^7 \nonumber \\
& & \mbox{\hglue0.6truecm} - \frac{3}{32} Q_1^7Q_2^6Q_3^4 + \frac{9}{1120} Q_1^7Q_2^7Q_3^5 + \frac{1}{56} \sqrt{3} Q_1^7Q_2^8Q_3^1 - \frac{57}{1120} \sqrt{3} Q_1^8Q_2^1Q_3^7 \nonumber \\
& & \mbox{\hglue0.6truecm} - \frac{57}{1120} \sqrt{3} Q_1^8Q_2^2Q_3^6 - \frac{57}{1120} \sqrt{3} Q_1^8Q_2^3Q_3^5 - \frac{57}{1120} \sqrt{3} Q_1^8Q_2^5Q_3^3 - \frac{9}{1120} Q_1^8Q_2^5Q_3^8 \nonumber \\
& & \mbox{\hglue0.6truecm} - \frac{57}{1120} \sqrt{3} Q_1^8Q_2^6Q_3^2 - \frac{57}{1120} \sqrt{3} Q_1^8Q_2^7Q_3^1 - \frac{9}{1120} Q_1^8Q_2^8Q_3^5,
\end{eqnarray}
\begin{eqnarray}
& & [\mathcal{P}^{(35+\overline{35})} Q_1Q_2Q_3]^{188} \nonumber \\
& & \mbox{\hglue0.2truecm} = \frac{1}{18} Q_1^1Q_2^2Q_3^2 + \frac{1}{18} Q_1^1Q_2^3Q_3^3 - \frac19 Q_1^1Q_2^4Q_3^4 - \frac19 Q_1^1Q_2^5Q_3^5 - \frac19 Q_1^1Q_2^6Q_3^6 - \frac19 Q_1^1Q_2^7Q_3^7 \nonumber \\
& & \mbox{\hglue0.6truecm} + \frac13 Q_1^1Q_2^8Q_3^8 - \frac{1}{36} Q_1^2Q_2^1Q_3^2 - \frac{1}{36} Q_1^2Q_2^2Q_3^1 - \frac{1}{36} Q_1^3Q_2^1Q_3^3 - \frac{1}{36} Q_1^3Q_2^3Q_3^1 + \frac{1}{18} Q_1^4Q_2^1Q_3^4 \nonumber \\
& & \mbox{\hglue0.6truecm} - \frac{1}{24} Q_1^4Q_2^2Q_3^5 + \frac{1}{24} Q_1^4Q_2^3Q_3^6 + \frac{1}{18} Q_1^4Q_2^4Q_3^1 - \frac{1}{24} Q_1^4Q_2^5Q_3^2 + \frac{1}{24} Q_1^4Q_2^6Q_3^3 \nonumber \\
& & \mbox{\hglue0.6truecm} - \frac{1}{12\sqrt{3}} Q_1^4Q_2^6Q_3^8 - \frac{1}{12\sqrt{3}} Q_1^4Q_2^8Q_3^6 + \frac{1}{18} Q_1^5Q_2^1Q_3^5 + \frac{1}{24} Q_1^5Q_2^2Q_3^4 + \frac{1}{24} Q_1^5Q_2^3Q_3^7 \nonumber \\
& & \mbox{\hglue0.6truecm} + \frac{1}{24} Q_1^5Q_2^4Q_3^2 + \frac{1}{18} Q_1^5Q_2^5Q_3^1 + \frac{1}{24} Q_1^5Q_2^7Q_3^3 - \frac{1}{12\sqrt{3}} Q_1^5Q_2^7Q_3^8 - \frac{1}{12\sqrt{3}} Q_1^5Q_2^8Q_3^7 \nonumber \\
& & \mbox{\hglue0.6truecm} + \frac{1}{18} Q_1^6Q_2^1Q_3^6 + \frac{1}{24} Q_1^6Q_2^2Q_3^7 - \frac{1}{24} Q_1^6Q_2^3Q_3^4 - \frac{1}{24} Q_1^6Q_2^4Q_3^3 - \frac{1}{12\sqrt{3}} Q_1^6Q_2^4Q_3^8 \nonumber \\
& & \mbox{\hglue0.6truecm} + \frac{1}{18} Q_1^6Q_2^6Q_3^1 + \frac{1}{24} Q_1^6Q_2^7Q_3^2 - \frac{1}{12\sqrt{3}} Q_1^6Q_2^8Q_3^4 + \frac{1}{18} Q_1^7Q_2^1Q_3^7 - \frac{1}{24} Q_1^7Q_2^2Q_3^6 \nonumber \\
& & \mbox{\hglue0.6truecm} - \frac{1}{24} Q_1^7Q_2^3Q_3^5 - \frac{1}{24} Q_1^7Q_2^5Q_3^3 - \frac{1}{12\sqrt{3}} Q_1^7Q_2^5Q_3^8 - \frac{1}{24} Q_1^7Q_2^6Q_3^2 + \frac{1}{18} Q_1^7Q_2^7Q_3^1 \nonumber \\
& & \mbox{\hglue0.6truecm} - \frac{1}{12\sqrt{3}} Q_1^7Q_2^8Q_3^5 - \frac16 Q_1^8Q_2^1Q_3^8 + \frac{1}{6\sqrt{3}} Q_1^8Q_2^4Q_3^6 + \frac{1}{6\sqrt{3}} Q_1^8Q_2^5Q_3^7 + \frac{1}{6\sqrt{3}} Q_1^8Q_2^6Q_3^4 \nonumber \\
& & \mbox{\hglue0.6truecm} + \frac{1}{6\sqrt{3}} Q_1^8Q_2^7Q_3^5 - \frac16 Q_1^8Q_2^8Q_3^1,
\end{eqnarray}
\begin{eqnarray}
& & [\mathcal{P}^{(35+\overline{35})} Q_1Q_2Q_3]^{288} \nonumber \\
& & \mbox{\hglue0.2truecm} = - \frac{1}{36} Q_1^1Q_2^1Q_3^2 - \frac{1}{36} Q_1^1Q_2^2Q_3^1 + \frac{1}{18} Q_1^2Q_2^1Q_3^1 + \frac{1}{18} Q_1^2Q_2^3Q_3^3 - \frac19 Q_1^2Q_2^4Q_3^4 - \frac19 Q_1^2Q_2^5Q_3^5 \nonumber \\
& & \mbox{\hglue0.6truecm} - \frac19 Q_1^2Q_2^6Q_3^6 - \frac19 Q_1^2Q_2^7Q_3^7 + \frac13 Q_1^2Q_2^8Q_3^8 - \frac{1}{36} Q_1^3Q_2^2Q_3^3 - \frac{1}{36} Q_1^3Q_2^3Q_3^2 + \frac{1}{24} Q_1^4Q_2^1Q_3^5 \nonumber \\
& & \mbox{\hglue0.6truecm} + \frac{1}{18} Q_1^4Q_2^2Q_3^4 - \frac{1}{24} Q_1^4Q_2^3Q_3^7 + \frac{1}{18} Q_1^4Q_2^4Q_3^2 + \frac{1}{24} Q_1^4Q_2^5Q_3^1 - \frac{1}{24} Q_1^4Q_2^7Q_3^3 \nonumber \\
& & \mbox{\hglue0.6truecm} + \frac{1}{12\sqrt{3}} Q_1^4Q_2^7Q_3^8 + \frac{1}{12\sqrt{3}} Q_1^4Q_2^8Q_3^7 - \frac{1}{24} Q_1^5Q_2^1Q_3^4 + \frac{1}{18} Q_1^5Q_2^2Q_3^5 + \frac{1}{24} Q_1^5Q_2^3Q_3^6 \nonumber \\
& & \mbox{\hglue0.6truecm} - \frac{1}{24} Q_1^5Q_2^4Q_3^1 + \frac{1}{18} Q_1^5Q_2^5Q_3^2 + \frac{1}{24} Q_1^5Q_2^6Q_3^3 - \frac{1}{12\sqrt{3}} Q_1^5Q_2^6Q_3^8 - \frac{1}{12\sqrt{3}} Q_1^5Q_2^8Q_3^6 \nonumber \\
& & \mbox{\hglue0.6truecm} - \frac{1}{24} Q_1^6Q_2^1Q_3^7 + \frac{1}{18} Q_1^6Q_2^2Q_3^6 - \frac{1}{24} Q_1^6Q_2^3Q_3^5 - \frac{1}{24} Q_1^6Q_2^5Q_3^3 - \frac{1}{12\sqrt{3}} Q_1^6Q_2^5Q_3^8 \nonumber \\
& & \mbox{\hglue0.6truecm} + \frac{1}{18} Q_1^6Q_2^6Q_3^2 - \frac{1}{24} Q_1^6Q_2^7Q_3^1 - \frac{1}{12\sqrt{3}} Q_1^6Q_2^8Q_3^5 + 
\frac{1}{24} Q_1^7Q_2^1Q_3^6 + \frac{1}{18} Q_1^7Q_2^2Q_3^7 \nonumber \\
& & \mbox{\hglue0.6truecm} + \frac{1}{24} Q_1^7Q_2^3Q_3^4 + \frac{1}{24} Q_1^7Q_2^4Q_3^3 + \frac{1}{12\sqrt{3}} Q_1^7Q_2^4Q_3^8 + \frac{1}{24} Q_1^7Q_2^6Q_3^1 + \frac{1}{18} Q_1^7Q_2^7Q_3^2 \nonumber \\
& & \mbox{\hglue0.6truecm} + \frac{1}{12\sqrt{3}} Q_1^7Q_2^8Q_3^4 - \frac16 Q_1^8Q_2^2Q_3^8 - \frac{1}{6\sqrt{3}} Q_1^8Q_2^4Q_3^7 + \frac{1}{6\sqrt{3}} Q_1^8Q_2^5Q_3^6 + \frac{1}{6\sqrt{3}} Q_1^8Q_2^6Q_3^5 \nonumber \\
& & \mbox{\hglue0.6truecm} - \frac{1}{6\sqrt{3}} Q_1^8Q_2^7Q_3^4 - \frac16 Q_1^8Q_2^8Q_3^2,
\end{eqnarray}
\begin{eqnarray}
& & [\mathcal{P}^{(35+\overline{35})} Q_1Q_2Q_3]^{488} \nonumber \\
& & \mbox{\hglue0.2truecm} = - \frac{1}{96} Q_1^1Q_2^1Q_3^4 + \frac{1}{32} Q_1^1Q_2^2Q_3^5 - \frac{1}{32} Q_1^1Q_2^3Q_3^6 - \frac{1}{96} Q_1^1Q_2^4Q_3^1 + \frac{1}{32} Q_1^1Q_2^5Q_3^2 - \frac{1}{32} Q_1^1Q_2^6Q_3^3 \nonumber \\
& & \mbox{\hglue0.6truecm} + \frac{5}{32\sqrt{3}} Q_1^1Q_2^6Q_3^8 + \frac{5}{32\sqrt{3}} Q_1^1Q_2^8Q_3^6 - \frac{1}{32} Q_1^2Q_2^1Q_3^5 - \frac{1}{96} Q_1^2Q_2^2Q_3^4 + \frac{1}{32} Q_1^2Q_2^3Q_3^7 \nonumber \\
& & \mbox{\hglue0.6truecm} - \frac{1}{96} Q_1^2Q_2^4Q_3^2 - \frac{1}{32} Q_1^2Q_2^5Q_3^1 + \frac{1}{32} Q_1^2Q_2^7Q_3^3 - \frac{5}{32\sqrt{3}} Q_1^2Q_2^7Q_3^8 - \frac{5}{32\sqrt{3}} Q_1^2Q_2^8Q_3^7 \nonumber \\
& & \mbox{\hglue0.6truecm} + \frac{1}{32} Q_1^3Q_2^1Q_3^6 - \frac{1}{32} Q_1^3Q_2^2Q_3^7 - \frac{1}{96} Q_1^3Q_2^3Q_3^4 - \frac{1}{96} Q_1^3Q_2^4Q_3^3 + \frac{5}{32\sqrt{3}} Q_1^3Q_2^4Q_3^8 \nonumber \\
& & \mbox{\hglue0.6truecm} + \frac{1}{32} Q_1^3Q_2^6Q_3^1 - \frac{1}{32} Q_1^3Q_2^7Q_3^2 + \frac{5}{32\sqrt{3}} Q_1^3Q_2^8Q_3^4 + \frac{1}{48} Q_1^4Q_2^1Q_3^1 + \frac{1}{48} Q_1^4Q_2^2Q_3^2 \nonumber \\
& & \mbox{\hglue0.6truecm} + \frac{1}{48} Q_1^4Q_2^3Q_3^3 - \frac{1}{8\sqrt{3}} Q_1^4Q_2^3Q_3^8 - \frac{1}{24} Q_1^4Q_2^5Q_3^5 - \frac{5}{48} Q_1^4Q_2^6Q_3^6 - \frac{5}{48} Q_1^4Q_2^7Q_3^7 \nonumber \\
& & \mbox{\hglue0.6truecm} - \frac{1}{8\sqrt{3}} Q_1^4Q_2^8Q_3^3 + \frac{3}{16} Q_1^4Q_2^8Q_3^8 + \frac{1}{48} Q_1^5Q_2^4Q_3^5 + \frac{1}{48} Q_1^5Q_2^5Q_3^4 - \frac{1}{8\sqrt{3}} Q_1^6Q_2^1Q_3^8 \nonumber \\
& & \mbox{\hglue0.6truecm} + \frac{5}{96} Q_1^6Q_2^4Q_3^6 + \frac{1}{32} Q_1^6Q_2^5Q_3^7 + \frac{5}{96} Q_1^6Q_2^6Q_3^4 + \frac{1}{32} Q_1^6Q_2^7Q_3^5 - \frac{1}{8\sqrt{3}} Q_1^6Q_2^8Q_3^1 \nonumber \\
& & \mbox{\hglue0.6truecm} + \frac{1}{8\sqrt{3}} Q_1^7Q_2^2Q_3^8 + \frac{5}{96} Q_1^7Q_2^4Q_3^7 - \frac{1}{32} Q_1^7Q_2^5Q_3^6 - \frac{1}{32} Q_1^7Q_2^6Q_3^5 + \frac{5}{96} Q_1^7Q_2^7Q_3^4 \nonumber \\
& & \mbox{\hglue0.6truecm} + \frac{1}{8\sqrt{3}} Q_1^7Q_2^8Q_3^2 - \frac{1}{32\sqrt{3}} Q_1^8Q_2^1Q_3^6 + \frac{1}{32\sqrt{3}} Q_1^8Q_2^2Q_3^7 - \frac{1}{32\sqrt{3}} Q_1^8Q_2^3Q_3^4 \nonumber \\
& & \mbox{\hglue0.6truecm} - \frac{1}{32\sqrt{3}} Q_1^8Q_2^4Q_3^3 - \frac{3}{32} Q_1^8Q_2^4Q_3^8 - \frac{1}{32\sqrt{3}} Q_1^8Q_2^6Q_3^1 + \frac{1}{32\sqrt{3}} Q_1^8Q_2^7Q_3^2 - \frac{3}{32} Q_1^8Q_2^8Q_3^4,
\end{eqnarray}
\begin{eqnarray}
& & [\mathcal{P}^{(35+\overline{35})} Q_1Q_2Q_3]^{588} \nonumber \\
& & \mbox{\hglue0.2truecm} = - \frac{1}{96} Q_1^1Q_2^1Q_3^5 - \frac{1}{32} Q_1^1Q_2^2Q_3^4 - \frac{1}{32} Q_1^1Q_2^3Q_3^7 - \frac{1}{32} Q_1^1Q_2^4Q_3^2 - \frac{1}{96} Q_1^1Q_2^5Q_3^1 - \frac{1}{32} Q_1^1Q_2^7Q_3^3 \nonumber \\
& & \mbox{\hglue0.6truecm} + \frac{5}{32\sqrt{3}} Q_1^1Q_2^7Q_3^8 + \frac{5}{32\sqrt{3}} Q_1^1Q_2^8Q_3^7 + \frac{1}{32} Q_1^2Q_2^1Q_3^4 - \frac{1}{96} Q_1^2Q_2^2Q_3^5 - \frac{1}{32} Q_1^2Q_2^3Q_3^6 \nonumber \\
& & \mbox{\hglue0.6truecm} + \frac{1}{32} Q_1^2Q_2^4Q_3^1 - \frac{1}{96} Q_1^2Q_2^5Q_3^2 - \frac{1}{32} Q_1^2Q_2^6Q_3^3 + \frac{5}{32\sqrt{3}} Q_1^2Q_2^6Q_3^8 + \frac{5}{32\sqrt{3}} Q_1^2Q_2^8Q_3^6 \nonumber \\
& & \mbox{\hglue0.6truecm} + \frac{1}{32} Q_1^3Q_2^1Q_3^7 + \frac{1}{32} Q_1^3Q_2^2Q_3^6 - \frac{1}{96} Q_1^3Q_2^3Q_3^5 - \frac{1}{96} Q_1^3Q_2^5Q_3^3 + \frac{5}{32\sqrt{3}} Q_1^3Q_2^5Q_3^8 \nonumber \\
& & \mbox{\hglue0.6truecm} + \frac{1}{32} Q_1^3Q_2^6Q_3^2 + \frac{1}{32} Q_1^3Q_2^7Q_3^1 + \frac{5}{32\sqrt{3}} Q_1^3Q_2^8Q_3^5 + \frac{1}{48} Q_1^4Q_2^4Q_3^5 + \frac{1}{48} Q_1^4Q_2^5Q_3^4 \nonumber \\
& & \mbox{\hglue0.6truecm} + \frac{1}{48} Q_1^5Q_2^1Q_3^1 + \frac{1}{48} Q_1^5Q_2^2Q_3^2 + \frac{1}{48} Q_1^5Q_2^3Q_3^3 - \frac{1}{8\sqrt{3}} Q_1^5Q_2^3Q_3^8 - \frac{1}{24} Q_1^5Q_2^4Q_3^4 \nonumber \\
& & \mbox{\hglue0.6truecm} - \frac{5}{48} Q_1^5Q_2^6Q_3^6 - \frac{5}{48} Q_1^5Q_2^7Q_3^7 - \frac{1}{8\sqrt{3}} Q_1^5Q_2^8Q_3^3 + \frac{3}{16} Q_1^5Q_2^8Q_3^8 - \frac{1}{8\sqrt{3}} Q_1^6Q_2^2Q_3^8 \nonumber \\
& & \mbox{\hglue0.6truecm} - \frac{1}{32} Q_1^6Q_2^4Q_3^7 + \frac{5}{96} Q_1^6Q_2^5Q_3^6 + \frac{5}{96} Q_1^6Q_2^6Q_3^5 - \frac{1}{32} Q_1^6Q_2^7Q_3^4 - \frac{1}{8\sqrt{3}} Q_1^6Q_2^8Q_3^2 \nonumber \\
& & \mbox{\hglue0.6truecm} - \frac{1}{8\sqrt{3}} Q_1^7Q_2^1Q_3^8 + \frac{1}{32} Q_1^7Q_2^4Q_3^6 + \frac{5}{96} Q_1^7Q_2^5Q_3^7 + \frac{1}{32} Q_1^7Q_2^6Q_3^4 + \frac{5}{96} Q_1^7Q_2^7Q_3^5 \nonumber \\
& & \mbox{\hglue0.6truecm} - \frac{1}{8\sqrt{3}} Q_1^7Q_2^8Q_3^1 - \frac{1}{32\sqrt{3}} Q_1^8Q_2^1Q_3^7 - \frac{1}{32\sqrt{3}} Q_1^8Q_2^2Q_3^6 - \frac{1}{32\sqrt{3}} Q_1^8Q_2^3Q_3^5 \nonumber \\
& & \mbox{\hglue0.6truecm} - \frac{1}{32\sqrt{3}} Q_1^8Q_2^5Q_3^3 - \frac{3}{32} Q_1^8Q_2^5Q_3^8 - \frac{1}{32\sqrt{3}} Q_1^8Q_2^6Q_3^2 - \frac{1}{32\sqrt{3}} Q_1^8Q_2^7Q_3^1 - \frac{3}{32} Q_1^8Q_2^8Q_3^5,
\end{eqnarray}
\begin{eqnarray}
& & [\mathcal{P}^{(64)} Q_1Q_2Q_3]^{188} \nonumber \\
& & \mbox{\hglue0.2truecm} = \frac{1}{42} Q_1^1Q_2^1Q_3^1 + \frac{1}{126} Q_1^1Q_2^2Q_3^2 + \frac{1}{126} Q_1^1Q_2^3Q_3^3 - \frac{5}{126} Q_1^1Q_2^4Q_3^4 - \frac{5}{126} Q_1^1Q_2^5Q_3^5 \nonumber \\
& & \mbox{\hglue0.6truecm} - \frac{5}{126} Q_1^1Q_2^6Q_3^6 - \frac{5}{126} Q_1^1Q_2^7Q_3^7 + \frac{5}{42} Q_1^1Q_2^8Q_3^8 + \frac{1}{126} Q_1^2Q_2^1Q_3^2 + \frac{1}{126} Q_1^2Q_2^2Q_3^1 \nonumber \\
& & \mbox{\hglue0.6truecm} + \frac{1}{126} Q_1^3Q_2^1Q_3^3 + \frac{1}{126} Q_1^3Q_2^3Q_3^1 - \frac{5}{126} Q_1^4Q_2^1Q_3^4 - \frac{5}{126} Q_1^4Q_2^4Q_3^1 - \frac{5}{42\sqrt{3}} Q_1^4Q_2^6Q_3^8 \nonumber \\
& & \mbox{\hglue0.6truecm} - \frac{5}{42\sqrt{3}} Q_1^4Q_2^8Q_3^6 - \frac{5}{126} Q_1^5Q_2^1Q_3^5 - \frac{5}{126} Q_1^5Q_2^5Q_3^1 - \frac{5}{42\sqrt{3}} Q_1^5Q_2^7Q_3^8 - \frac{5}{42\sqrt{3}} Q_1^5Q_2^8Q_3^7 \nonumber \\
& & \mbox{\hglue0.6truecm} - \frac{5}{126} Q_1^6Q_2^1Q_3^6 - \frac{5}{42\sqrt{3}} Q_1^6Q_2^4Q_3^8 - \frac{5}{126} Q_1^6Q_2^6Q_3^1 - \frac{5}{42\sqrt{3}} Q_1^6Q_2^8Q_3^4 - \frac{5}{126} Q_1^7Q_2^1Q_3^7 \nonumber \\
& & \mbox{\hglue0.6truecm} - \frac{5}{42\sqrt{3}} Q_1^7Q_2^5Q_3^8 - \frac{5}{126} Q_1^7Q_2^7Q_3^1 - \frac{5}{42\sqrt{3}} Q_1^7Q_2^8Q_3^5 + \frac{5}{42} Q_1^8Q_2^1Q_3^8 - \frac{5}{42\sqrt{3}} Q_1^8Q_2^4Q_3^6 \nonumber \\
& & \mbox{\hglue0.6truecm} - \frac{5}{42\sqrt{3}} Q_1^8Q_2^5Q_3^7 - \frac{5}{42\sqrt{3}} Q_1^8Q_2^6Q_3^4 - \frac{5}{42\sqrt{3}} Q_1^8Q_2^7Q_3^5 + \frac{5}{42} Q_1^8Q_2^8Q_3^1,
\end{eqnarray}
\begin{eqnarray}
& & [\mathcal{P}^{(64)} Q_1Q_2Q_3]^{288} \nonumber \\
& & \mbox{\hglue0.2truecm} = \frac{1}{126} Q_1^1Q_2^1Q_3^2 + \frac{1}{126} Q_1^1Q_2^2Q_3^1 + \frac{1}{126} Q_1^2Q_2^1Q_3^1 + \frac{1}{42} Q_1^2Q_2^2Q_3^2 + \frac{1}{126} Q_1^2Q_2^3Q_3^3 - \frac{5}{126} Q_1^2Q_2^4Q_3^4 \nonumber \\
& & \mbox{\hglue0.6truecm} - \frac{5}{126} Q_1^2Q_2^5Q_3^5 - \frac{5}{126} Q_1^2Q_2^6Q_3^6 - \frac{5}{126} Q_1^2Q_2^7Q_3^7 + \frac{5}{42} Q_1^2Q_2^8Q_3^8 + \frac{1}{126} Q_1^3Q_2^2Q_3^3 \nonumber \\
& & \mbox{\hglue0.6truecm} + \frac{1}{126} Q_1^3Q_2^3Q_3^2 - \frac{5}{126} Q_1^4Q_2^2Q_3^4 - \frac{5}{126} Q_1^4Q_2^4Q_3^2 + \frac{5}{42\sqrt{3}} Q_1^4Q_2^7Q_3^8 + \frac{5}{42\sqrt{3}} Q_1^4Q_2^8Q_3^7 \nonumber \\
& & \mbox{\hglue0.6truecm} - \frac{5}{126} Q_1^5Q_2^2Q_3^5 - \frac{5}{126} Q_1^5Q_2^5Q_3^2 - \frac{5}{42\sqrt{3}} Q_1^5Q_2^6Q_3^8 - \frac{5}{42\sqrt{3}} Q_1^5Q_2^8Q_3^6 - \frac{5}{126} Q_1^6Q_2^2Q_3^6 \nonumber \\
& & \mbox{\hglue0.6truecm} - \frac{5}{42\sqrt{3}} Q_1^6Q_2^5Q_3^8 - \frac{5}{126} Q_1^6Q_2^6Q_3^2 - \frac{5}{42\sqrt{3}} Q_1^6Q_2^8Q_3^5 - \frac{5}{126} Q_1^7Q_2^2Q_3^7 + \frac{5}{42\sqrt{3}} Q_1^7Q_2^4Q_3^8 \nonumber \\
& & \mbox{\hglue0.6truecm} - \frac{5}{126} Q_1^7Q_2^7Q_3^2 + \frac{5}{42\sqrt{3}} Q_1^7Q_2^8Q_3^4 + \frac{5}{42} Q_1^8Q_2^2Q_3^8 + \frac{5}{42\sqrt{3}} Q_1^8Q_2^4Q_3^7 - \frac{5}{42\sqrt{3}} Q_1^8Q_2^5Q_3^6 \nonumber \\
& & \mbox{\hglue0.6truecm} - \frac{5}{42\sqrt{3}} Q_1^8Q_2^6Q_3^5 + \frac{5}{42\sqrt{3}} Q_1^8Q_2^7Q_3^4 + \frac{5}{42} Q_1^8Q_2^8Q_3^2,
\end{eqnarray}
\begin{eqnarray}
& & [\mathcal{P}^{(64)} Q_1Q_2Q_3]^{488} \nonumber \\
& & \mbox{\hglue0.2truecm} = \frac{1}{42} Q_1^1Q_2^1Q_3^4 + \frac{1}{42} Q_1^1Q_2^4Q_3^1 + \frac{1}{14\sqrt{3}} Q_1^1Q_2^6Q_3^8 + \frac{1}{14\sqrt{3}} Q_1^1Q_2^8Q_3^6 + \frac{1}{42} Q_1^2Q_2^2Q_3^4 + \frac{1}{42} Q_1^2Q_2^4Q_3^2 \nonumber \\
& & \mbox{\hglue0.6truecm} - \frac{1}{14\sqrt{3}} Q_1^2Q_2^7Q_3^8 - \frac{1}{14\sqrt{3}} Q_1^2Q_2^8Q_3^7 + \frac{1}{42} Q_1^3Q_2^3Q_3^4 + \frac{1}{42} Q_1^3Q_2^4Q_3^3 + \frac{1}{14\sqrt{3}} Q_1^3Q_2^4Q_3^8 \nonumber \\
& & \mbox{\hglue0.6truecm} + \frac{1}{14\sqrt{3}} Q_1^3Q_2^8Q_3^4 + \frac{1}{42} Q_1^4Q_2^1Q_3^1 + \frac{1}{42} Q_1^4Q_2^2Q_3^2 + \frac{1}{42} Q_1^4Q_2^3Q_3^3 + \frac{1}{14\sqrt{3}} Q_1^4Q_2^3Q_3^8 \nonumber \\
& & \mbox{\hglue0.6truecm} - \frac17 Q_1^4Q_2^4Q_3^4 - \frac{1}{21} Q_1^4Q_2^5Q_3^5 - \frac{1}{21} Q_1^4Q_2^6Q_3^6 - \frac{1}{21} Q_1^4Q_2^7Q_3^7 + \frac{1}{14\sqrt{3}} Q_1^4Q_2^8Q_3^3 + \frac{3}{14} Q_1^4Q_2^8Q_3^8 \nonumber \\
& & \mbox{\hglue0.6truecm} - \frac{1}{21} Q_1^5Q_2^4Q_3^5 - \frac{1}{21} Q_1^5Q_2^5Q_3^4 + \frac{1}{14\sqrt{3}} Q_1^6Q_2^1Q_3^8 - \frac{1}{21} Q_1^6Q_2^4Q_3^6 - \frac{1}{21} Q_1^6Q_2^6Q_3^4 \nonumber \\
& & \mbox{\hglue0.6truecm} + \frac{1}{14\sqrt{3}} Q_1^6Q_2^8Q_3^1 - \frac{1}{14\sqrt{3}} Q_1^7Q_2^2Q_3^8 - \frac{1}{21} Q_1^7Q_2^4Q_3^7 - \frac{1}{21} Q_1^7Q_2^7Q_3^4 - \frac{1}{14\sqrt{3}} Q_1^7Q_2^8Q_3^2 \nonumber \\
& & \mbox{\hglue0.6truecm} + \frac{1}{14\sqrt{3}} Q_1^8Q_2^1Q_3^6 - \frac{1}{14\sqrt{3}} Q_1^8Q_2^2Q_3^7 + \frac{1}{14\sqrt{3}} Q_1^8Q_2^3Q_3^4 + \frac{1}{14\sqrt{3}} Q_1^8Q_2^4Q_3^3 + \frac{3}{14} Q_1^8Q_2^4Q_3^8 \nonumber \\
& & \mbox{\hglue0.6truecm} + \frac{1}{14\sqrt{3}} Q_1^8Q_2^6Q_3^1 - \frac{1}{14\sqrt{3}} Q_1^8Q_2^7Q_3^2 + \frac{3}{14} Q_1^8Q_2^8Q_3^4,
\end{eqnarray}
\begin{eqnarray}
& & [\mathcal{P}^{(64)} Q_1Q_2Q_3]^{588} \nonumber \\
& & \mbox{\hglue0.2truecm} = \frac{1}{42} Q_1^1Q_2^1Q_3^5 + \frac{1}{42} Q_1^1Q_2^5Q_3^1 + \frac{1}{14\sqrt{3}} Q_1^1Q_2^7Q_3^8 + \frac{1}{14\sqrt{3}} Q_1^1Q_2^8Q_3^7 + \frac{1}{42} Q_1^2Q_2^2Q_3^5 + \frac{1}{42} Q_1^2Q_2^5Q_3^2 \nonumber \\
& & \mbox{\hglue0.6truecm} + \frac{1}{14\sqrt{3}} Q_1^2Q_2^6Q_3^8 + \frac{1}{14\sqrt{3}} Q_1^2Q_2^8Q_3^6 + \frac{1}{42} Q_1^3Q_2^3Q_3^5 + \frac{1}{42} Q_1^3Q_2^5Q_3^3 + \frac{1}{14\sqrt{3}} Q_1^3Q_2^5Q_3^8 \nonumber \\
& & \mbox{\hglue0.6truecm} + \frac{1}{14\sqrt{3}} Q_1^3Q_2^8Q_3^5 - \frac{1}{21} Q_1^4Q_2^4Q_3^5 - \frac{1}{21} Q_1^4Q_2^5Q_3^4 + \frac{1}{42} Q_1^5Q_2^1Q_3^1 + \frac{1}{42} Q_1^5Q_2^2Q_3^2 + \frac{1}{42} Q_1^5Q_2^3Q_3^3 \nonumber \\
& & \mbox{\hglue0.6truecm} + \frac{1}{14\sqrt{3}} Q_1^5Q_2^3Q_3^8 - \frac{1}{21} Q_1^5Q_2^4Q_3^4 - \frac17 Q_1^5Q_2^5Q_3^5 - \frac{1}{21} Q_1^5Q_2^6Q_3^6 - \frac{1}{21} Q_1^5Q_2^7Q_3^7 \nonumber \\
& & \mbox{\hglue0.6truecm} + \frac{1}{14\sqrt{3}} Q_1^5Q_2^8Q_3^3 + \frac{3}{14} Q_1^5Q_2^8Q_3^8 + \frac{1}{14\sqrt{3}} Q_1^6Q_2^2Q_3^8 - \frac{1}{21} Q_1^6Q_2^5Q_3^6 - \frac{1}{21} Q_1^6Q_2^6Q_3^5 \nonumber \\
& & \mbox{\hglue0.6truecm} + \frac{1}{14\sqrt{3}} Q_1^6Q_2^8Q_3^2 + \frac{1}{14\sqrt{3}} Q_1^7Q_2^1Q_3^8 - \frac{1}{21} Q_1^7Q_2^5Q_3^7 - \frac{1}{21} Q_1^7Q_2^7Q_3^5 + \frac{1}{14\sqrt{3}} Q_1^7Q_2^8Q_3^1 \nonumber \\
& & \mbox{\hglue0.6truecm} + \frac{1}{14\sqrt{3}} Q_1^8Q_2^1Q_3^7 + \frac{1}{14\sqrt{3}} Q_1^8Q_2^2Q_3^6 + \frac{1}{14\sqrt{3}} Q_1^8Q_2^3Q_3^5 + \frac{1}{14\sqrt{3}} Q_1^8Q_2^5Q_3^3 + \frac{3}{14} Q_1^8Q_2^5Q_3^8 \nonumber \\
& & \mbox{\hglue0.6truecm} + \frac{1}{14\sqrt{3}} Q_1^8Q_2^6Q_3^2 + \frac{1}{14\sqrt{3}} Q_1^8Q_2^7Q_3^1 + \frac{3}{14} Q_1^8Q_2^8Q_3^5.
\end{eqnarray}


\begin{thebibliography}{99}

\bibitem{gell}
M.~Gell-Mann and Y.~Ne'eman,
{\it The eightfold way}
(W.\ A.\ Benjamin, New York, 1964).

\bibitem{part}
S.~Navas \textit{et al.} [Particle Data Group],
``Review of particle physics,''
Phys.\ Rev.\ D \textbf{110}, 030001 (2024).

\bibitem{oh1}
K.~Ohya, M.~Imoto and S.~Kawasaki,
``Phenomenological approach to non-leptonic decays. II. $\Omega^-$ decays under parameters fit to decays of octet baryons,''
Prog.\ Theor.\ Phys.\ \textbf{39}, 1268 (1968).

\bibitem{tan1}
J.~Tandean and G.~Valencia,
``CP violation in non-leptonic $\Omega^-$ decays,''
Phys.\ Lett.\ B \textbf{451}, 382 (1999).

\bibitem{abli1}
M.~Ablikim \textit{et al.} [BESIII],
``Measurements of the absolute branching fractions of $\Omega^-$ decays and test of the $\Delta I=1/2$ rule,''
Phys.\ Rev.\ D \textbf{108}, L091101 (2023).

\bibitem{bou1}
M.~Bourquin \textit{et al.} [Bristol-Geneva-Heidelberg-Orsay-Rutherford-Strasbourg Collaboration]
``Measurement of $\Omega^-$ decay properties in the CERN SPS hyperon beam,''
Nucl.\ Phys.\ B \textbf{241}, 1 (1984).

\bibitem{leu1}
C.~J.~G.~Mommers and S.~Leupold,
``Estimates for rare three-body decays of the $\Omega^-$ baryon using chiral symmetry and the $\Delta I=1/2$ rule,''
Phys.\ Rev.\ D \textbf{106}, 093001 (2022).

\bibitem{leu2}
M.~Bertilsson and S.~Leupold,
``Goldberger-Treiman relation and Wu-type experiment in the decuplet sector,''
Phys.\ Rev.\ D \textbf{109}, 034028 (2024).

\bibitem{rfm25}
R.~Flores-Mendieta and G.~Sanchez-Almanza,
``Universality of the baryon axial vector current operator in large-$N_c$ chiral perturbation theory,''
[arXiv:2411.19838 [hep-ph]].

\bibitem{djm94}
R.~F.~Dashen, E.~E.~Jenkins and A.~V.~Manohar,
``The $1/N_c$ expansion for baryons,''
Phys.\ Rev.\ D \textbf{49}, 4713 (1994);
[erratum: Phys.\ Rev.\ D \textbf{51}, 2489 (1995).]

\bibitem{djm95}
R.~F.~Dashen, E.~E.~Jenkins and A.~V.~Manohar,
``Spin flavor structure of large $N_c$ baryons,''
Phys.\ Rev.\ D \textbf{51}, 3697 (1995).

\bibitem{jen96}
E.~E.~Jenkins,
``Chiral Lagrangian for baryons in the $1/N_c$ expansion,''
Phys.\ Rev.\ D \textbf{53}, 2625 (1996).

\bibitem{dai}
J.~Dai, R.~F.~Dashen, E.~E.~Jenkins and A.~V.~Manohar,
``Flavor symmetry breaking in the $1/N_c$ expansion,''
Phys.\ Rev.\ D \textbf{53}, 273 (1996).

\bibitem{rfm98}
R.~Flores-Mendieta, E.~E.~Jenkins and A.~V.~Manohar,
``$SU(3)$ symmetry breaking in hyperon semileptonic decays,''
Phys.\ Rev.\ D \textbf{58}, 094028 (1998).

\bibitem{rfm04}
R.~Flores-Mendieta,
``$V_{us}$ from hyperon semileptonic decays,''
Phys.\ Rev.\ D \textbf{70}, 114036 (2004).

\bibitem{rfm17}
R.~Flores-Mendieta and R.~Padr{\'o}n-Stevens,
``Sum rules for leading vector form factors in hyperon semileptonic decays,''
Phys.\ Rev.\ D \textbf{95}, 076018 (2017).

\bibitem{kra}
A.~Krause,
``Baryon matrix elements of the vector current in chiral perturbation theory,''
Helv.\ Phys.\ Acta {\bf 63}, 3 (1990).

\bibitem{and}
J.~Anderson and M.~A.~Luty,
``Chiral corrections to hyperon vector form-factors,''
Phys.\ Rev.\ D {\bf 47}, 4975 (1993).

\bibitem{villa}
G.~Villadoro,
``Chiral corrections to the hyperon vector form factors,''
Phys.\ Rev.\ D {\bf 74}, 014018 (2006).

\bibitem{meiss}
A.~Lacour, B.~Kubis and U.~-G.~Meissner,
``Hyperon decay form-factors in chiral perturbation theory,''
J.\ High Energy Phys.\ {\bf 0710}, 083 (2007).

\bibitem{geng}
L.~S.~Geng, J.~Martin Camalich and M.~J.~Vicente Vacas,
``$SU(3)$-breaking corrections to the hyperon vector coupling $f_1(0)$ in covariant baryon chiral perturbation theory,''
Phys.\ Rev.\ D \textbf{79}, 094022 (2009).

\bibitem{rfm14}
R.~Flores-Mendieta and J.~L.~Goity,
``Baryon vector current in the chiral and $1/N_c$ expansions,''
Phys.\ Rev.\ D \textbf{90}, 114008 (2014).

\bibitem{fer}
I.~P.~Fernando and J.~L.~Goity,
``$SU(3)$ vector currents in baryon chiral perturbation theory combined with the $1/N_c$ expansion,''
Phys.\ Rev.\ D \textbf{101}, 054026 (2020).

\bibitem{gua}
D.~Guadagnoli, V.~Lubicz, M.~Papinutto and S.~Simula,
``First lattice QCD study of the $\Sigma^- \to n$ axial and vector form factors with $SU(3)$ breaking corrections,''
Nucl.\ Phys.\ B \textbf{761}, 63 (2007).

\bibitem{sas1}
S.~Sasaki and T.~Yamazaki,
``Lattice study of flavor $SU(3)$ breaking in hyperon beta decay,''
Phys.\ Rev.\ D \textbf{79}, 074508 (2009).

\bibitem{sas2}
S.~Sasaki,
``Hyperon vector form factor from 2+1 flavor lattice QCD,''
Phys.\ Rev.\ D \textbf{86}, 114502 (2012).

\bibitem{sas3}
S.~Sasaki,
``Continuum limit of hyperon vector coupling $f_1(0)$ from 2+1 flavor domain wall QCD,''
Phys.\ Rev.\ D \textbf{96}, 074509 (2017).

\bibitem{bs}
R.~E.~Behrends and A.~Sirlin,
``Effect of mass splittings on the conserved vector current,''
Phys.\ Rev.\ Lett.\ \textbf{4}, 186 (1960).

\bibitem{ag}
M.~Ademollo and R.~Gatto,
``Nonrenormalization theorem for the strangeness violating vector currents,''
Phys.\ Rev.\ Lett.\ \textbf{13}, 264 (1964).

\bibitem{banda1}
V.~M.~Banda Guzm{\'a}n, R.~Flores-Mendieta, J.~Hern{\'a}ndez and F.~J.~Rosales-Aldape,
``Spin and flavor projection operators in the $SU(2N_f)$ spin-flavor group,''
Phys.\ Rev.\ D \textbf{102}, 036010 (2020).

\bibitem{banda2}
V.~M.~Banda~Guzm{\'a}n, R.~Flores-Mendieta and J.~Hernandez,
``Baryon-meson scattering amplitude in the $1/N_c$ expansion,''
[arXiv:2305.00879 [hep-ph]].

\bibitem{dm1}
R.~F.~Dashen and A.~V.~Manohar,
``Baryon-pion couplings from large $N_c$ QCD,''
Phys.\ Lett.\ B \textbf{315}, 425-430 (1993).

\bibitem{dm2}
R.~F.~Dashen and A.~V.~Manohar,
``$1/N_c$ corrections to the baryon axial currents in QCD,''
Phys.\ Lett.\ B \textbf{315}, 438-440 (1993).

\bibitem{gs1}
J.~Gervais and B.~Sakita,
``Large $N$ QCD baryon dynamics: Exact results from its relation to the static strong coupling theory,''
Phys.\ Rev.\ Lett.\ \textbf{52}, 87 (1984).

\bibitem{gs2}
J.~Gervais and B.~Sakita,
``Large $N$ baryonic soliton and quarks,''
Phys.\ Rev.\ D \textbf{30}, 1795 (1984).

\bibitem{rfm24}
R.~Flores-Mendieta, S.~A.~Garcia-Monreal, L.~R.~Ruiz-Robles and F.~A.~Torres-Bautista,
``Alternative approach to baryon masses in the $1/N_c$ expansion of QCD,''
Phys.\ Rev.\ D \textbf{109}, 114014 (2024).

\bibitem{fubini}
S.~Fubini and G.~Furlan,
``Renormalization effects for partially conserved currents,''
Physics \textbf{1}, 229 (1965).

\bibitem{jm255}
E.~Jenkins and A.~V.~Manohar,
``Baryon chiral perturbation theory using a heavy fermion Lagrangian,''
Phys.\ Lett.\ B {\bf 255}, 558 (1991).

\bibitem{jm259}
E.~Jenkins and A.~V.~Manohar,
``Chiral corrections to the baryon axial currents,''
Phys.\ Lett.\ B {\bf 259}, 353 (1991).

\end{thebibliography}
\end{document}